\documentclass[a4paper,12pt]{article}
\pdfoutput=1
\usepackage{geometry}
\usepackage{amsmath,amssymb,amsfonts}
\usepackage{xcolor,graphicx,soul}
\usepackage{array,booktabs,longtable}
\usepackage{caption,microtype}
\usepackage{subcaption}
\usepackage[colorlinks=true, allcolors=teal]{hyperref}
\usepackage{datetime}
\usepackage{appendix}
\usepackage{mathtools}
\usepackage{makecell}
\usepackage{multirow}

\usepackage[style=ieee,citestyle=numeric-comp]{biblatex} 
\AtEveryBibitem{%
  \clearfield{doi} 
}

\newcommand{\lsim}
{\;\raisebox{-.3em}{$\stackrel{\displaystyle <}{\sim}$}\;}

\newcommand\tb{\tan\beta}

\newcommand\CBA{c_{\beta - \alpha}}
\newcommand\SBA{s_{\beta - \alpha}}

\newcommand\LP{\left(}
\newcommand\RP{\right)}

\renewcommand\Re{\mathop{\mathrm{Re}}}

\newcommand\ReDiag{\mathop{%
  \raise .5pt\hbox{[}%
  \widetilde{\mathrm{Re}}%
  \raise .5pt\hbox{]}}}
\newcommand\ReOffDiag{\mathop{%
  \raise .5pt\hbox{$\llbracket$}%
  \widetilde{\mathrm{Re}}%
  \raise .5pt\hbox{$\rrbracket$}}}

\newcommand\MSbar{\ensuremath{\overline{\mathrm{MS}}}}

\newcommand\SW{s_\mathrm{w}}

\newcommand\MW{m_W}
\newcommand\MZ{m_Z}
\newcommand\Mh{m_h}
\newcommand\MH{m_H}
\newcommand\MA{m_A}
\newcommand\MHp{m_{H^\pm}}

\newcommand\msq{m_{12}^{2}}
\newcommand\mbar{\bar m}
\newcommand\mbarsq{\bar m^2}

\renewcommand\refeq[1]{Eq.~(\ref{#1})}
\newcommand\refeqs[1]{Eqs.~(\ref{#1})}
\newcommand\refta[1]{Tab.~\ref{#1}}
\newcommand\refse[1]{Sect.~\ref{#1}}

\newcommand\citere[1]{Ref.~\cite{#1}}
\newcommand\citeres[1]{Refs.~\cite{#1}}
\newcommand\refap[1]{App.~\ref{#1}}

\newcommand\eg{e.g.\ }
\newcommand\ie{i.e.\ }
\newcommand\wrt{w.r.t.\ }
\newcommand\vs{vs.\ }

\newcommand{\CP}{{\cal CP}}
\newcommand{\cp}{{\CP}}

\newcommand{\tev}{\,\, \mathrm{TeV}}
\newcommand{\gev}{\,\, \mathrm{GeV}}

\newcommand\ab{\ensuremath{\mbox{ab}}}

\newcommand\iab{\ensuremath{\ab^{-1}}}

\newcommand{\br}{\text{BR}}

\def\reffi#1{\mbox{Fig.~\ref{#1}}}
\def\reffis#1{\mbox{Figs.~\ref{#1}}}

\def\la{\lambda}
\newcommand\kala{\ensuremath{\kappa_{\lambda}}}
\newcommand\laSM{\ensuremath{\lambda_{\mathrm{SM}}}}
\newcommand{\lahhh}{\ensuremath{\la_{hhh}}}
\newcommand{\lahhH}{\ensuremath{\la_{hhH}}}
\newcommand{\lahHH}{\ensuremath{\la_{hHH}}}

\newcommand{\Gacorr}{\Gamma^{\mathrm{corr}}}

\definecolor{Orange}{named}{orange}
\definecolor{Purple}{named}{purple}
\definecolor{Lightblue}{cmyk}{0.9,0.1,0.1,0.3}
\definecolor{dgelborange}{cmyk}{0.,0.3,0.5, 0.}
\definecolor{Lila}{rgb}{0.5,0.,1}
\definecolor{Darkgreen}{rgb}{0.,.75,0.}

\newcommand{\Delone}{\Delta^{\!\left(1\right)}}
\newcommand{\lahhhzero}{\ensuremath{\lahhh^{\mathrm{\!\left(0\right)}}}}
\newcommand{\lahhhone}{\ensuremath{\lahhh^{\mathrm{\!\left(1\right)}}}}
\newcommand{\kalazero}{\ensuremath{\kala^{\mathrm{\left(0\right)}}}}
\newcommand{\kalaone}{\ensuremath{\kala^{\mathrm{\left(1\right)}}}}
\newcommand{\lahhHzero}{\ensuremath{\lahhH^{\mathrm{\!\left(0\right)}}}}
\newcommand{\lahhHone}{\ensuremath{\lahhH^{\mathrm{\!\left(1\right)}}}}

\newcommand{\SM}{\mathrm{SM}}

\newcommand{\minv}{m_{hh}}
\newcommand{\mhh}{\ensuremath{m_{hh}}}

\newcommand{\Zdiffzero}{Z_{\mathrm{diff}}^{\left(0\right)}}
\newcommand{\Zdiffone}{Z_{\mathrm{diff}}^{\left(1\right)}}
\newcommand{\Zzero}{Z^{\left(0\right)}}
\newcommand{\Zone}{Z^{\left(1\right)}}

\newcommand{\sigeff}{\sigma_{\mathrm{Eff.Pot.}}^{\left(1\right)}}
\newcommand{\sigmom}{\sigma_{\mathrm{diag.}}^{\left(1\right)}}
\newcommand{\sigtree}{\sigma_{\mathrm{2HDM}}^{\left(0\right)}}
\newcommand{\sigSM}{\sigma_{\mathrm{SM}}^{\left(0\right)}}

\newcommand{\eeZhh}{\ensuremath{e^+e^- \to Zhh}}
\newcommand{\eeZhhbb}{\ensuremath{e^+e^- \to Zhh \to Zb\bar bb\bar b}}
\newcommand{\eenunuhh}{\ensuremath{e^+e^- \to \nu\nu hh}}
\newcommand{\vev}{vev}
\graphicspath{{figs/}}
\allowdisplaybreaks
\begin{document}
\thispagestyle{empty}

\def\thefootnote{\fnsymbol{footnote}}

\begin{flushright}
\mbox{}
DESY-25-073\\
IFT--UAM/CSIC-25-016\\
KA-TP-13-2025
\end{flushright}

\vspace{0.5cm}

\begin{center}

{\large\sc\boldmath
{\bf Large One-Loop Effects of BSM Triple Higgs Couplings on\\[.5em] Double Higgs Production at $e^+e^-$ Colliders}
}

\vspace{1cm}

{\sc
F.~Arco$^{1,2}$%
\footnote{\label{former} Former address.}%
$^{,3}$%
$^{\ref{former}}$%
\footnote{email: Francisco.Arco@desy.de}%
, S.~Heinemeyer$^{4}$%
\footnote{email: Sven.Heinemeyer@cern.ch}%
~and M.~Mühlleitner$^{3}$%
\footnote{email: Milada.Muehlleitner@kit.edu}%
}

\vspace*{.7cm}

{\sl
$^1$
Deutsches Elektronen-Synchrotron DESY,
Notkestr.~85, 22607 Hamburg, Germany

\vspace*{0.1cm}

$^2$
Institute for Astroparticle Physics, Karlsruhe Institute of Technology,
Hermann-von-Helmholtz-Platz 1, 76344 Eggenstein-Leopoldshafen, Germany

\vspace*{0.1cm}

$^3$ 
Institute for Theoretical Physics, Karlsruhe Institute of Technology, \\
Wolfgang-Gaede-Str. 1, 76131 Karlsruhe, Germany

\vspace*{0.1cm}

$^4$Instituto de F\'isica Te\'orica (UAM/CSIC), 
Universidad Aut\'onoma de Madrid, \\
Cantoblanco, 28049, Madrid, Spain

}

\end{center}

\vspace*{0.1cm}

\begin{abstract}
\noindent
The measurement of the Higgs boson self-coupling is crucial 
for our understanding of the nature of electroweak symmetry breaking and potential physics beyond the Standard Model (BSM).
In this work, we study in the framework of the 2-Higgs-Doublet Model (2HDM) the impact of one-loop corrections to triple Higgs couplings (THCs) on the pair production of two Standard Model (SM)-like Higgs bosons $h$ at future high-energy $e^+ e^-$ colliders, focusing on the $e^+ e^- \to Zhh$ process.
By including the one-loop corrections to the THCs relevant for this process, i.e.~the coupling between three SM-like Higgs bosons, $\lambda_{hhh}$, and between the non-SM-like Higgs $H$, assumed to be heavier, and two SM-like Higgs bosons, $\lambda_{hhH}$, we account for the leading one-loop corrections to the di-Higgs production cross section.
We show that the one-loop corrected THC $\lambda_{hhh}$ can be enhanced up to nearly six times its SM value, which substantially enhances the di-Higgs production cross section w.r.t.\ the tree-level prediction, even in the alignment limit.
On the other hand, one-loop corrections to $\lambda_{hhH}$ can also enhance its value, potentially yielding to more prominent heavy Higgs $H$ resonant production. 
We explore the sensitivity to the loop-corrected $\lambda_{hhh}$ and the possible access to $\lambda_{hhH}$ via the $H$ resonant peak at a  future high-energy $e^+e^-$ collider, such as the ILC.
We highlight the fact that including the one-loop corrected THCs can enhance the sensitivity to the $H$ resonant peak, and therefore to $\lambda_{hhH}$.
Finally, we discuss the required experimental precision at future $e^+e^-$ colliders necessary to achieve these sensitivities.

\end{abstract}

\def\thefootnote{\arabic{footnote}}
\setcounter{page}{0}
\setcounter{footnote}{0}

\newpage


\section{Introduction}
\label{sec:intro}

The discovery of the Higgs boson in 2012 by the ATLAS and CMS collaborations~\cite{ATLAS:2012yve,CMS:2012qbp} 
constituted a milestone in particle physics and structurally completed the Standard Model (SM).
To date, all the measured signal strengths for the discovered Higgs boson are consistent with the SM predictions 
within the experimental and theoretical uncertainties~\cite{ATLAS:2022vkf,CMS:2022dwd}.
However, the access to the Higgs-boson self interactions at the LHC is very challenging and only upper
limits could be set on the di-Higgs production cross section, which is used to measure the triple Higgs coupling (THC).
In particular, the ratio of the triple Higgs self-coupling to its SM value at the tree level, 
denoted by $\kala$, is only constrained to be within the interval $-1.2 < \kala < 7.2 $ at 95\% C.L.\  as reported by ATLAS~\cite{ATLAS:2024ish}, and within $-1.4 < \kala < 7.8$ at 95\% C.L.\ as reported by CMS~\cite{CMS:2024awa}.
Therefore, the Higgs-boson potential is so far largely unconstrained, leaving plenty of room for new
physics in the Higgs sector.
Consequently, there have been many phenomenological studies of possible new scalar sectors from extended 
beyond the SM (BSM) Higgs sectors. 
Such models predict the existence of new Higgs bosons and new scalar interactions among them.
A well studied extension of the SM Higgs-boson sector is the 2-Higgs-Doublet Model 
(2HDM)~\cite{Lee:1973iz,Gunion:1989we,Aoki:2009ha,Branco:2011iw}, which consists of adding a second complex
Higgs doublet to the SM  Higgs sector implying the existence of five physical Higgs bosons: two charged Higgs bosons $H^\pm$, two neutral 
$\CP$-even Higgs bosons $h$ and $H$ 
(with $\Mh<\MH$), and one neutral $\CP$-odd scalar $A$.
In this work, the Higgs boson $h$ is identified with the discovered Higgs boson at the LHC, and the 
remaining Higgs bosons are assumed to be heavier. The Higgs boson $h$ will be referred to as SM-like Higgs boson in the following.

In particular, the THC of the  SM-like Higgs-boson, $\lahhh$, in the context of the 2HDM has been 
studied extensively.
At tree level, this coupling can only present deviations of at most -100\% and +20\% w.r.t.\ its SM prediction 
in specific regions of the
parameter space allowed by all theoretical and experimental constraints~\cite{Bernon:2015qea,Arco:2020ucn,Abouabid:2021yvw,Arco:2022xum}. 
On the contrary, the couplings of the SM-like Higgs boson to two other heavier Higgs bosons can be large,
even in the alignment limit. 
In particular, in the 2HDM large mass differences between the heavy Higgs bosons can induce large triple 
and quartic scalar couplings
between Higgs bosons, which are only constrained by the requirement of unitarity~\cite{Arco:2020ucn,Arco:2022xum}.
In turn, such large values of the scalar couplings can induce large loop corrections to the Higgs boson 
self-coupling in extended Higgs sectors.
These higher-order corrections are known and have been studied at the one-loop
level~\cite{Kanemura:2002vm,Kanemura:2004mg,Bahl:2023eau}, 
and even the two-loop level~\cite{Braathen:2019pxr,Braathen:2019zoh,Bahl:2022jnx} in the context of the 2HDM.
Although higher-order corrections to other scalar couplings have been less studied in the literature, they could 
potentially exhibit 
similar higher-order effects as $\lahhh$, since a comparable set of scalar couplings enters in their predictions.
Other extended Higgs sectors are expected to exhibit large higher-order corrections in the case that large scalar
couplings can be realized (see for instance \cite{Aoki:2012jj,Arhrib:2015hoa,Kanemura:2016lkz,Chiang:2018xpl,Bahl:2023eau} and references therein). 
The higher-order corrections to the Higgs potential can have significant phenomenological consequences.
As an example, they are crucial to study the thermal cosmological evolution of BSM models in the 
search for a possible first-order electroweak phase
transition (FOEWPT). 
For instance, it was found in \citeres{Basler:2017uxn,Biekotter:2022kgf} that the region of the parameter 
space in the 2HDM
that leads to a FOEWPT requires values of $\kala$ around~2, which can only be realized at the loop level.

In the investigation of the Higgs sector, not only hadron 
colliders, such as the LHC and the 
HL-LHC will be relevant, but also future $e^+e^-$ colliders will play a crucial 
role~\cite{deBlas:2019rxi,DiMicco:2019ngk}.  
In particular, at energies below about $\sqrt{s} = 1 \tev$ the most relevant process is the double Higgs-strahlung process, \eeZhh. 
At energies above $\sim 1\tev$ $W$ boson fusion into Higgs pairs, \eenunuhh, is the main production channel~\cite{Ilyin:1995iy,Osland:1998hv,Djouadi:1999gv,Muhlleitner:2000jj,Moortgat-Pick:2015lbx,CLICdp:2018cto,Arco:2021bvf}. 
There are several proposals for high-energy $e^+e^-$ colliders, for example, the International Linear Collider
(ILC)~\cite{Bambade:2019fyw}, the Compact Linear Collider (CLIC)~\cite{CLICdp:2018cto}, the Cool Cooper 
Collider (C$^3$)~\cite{Bai:2021rdg}
or the Linear Collider Facility (LCF) at CERN~\cite{LinearCollider:2025lya,LinearColliderVision:2025hlt}. 
Overall, these proposed high-energy linear colliders could  reach a 10-20\% accuracy in the 
measurement of \kala, for the
case $\kala = 1$, see for instance~\citeres{Durig:2016jrs,Abramowicz:2016zbo,Roloff:2019crr}. 
The sensitivity projections for the measurement of a non-SM-like triple Higgs coupling where $\kappa_\lambda \ne 1$ can be found in
\citeres{Torndal:2023fky,Torndal:2023mmr,LinearColliderVision:2025hlt}.%
\footnote{Recently, the ATLAS and CMS collaborations published the projection on the precision of the determination of $\kala$ at the HL-LHC for the case $\kala\neq1$~\cite{CMS:2025hfp}.}
It should be noted that other planned $e^+e^-$ circular colliders, such as the Future Circular Collider (FCC-ee)~\cite{FCC:2018evy} 
or the Circular Electron Positron Collider (CEPC)~\cite{CEPCStudyGroup:2018ghi,CEPCStudyGroup:2023quu}, have only limited 
access to \lahhh,
since it can only be accessed indirectly via higher-order loop corrections to the single Higgs production cross section 
(see for instance \citeres{deBlas:2019rxi,DiMicco:2019ngk}).
Consequently, we focus on high-energy linear $e^+e^-$ colliders in this work.

In this paper, we study the impact of the one-loop corrections to the THCs. In a first step we determine 
the intervals of \lahhh\ and \lahhH, evaluated at the one-loop level, that are allowed by all relevant experimental 
and theoretical constraints (updating the corresponding analyses in \citeres{Arco:2020ucn,Arco:2022xum,Abouabid:2021yvw}).
The main part of our work analyzes the impact of the loop-corrected THCs
on double Higgs production in the channel \eeZhh.
We assume the foreseen ILC operating stages at the center-of-mass energies of 500~GeV and 1~TeV, with their 
anticipated integrated luminosity and polarization of the electron and positron 
beams~\cite{Barklow:2015tja,Bambade:2019fyw}.
The production of two Higgs bosons at $e^+e^-$ colliders at tree-level and the potential access to THCs 
in the 2HDM (and the MSSM) has been studied before in \citeres{Ilyin:1995iy,Osland:1998hv,Djouadi:1999gv,Muhlleitner:2000jj,Arhrib:2008jp,Lopez-Val:2009xtx,Asakawa:2010xj,Kon:2018vmv,Sonmez:2018smv,Arco:2021bvf,Ahmed:2021crg}.
Our computation uses the analytic tree-level formulas from \citeres{Djouadi:1999gv,Muhlleitner:2000jj} 
for the di-Higgs production cross section, adapted to the 2HDM, with the inclusion of the THCs $\lahhh$ and $\lahhH$
evaluated at the one loop level.
We employ the Coleman-Weinberg effective potential approach~\cite{Coleman:1973jx,Weinberg:1973am} 
to obtain the one-loop corrections to the THCs.
In particular, we use the results obtained from the implementation of the 2HDM effective potential in the public code 
{\tt BSMPT}~\cite{Basler:2018cwe,Basler:2020nrq,Basler:2024aaf}.
Moreover, in the case of $\lahhh$, we also perform a diagrammatic one-loop computation  using the public code {\tt anyH3/anyBSM}~\cite{Bahl:2023eau}.
With this full-diagrammatic approach, we can evaluate the effect of the finite-momentum dependence of the one-loop
corrected $\lahhh$ entering the cross section prediction, which is neglected in the effective potential
computation.
The inclusion of the couplings $\lahhh$ and $\lahhH$ at the one-loop level, denoted as
\lahhhone\ and \lahhHone, respectively, in the calculation of \eeZhh\ 
takes into account
the main one-loop electroweak (EW) corrections. We demonstrate in this work that they can 
significantly enhance the production cross section by up to a factor of five w.r.t.\ the tree-level prediction. 
In the context of the (HL-)LHC the effect of BSM one-loop corrected $\lahhhone$ and \lahhHone\ on di-Higgs 
production has been considered in~\cite{Heinemeyer:2024hxa,Heinemeyer:2024vqw}.

To study the sensitivity to the one-loop corrected THCs, we analyze the differential cross section as a 
function of the di-Higgs invariant mass \mhh.
Here we take into account the main decay channel into four $b$-quarks, as well as detector cuts and $b$-tagging 
efficiencies. 
We find that the large one-loop corrections to $\lahhh$ can strongly enhance the non-resonant contributions to the di-Higgs production cross section, even in the alignment limit.
Furthermore,
the obtained \mhh\ distributions exhibit, as expected, a clear peak-dip (or dip-peak) structure around $\mhh = \MH$,
which may yield access to \lahhH.
In this context, several experimental uncertainties have to be considered. The experimental uncertainty in the \mhh\ 
measurement is taken into account by a Gaussian smearing of our theoretical cross section distributions.
On top of that a finite resolution in \mhh\ (i.e.\ a binning in \mhh) has to be considered. 
We quantify the possible sensitivity to the $H$ resonant peak, and thus to \lahhH, by means of the 
significance of the $H$ resonant peak against the no-resonance hypothesis as given by a likelihood profile
ratio~\cite{Cowan:2010js}. Our results indicate that indeed a high-energy $e^+e^-$ collider may give access to 
the BSM THC \lahhH.

\bigskip
The paper is organized as follows:
In \refse{sec:2HDM} we introduce the 2HDM and its scalar couplings, fix our notation, and we discuss how we compute the THCs $\lahhh$ and $\lahhH$ at the one-loop level.
In \refse{sec:scan} we give the currently allowed ranges for the tree-level and one-loop corrected THCs $\lahhh$ and $\lahhH$, taking into account all relevant theoretical and current experimental constraints. 
In \refse{sec:eehhZ} we describe how we compute the di-Higgs cross section in the 2HDM and discuss possible sources of experimental uncertainty, such as detector acceptance, reconstruction efficiency, detector resolution, and cross section binning.
In \refse{sec:diffxs} we analyze the differential distributions of the double Higgs production cross section including one-loop THCs and analyze the potential experimental sensitivity to them.
Finally, in \refse{sec:conclusions} we present our conclusions.


\section{The 2HDM and its Scalar Couplings}
\label{sec:2HDM}

The 2HDM~\cite{Gunion:1989we,Aoki:2009ha,Branco:2011iw} adds a 
second complex Higgs doublet  to the SM Higgs sector.
The renormalizable scalar potential for the two $SU(2)_L$ Higgs doublets, $\Phi_1$ and $\Phi_2$, reads
\begin{equation}
\begin{gathered}
    V^{\left(0\right)}_{\rm 2HDM} = m_{11}^2 \left(\Phi_1^\dagger\Phi_1\right) + m_{22}^2 \left(\Phi_2^\dagger\Phi_2\right) - \left[ m_{12}^2 \left(\Phi_1^\dagger \Phi_2\right) + \mathrm{h.c.}\right]
+ \frac{\la_1}{2} \left(\Phi_1^\dagger \Phi_1\right)^2  + \\\frac{\la_2}{2} \left(\Phi_2^\dagger \Phi_2\right)^2 
 + \la_3 \left(\Phi_1^\dagger \Phi_1\right) \left(\Phi_2^\dagger \Phi_2\right)
  + \la_4 \left(\Phi_1^\dagger \Phi_2\right) \left(\Phi_2^\dagger \Phi_1\right) + 
\left[ \frac{\la_5}{2} \left(\Phi_1^\dagger \Phi_2\right)^2 + \mathrm{h.c.} \right] 
\,.
    \label{eq:pottree}
\end{gathered}
\end{equation}
We assume $\CP$ conservation in the Higgs sector so that all mass and coupling parameters in the potential are real.
Furthermore, the potential in \refeq{eq:pottree} respects a $Z_2$ symmetry 
(with $\Phi_1\to\Phi_1$ and $\Phi_2\to-\Phi_2$), 
which is imposed to avoid flavor-changing neutral currents at tree-level~\cite{Glashow:1976nt,Paschos:1976ay}.

The 2HDM predicts five physical Higgs bosons,  two $\CP$-even Higgs bosons, $h$ and $H$ (with~$\Mh<\MH$), 
one $\CP$-odd Higgs boson, $A$, and a pair of charged Higgs bosons, $H^\pm$.
The angle~$\alpha$ diagonalizes the $\CP$-even sector of the model, while the angle~$\beta$ diagonalizes the 
$\CP$-odd and the charged sectors,
with $\tb$ given by the ratio of the two vacuum expectation values (\vev s) of the two Higgs doublets,
$\tan\beta=v_2/v_1$.
The \vev s satisfy the relation $v_1^2 + v_2^2 =v^2$, where $v=\LP \sqrt 2 \, G_F \RP^{-1/2} \simeq 246.22 \gev$ is the SM \vev.

The 2HDM Higgs couplings to the SM particles are modified w.r.t.~to the corresponding SM Higgs couplings. The modification factors for the $h,H,A$ couplings to the massive massive gauge bosons $V=Z,W$ are given by
\begin{equation}
    \xi_V^h = \SBA\,, \quad	\xi_V^H = \CBA\,, \quad	\xi_V^A=0\,,
    \label{eq:xi-gauge}
\end{equation}
where we have introduced the short-hand notation $s_x=\sin x$ and $c_x=\cos x$.
Due to the assumed $\CP$ conservation, the pseudoscalar does not couple to the massive gauge bosons.
In the Yukawa sector, the imposed $Z_2$ symmetry leads to four distinct Yukawa sectors. The values of the Higgs coupling modification factors $\xi_f^{h,H,A}$ to the fermions (modulo a factor $\gamma_5$ in the pseudoscalar couplings to the fermions) can be expressed as 
\begin{equation}
	\xi_f^h = s_{\beta-\alpha}+\xi_f c_{\beta-\alpha}\,, \quad	\xi_f^H = c_{\beta-\alpha}-\xi_f s_{\beta-\alpha}\,,  \quad \xi_u^A = -i \xi_u \;, \quad	\xi_{d,l}^A = i\xi_{d,l}\;.
\label{eq:xi-yuk}	
\end{equation}
The corresponding $\xi_f$ ($f=u,d,l$) for the up-type, down-type and lepton couplings, respectively, in the four 2HDM Yukawa types are given in \refta{tab:xi}.

\begin{table}[t!]
	\begin{center}
		\begin{tabular}{cccccc}
			\hline
			&& Type I & Type II & \makecell{Type III, Y\\ or Flipped (FL)} 
            & \makecell{Type IV, X\\ or Lepton-Specific (LS)} \\
			\hline 
			$\xi_{u}$ && $\cot\beta$ & $\cot\beta$ & $\cot\beta$ & $\cot\beta$ \\
			$\xi_{d}$ && $\cot\beta$ & $-\tan\beta$ & $-\tan\beta$ & $\cot\beta$ \\ 
			$\xi_{l}$ && $\cot\beta$ & $-\tan\beta$ & $\cot\beta$ & $-\tan\beta$ \\ 
			\hline
		\end{tabular}
		\caption{Values for $\xi_{f}$ in the four $Z_2$ conserving 2HDM Yukawa types. }
		\label{tab:xi}
	\end{center}
\end{table}

As it can be inferred from \refeqs{eq:xi-gauge} and (\ref{eq:xi-yuk}), in the limit $\CBA\to0$ the
light $\CP$-even Higgs boson has the same couplings (at tree-level) to the SM particle content as predicted by the
SM~\cite{Gunion:2002zf}.
Therefore, by keeping $\CBA$ close to zero, one can get a $h$ Higgs boson with similar properties as the SM one.
However, one should keep in mind that there are still non-SM couplings that do not vanish in this so-called {\it alignment limit}, 
such as $ZHA$ or $\gamma H^+H^-$, and even couplings involving the $h$ Higgs boson, such as $hHH$, $hAA$ or $hH^+H^-$.

The potential in \refeq{eq:pottree} introduces eight free parameters in the model.
The minimization conditions allow to relate the parameters $m_{11}^2$ and $m_{22}^2$ with $\tb$ and $v$.
The masses of the physical Higgs bosons can be related to the remaining parameters of the potential, the mixing angles 
$\alpha$ and $\beta$, and the soft-breaking parameter $\msq$  (see e.g.~\citere{Arco:2020ucn} for the explicit expressions).
Therefore, the input parameters of the 2HDM can be given in the so-called ``physical'' basis 
\begin{equation}
    v\,, \quad \Mh\,, \quad	\MH\,, \quad	\MA\,, \quad	\MHp\,, \quad	\tb\,, \quad	\CBA\,, \quad	 \msq\,.
\end{equation}
For later convenience, we define the parameter $\bar m^2$, which is related to the soft-breaking parameter $\msq$ as
\begin{equation}
    \bar m^2 = \frac{\msq}{\sin\beta\cos\beta}\,.
\end{equation}
In this paper, we always identify the light $\CP$-even Higgs boson with the Higgs boson discovered at the LHC, and therefore we set $\Mh=125.25\gev$~\cite{ParticleDataGroup:2022pth}.
This, together with the fact that $v$ is also fixed, leads to a total of six free parameters in our analysis.

\subsection{Tree-Level Scalar Higgs Couplings}
\label{sec:treelevel}

The potential from \refeq{eq:pottree} leads to new triple scalar interactions among the physical Higgs bosons that we denote generically as:
\begin{equation}
    \mathcal{L}_{\mathrm{scalar}} \supset  - v \sum_{i,j,k} \lambda_{h_ih_jh_k}^{\LP0\RP} h_ih_jh_k\,,
    \label{eq:THC-lag}
\end{equation}
where $h_{i,j,k}$ refers to the five physical Higgs bosons present in the 2HDM.
We will refer to the parameter $\lambda_{h_ih_jh_k}^{(0)}$ as the tree-level triple Higgs coupling (THC) between the Higgs bosons $h_i$, $h_j$ and $h_k$.
We adopt this convention to resemble the SM Higgs self-interaction, where $ \mathcal{L}_{\mathrm{scalar}}^{\SM} \supset - v \lambda_{\SM}^{\LP0\RP}  h_\SM^3$ with $h_\SM$ being the SM Higgs boson and 
\begin{equation}
    \lambda_\SM^{(0)}=\frac{m_{h_\SM}^2}{2v^2}\simeq0.13\,,
\end{equation}
where we assume $\Mh=m_{h_\SM}$ in this work.
The Feynman rule for the interaction between three Higgs bosons in terms of this THC then reads
\begin{equation}
    h_ih_jh_k:
    \qquad -ivn!\;\lambda^{(0)}_{h_ih_jh_k}\,,
\label{eq:THC-tree}
\end{equation}
where $n$ is the number of identical particles in the vertex.

We provide the 2HDM tree-level predictions for the couplings $\lahhhzero$ and $\lahhHzero$ since they  enter the $\eeZhh$ cross section,
\begin{gather}
	\begin{gathered}
	\lahhhzero =  \frac{1}{2v^2}\left(\SBA^3 m_h^2+\SBA \CBA^2 \left(3 m_h^2-2 \bar{m}^2\right)  + 2 \CBA^3 \cot2\beta \left(m_h^2-\bar{m}^2\right)\right), 
	\label{eq:lahhh}
	\end{gathered}
        \\
	\begin{gathered}
		\lahhHzero  = \frac{-\CBA }{2v^2}  \left(\SBA^2 \left(2 m_h^2+m_H^2-4 \bar{m}^2\right)  +  
		2 \SBA \CBA \cot2\beta  \left(2 m_h^2+m_H^2-3 \bar{m}^2\right)  \right. \\  \left.  -\CBA^2 \left(2 m_h^2+m_H^2-2 \bar{m}^2\right)\right)\,.
		\label{eq:lahhH}
	\end{gathered}
\end{gather}
It should be noted that in the alignment limit, one finds $\lahhhzero=\lambda_\SM^{\LP0\RP}$ and 
$\lahhHzero=0$. 
This can change, however, when one-loop corrections to these couplings are considered,
as 
will be discussed in the following sections.

In our evaluation of the one-loop corrected THCs, denoted by $\lahhhone$ and $\lahhHone$ in the following, all tree-level triple and quartic scalar couplings involving the $h$ and $H$ bosons can play a relevant role in the prediction. 
Therefore, for reference, we provide the Feynman rules of such interactions in the alignment limit (\ie $\CBA=0$), 
\begin{align}
    hhh/v = hhhh:&\qquad \frac{-3i\Mh^2}{v^2}=-6i\lambda_\mathrm{SM}^{\left(0\right)} \,,\label{eq:THC1}\\
    hhH=hhhH:&\qquad 0\,,\label{eq:THC2}\\
    hHH/v=hhHH:&\qquad \frac{-i\left(m_h^2 + 2m_H^2 - 2\bar m^2 \right)}{v^2} \,,\label{eq:THC3}\\
    h\phi\phi/v=hh\phi\phi:&\qquad \frac{-i\left(m_h^2 + 2m_\phi^2 - 2\bar m^2 \right)}{v^2} \,,\label{eq:THC4}\\
    HHH/(3v)=hHHH/3=H\phi\phi/v=hH\phi\phi:&\qquad \frac{2i\left(\MH^2-\bar m^2\right)\cot2\beta}{v^2}\,,\label{eq:THC5}
\end{align}
where $\phi$ refers to $A$ or $H^\pm$.
For a complete set of Feynman rules for triple and quartic Higgs couplings in the $\CP$-conserving 2HDM outside the alignment limit, see App.~A of \citere{ArcoGarcia:2023zjz}.


\subsection{One-Loop Triple Higgs Couplings}
\label{sec:1LTHCs}
The one-loop corrected triple Higgs couplings in the 2HDM can be obtained in the limit of vanishing external momenta from the effective potential, or alternatively, in the diagrammatic approach which includes the finite momentum effects. We use the loop-corrected couplings obtained in both approaches. 

In order to get the loop-corrected THCs in the effective potential approach we use the public code {\tt BSMPTv3}~\cite{Basler:2018cwe,Basler:2020nrq,Basler:2024aaf}, where the trilinear Higgs self-couplings are calculated from the third derivatives of the one-loop corrected effective potential with respect to the corresponding scalar fields. In {\tt BSMPT}, the $\overline{\mbox{MS}}$-renormalized effective potential at the renormalization scale $\mu = v \approx 246.22$~GeV receives additional finite counterterms, which in an on-shell-like renormalization scheme fix the mixing matrix elements in the Higgs mass matrices to their tree-level values, such that the masses and Higgs coupling modification factors are renormalized to their leading-order values. The trilinear Higgs self-couplings, however, obtain non-zero loop corrections. For details, we refer the reader to \citere{Basler:2018cwe}. 

The computation of the loop-corrected THCs in the diagrammatic approach is done with the code {\tt anyH3/anyBSM}~\cite{Bahl:2023eau}, which evaluates in a semi-automatic way
the THC of the SM-like Higgs boson in a plethora of BSM models with extended Higgs sectors, including the 2HDM.
The result is obtained in the 't~Hooft-Feynman gauge. The masses are renormalized on-shell, and for the mixing angles $\alpha$ and $\beta$ the scheme described in \cite{Kanemura:2004mg} is applied.
The parameter $\msq$ is renormalized in the $\MSbar$ scheme.
The VEV counterterm is derived from the counteterms of $m_W$, $m_Z$ and the electric charge $e$, applying its on-shell relation to these quantities. The electric charge is renormalized on-shell in the Thomson limit.
For details, we refer to \citere{Bahl:2023eau}.

We will use the definition $\kala^{(1)}$ as the value of the one-loop corrected coupling $\lahhh^{(1)}$ with respect to the SM tree-level value $\lambda_{\text{SM}}^{(0)}$, 
\begin{equation}
    \kalaone = \frac{\lahhh^{\left(1\right)}}{\lambda^{\left(0\right)}_{\rm SM}}\,,
    \label{eq:kala}
\end{equation}
which corresponds to the variable used in the experimental analyses.

Due to the simplicity of the formulae, we give some sample results here in the effective potential approach. In the SM, the main one-loop correction to the $h_{\rm SM}$ triple self-coupling is the 
top-mediated loop diagram, which leads to~\cite{Kanemura:2002vm}
\begin{equation}
    \lambda_{\mathrm{SM}}^{\left(1\right)} =  \lambda_{\mathrm{SM}}^{\left(0\right)} \left( 1 -\frac{m_t^4}{\pi^2m_{h_\text{SM}}^2v^2} \right)\,.
\end{equation}
This implies a correction of about -9\% for $\kala^{(1)}$ in the SM.
In addition to this, in the 2HDM this coupling can receive new loop-corrections from the new scalar interactions between three and four Higgs bosons.
The main contributions in the alignment limit can be written as~\cite{Kanemura:2002vm}
\begin{equation}
    \lahhhone \simeq \lambda_\SM^{\left(0\right)} \left( 1 - \frac{m_t^4}{\pi^2 m_{h}^2 v^2} + \sum_{\phi=H,A,H^\pm} \frac{n_\phi m_\phi^4}{12 \pi^2 m_{h}^2 v^2} \left( 1 - \frac{\bar m^2}{m_\phi^2} \right)^3  \right) \,,
\end{equation}
with $n_H=n_A=1$ and $n_{H^\pm}=2$. 
The first term is the top loop contribution, as in the SM, but the remaining terms are pure Higgs contributions.
In particular, in the limit where $\MH,\MA,\MHp\gg\bar m$, corresponding to large scalar couplings 
(see \refse{sec:treelevel}), 
these purely scalar corrections can become very large and be well above the size of the top-quark contributions.


\section{Currently Allowed Values for the THCs \texorpdfstring{\boldmath{$\lahhh$}}{lahhh} and 
\texorpdfstring{\boldmath{$\lahhH$}}{lahhH}}
\label{sec:scan}

In this section, we explore how large the one-loop corrections to the THCs $\lahhh$ and $\lahhH$ can be in the 2HDM 
while respecting all relevant theoretical and experimental constraints of the model.
Moreover, we discuss in which regions of the parameter space we find  large
one-loop corrections, and which class of loop corrections are responsible for them.
Concerning the higher-order corrections, in this section, we concentrate on the results for the one-loop 
THCs obtained in the effective potential approach (as described in \refse{sec:1LTHCs}).


\subsection{Theoretical and Experimental Constraints}
\label{sec:constraints}

To evaluate the currently allowed ranges for the THCs $\lahhh$ and $\lahhH$ (at tree-level and at  one-loop), 
we performed a scan of the 2HDM parameter space by using the public code {\tt ScannerS}~\cite{Coimbra:2013qq,Muhlleitner:2020wwk}.
In the following, we summarize the main constraints on the 2HDM and how they are applied by {\tt ScannerS}:

\begin{itemize}
    \item {\bf Electroweak precision data:} 
    The oblique parameters $S$, $T$ and $U$ are a common way to parametrize BSM radiative corrections to the EW gauge boson self-energies.
    We use the $2\sigma$ allowed region given by the reported values by the PDG23 fit~\cite{ParticleDataGroup:2022pth}:
    $S=-0.02\pm0.10$,
    $T=0.03\pm0.12$ and
    $U=0.01\pm0.11$, 
    with correlations 
    $\rho_{ST}=0.92$,
    $\rho_{SU}=-0.80$ and
    $\rho_{TU}=-0.93$.
    The most constraining parameter is $T$, which can receive large corrections in the 2HDM.
    To keep these corrections small, the mass of one of the neutral Higgs bosons, $\MH$ and/or $\MA$, should be close to the mass of the charged Higgs boson, \ie $\MH\sim\MHp$ and/or $\MA\sim\MHp$~\cite{Bertolini:1985ia}.

    \item {\bf Tree-level perturbative unitarity:} We require that the eigenvalues of the $2\to 2$ scalar scattering
    matrix for the lowest term in the partial wave expansion are below the unitarity limit in the large energy limit.
    This leads to the condition $\left|\Re\LP a_0\RP \right| < 1/2$ (see~\citeres{Akeroyd:2000wc,Ginzburg:2005dt} for 
    the explicit expressions for $a_0$).

    \item {\bf Potential stability:} We impose the bounded-from-below condition to the potential, 
    i.e.\ we require that it does not go to minus infinity in any direction in the $\Phi_1$--$\Phi_2$ 
    plane~\cite{Deshpande:1977rw}.
    In addition, we also require that the EW minimum is a global minimum of the potential~\cite{Barroso:2013awa}.
    We do not consider the possibility that the EW minimum is meta-stable with a tunneling time larger than the age of the universe, which could enlarge the allowed region of the model.

    \item {\bf BSM Higgs boson searches:}  We consider all current searches for BSM Higgs bosons at the LHC, Tevatron
    and LEP, which typically impose bounds on the production cross sections times branching ratio 
    of these new states at the 95\% confidence level (CL).
    For each point in the parameter space, we identify the most sensitive channel according to the expected experimental bound and then check whether or not it is excluded by the experimental result.
    We impose these bounds with the help of the public code 
    {\tt HiggsBounds}~\cite{Bechtle:2008jh,Bechtle:2011sb,Bechtle:2013wla,Bechtle:2015pma,Bechtle:2020pkv}, 
    as part of {\tt HiggsTools}~\cite{Bahl:2022igd}. 

    \item {\bf Signal strength measurements for the SM-like Higgs boson:} We require statistical compatibility between 
    the signal strength measurements for the discovered Higgs boson to date, and the model predictions for the
    light $\CP$-even Higgs boson $h$.
    We perform a $\chi^2$ statistical analysis between the 2HDM predicted and experimentally measured signal strengths 
    of the SM-like Higgs boson, and for simplicity we require compatibility with the SM prediction at the $2\sigma$ level
    (\ie $\Delta\chi^2\leq6.18$ in the case of a two-dimensional plane), where the result of the SM fit 
    is $\chi_\SM^2=152.54$ with 159 observables.
    For this we use the code {\tt HiggsSignals}~\cite{Bechtle:2013xfa,Bechtle:2014ewa,Bechtle:2020uwn}, 
    as a part of {\tt HiggsTools}~\cite{Bahl:2022igd}.

    The 2HDM predictions for the branching ratios of the 2HDM Higgs bosons required by {\tt HiggsTools} as input are obtained with the public code {\tt HDECAY6.60}~\cite{Djouadi:1997yw,Djouadi:2018xqq}, including the state-of-the-art higher-order QCD corrections and off-shell decays. The production cross sections are computed internally in \texttt{HiggsSignals}, respectively, \texttt{HiggsTools}, applying the appropriate coupling rescaling factors.

    \item {\bf Flavor Observables:} 
    Flavor-changing neutral-current processes can receive additional contributions in the 2DHM from the charged Higgs bosons in the loops.
    To take them into account we use the results of \citere{Haller:2018nnx} at the $2\sigma$ level in the 
    $\MHp$--$\tb$ plane.
    
\end{itemize}


\subsection{Allowed Ranges for \texorpdfstring{\boldmath{$\lahhh$}}{lahhh} and 
\texorpdfstring{\boldmath{$\lahhH$}}{lahhH}}
\label{sec:allranges}

In this section we present and discuss the allowed ranges for the THCs $\lahhh$ and $\lahhH$ at tree level and at one-loop level. 
We generated 400,000 points allowed by the previously discussed constraints, where the input parameters of the 2HDM  were varied randomly inside the following ranges,
\begin{equation}
\begin{gathered}
    \MH \in [125.25, 1500]\gev\,, \quad \MA \in [10, 1500]\gev\,, \quad \MHp \in [m_{\rm min}, 1500]\gev\,, \\
    \tb \in [0.1, 30]\,, \quad \CBA \in [-c_{\rm lim},c_{\rm lim}]\,, \quad \msq \in [0, 4\times 10^6] \gev\,.
\end{gathered}
\end{equation}
The value of $m_{\rm min}$ is set to 590 GeV for types~II and~FL to  apply the constraint of 
$\MHp\gtrsim 600\gev$ from the flavor-changing decay $b\to s\gamma$ \cite{Haller:2018nnx,Hermann:2012fc,Misiak:2015xwa,Misiak:2017bgg,Misiak:2020vlo}. 
For types~I and~LS, we set $m_{\rm min}=10\gev$ (even though these low values for the mass of the charged Higgs boson 
are strongly disfavored by direct searches).
We set $c_{\rm lim}=0.35$ for type~I, since in this type larger deviations from the alignment limits are 
experimentally allowed,
while we set $c_{\rm lim}=0.2$ in type~LS and $c_{\rm lim}=0.1$ for the remaining types~II and~FL.
We also performed a dedicated scan where we generated additional 150,000 points with $\CBA=0$ to fully 
explore the  alignment limit. 
This increases the number of total points in our scan to 550,000 for each 2HDM type.

\begin{table}[t!]
\begin{center}
    \begin{tabular}{c||c|c|c||c|c|c} 
		Type	&	$\kalazero$	&  $\kalaone$  &   $\Delone\kala$    &	$\lahhHzero$ &   $\lahhHone$ &   $\Delone\lahhH$  \\ \hline
        I		& 	[-0.2, 1.2]	&	[0.2, 6.8]	&  [-0.3, 5.8]  &	[-1.6, 1.5]	&	[-2.1, 1.9]  & [-1.0, 1.1] \\ 
		II		&	[0.6, 1.0]	&	[0.7, 5.8]	&   [-0.1, 4.7]  &	[-1.5, 1.6]	&	[-1.7, 2.0]    &   [-0.4, 0.6] \\ 
		LS		&	[0.5, 1.0]	&	[0.6, 6.3] &   [-0.1, 5.3]	&	[-1.7, 1.7]	&	[-2.2, 2.1] &   [-0.6, 0.7] \\ 
		FL		&	[0.7, 1.0]	&	[0.8, 5.8]	&   [-0.1, 4.8]  &	[-1.6, 1.3]	&	[-1.9, 1.5] &   [-0.5, 0.4]
	\end{tabular}
\caption{Allowed ranges for $\kappa_\lambda$ and $\lambda_{hhH}$ at tree level and at one-loop level and the differences between their respective loop-corrected and tree-level values as defined in the text, for the four 2HDM types. }
        \label{tab:THCranges}
\end{center}
\end{table}

The allowed ranges for $\kala$ and $\lahhH$ obtained from the scan taking into account the previously described constraints, are shown in \refta{tab:THCranges}.
We also show the allowed ranges for the difference between the one-loop and the tree-level predictions, namely 
\begin{eqnarray}
\Delone\kala=\kalaone-\kalazero \quad \mbox{and} \quad \Delone\lahhH=\lahhHone-\lahhHzero\,,
\end{eqnarray} to quantify the size of the one-loop 
corrections.

The obtained allowed ranges for the tree-level couplings $\lahhhzero$ and $\lahhHzero$ are compatible with the results
published in \citeres{Arco:2020ucn,Arco:2022xum,Abouabid:2021yvw}, with subtle differences due to the slight changes in
the constraints applied in each work.
These results were extensively discussed in those works, but we briefly review them here for completeness.
Type~I is the only type that allows negative values for \kalazero\ down to $-0.2$, along with values above 1, up to 1.2. 
This is because in type I, relatively large values of $\CBA$ and small masses are still allowed for the non-SM-like Higgs boson.
In the other types, the allowed range for $\kalazero$ is between $\sim0.5$ and exactly 1.0 (corresponding to the alignment limit).
For $\lahhHzero$, values between $\sim\pm1.5$ are allowed in all types.
These minimal and maximal values are obtained for Higgs boson masses above 1.2 TeV, $\CBA\sim\pm0.05$ and values of $\tb\sim1$.

The allowed values for \kala\ change in a relevant way due to the one-loop corrections to $\kala$. 
It can be seen that very large values for $\kalaone$, up to $\sim5.8$, can be realized in all types.
These large corrections are found close to the alignment limit for $\MH\sim\mbar\ll\MA\sim\MHp$,  which leads to large
triple and quartic couplings between the SM-like Higgs boson $h$ and the heavy Higgs bosons $A$ and $H^\pm$ 
(see \refeqs{eq:THC3} to  (\ref{eq:THC5})).
In type~I and~LS, even larger values of up to $\sim6.5$ are also allowed for $\kalaone$,
since in these types even lower values of $\MH$ are still allowed, around $200\gev$.
This in turn leads to large values for the relevant scalar couplings for smaller values of $\MA$ and $\MHp$. 
On the other hand, no negative values of \kalaone\ are found, as the one-loop corrections are in general positive.
In principle, negative one-loop contributions to $\kala$ are possible, but they are found to be small.
The reason is that negative values of the relevant scalar couplings for the one-loop corrections to $\kala$ in the $\CP$-conserving 2HDM usually lead to  an unstable potential,
and the THCs appear in the third power.
Therefore, only positive sizable corrections can be realized for $\kalaone$.

Let us highlight that values of $\CBA$ far from the alignment limit do not imply larger one-loop corrections to $\kala$,
since the maximum values allowed for the relevant triple and quartic Higgs couplings depend only mildly on this
parameter. Thus, the 2HDM can predict a very SM-like boson $h$ in all the couplings to fermions and the EW gauge bosons,
while yielding a value for $\kalaone$ much larger than~1~\cite{Kanemura:2002vm,Kanemura:2004mg}. 

Finally, we discuss the one-loop predictions for $\lahhH$.
We find the maximum and minimum allowed values of $\lahhHone$ for very heavy nearly degenerate Higgs boson masses 
(close to our scan limit) and slightly away from the alignment limit, with values of $\CBA\sim\pm0.03$.
The value of $\mbar$ is close to the values of the heavy Higgs boson masses, but a small splitting between $\mbar$ 
and $\MH$ helps to enhance the size of the one-loop corrections.
The reason is that in the alignment limit the relevant couplings of $H$ to other Higgs bosons depend on 
$\MH^2-\mbarsq$. Since these contributions are found close to the alignment limit, these large corrections can be realized in all four 
2HDM types. These large values of $\lahhHone$ are also a consequence of large tree-level
values $\lahhHzero$.
In other words, a large value of $\lahhHone$ does not have to coincide with a large value of $\Delone\lahhH$.
For example, in type~I, the relative one-loop corrections $\Delone\lahhH$ can reach values between  $\pm 1$.
This happens at low values for $\MH \sim \MHp \sim \mbar < 200 \gev$ and large and negative values of $\CBA$.
In the other types, $\Delone\lahhH$ only ranges between approximately $\pm 0.5$, where the limits are found to be 
slightly away from the alignment limit, but with lighter nearly degenerate heavy Higgs bosons with masses around 1~TeV.
It is worth mentioning that the largest one-loop corrections to $\lahhHone$ occur in a different region of the 2HDM parameter space compared to the one-loop corrections to $\kalaone$.
This is due to the fact that the set of (scalar) couplings producing the largest corrections is different. For $\kalaone$ the important couplings are $h\phi\phi$ and $hh\phi\phi$, with $\phi=A$ or $H^\pm$, which can be large for $\MH\sim\mbar<\MA\sim\MHp$, while the computation of the correction $\lahhHone$ always contains a coupling to $H$,  like $hhH$, $HHH$ or $hH\phi\phi$, which are small if $\MH\sim\mbar$.

It should be noticed that, even though the relative one-loop corrections $\Delta^{(1)} \lahhH$ seem smaller than $\Delta^{(1)} \kala$, this is an artifact of the definition of $\kalaone$ \vs $\lahhHone$. 
While the corrections to $\lahhh$ are usually larger than to $\lahhH$, the shift between the tree-level and one-loop
predictions for $\lahhh$ and $\lahhH$ are comparable in many cases.
For instance, one can compare the allowed maximum and minimum values for $\Delone\lahhh$ and $\Delone\lahhH$.
By looking at the allowed ranges for $\Delone\lahhh$ (which would mean to multiply $\Delone\kala$ by $\laSM^{\LP0\RP}\simeq 0.13$),
one can see that they are between $\sim 0$ and $\sim 0.7$, which is comparable to the maximum allowed relative one-loop 
corrections $\Delone\lahhH$.


\section{Di-Higgs Production with One-Loop Corrected THCs}
\label{sec:eehhZ} 

\subsection{Cross Section Calculation}
\label{sec:calc}

\begin{figure}[t!]
    \centering
    \subfloat{
    \includegraphics[scale=0.7]{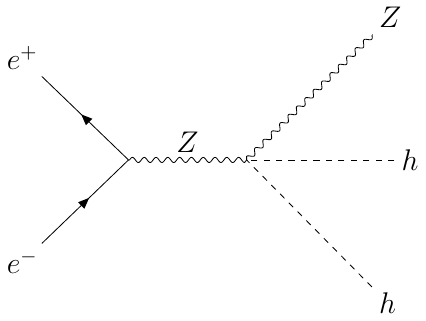} \vspace{-4.5mm}
    }\hspace{1mm}%
    \subfloat{
    \includegraphics[scale=0.7]{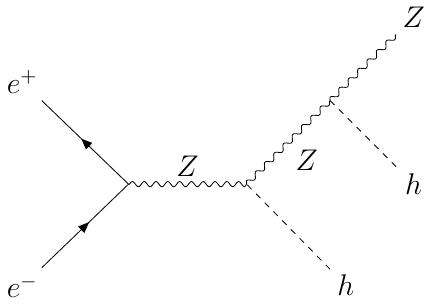} }\hspace{1mm}%
    \subfloat{
    \includegraphics[scale=0.7]{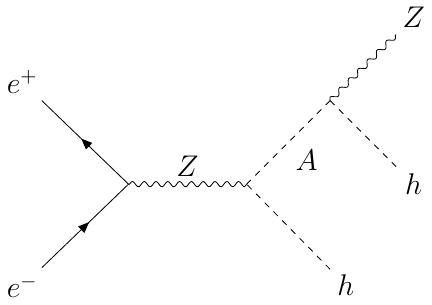} 
    } \\
    \includegraphics[scale=0.9]{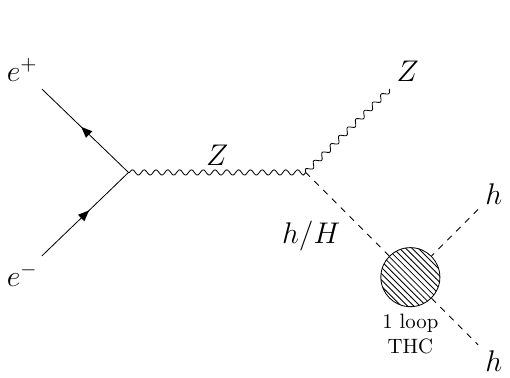}
    \caption{Feynman diagrams contributing to the \eeZhh\ cross section in the 2HDM. 
    The blob represents the one-loop corrected triple Higgs couplings $\kalaone$ and $\lahhHone$ for the $h$ and $H$-mediated diagrams, respectively. }
    \label{fig:diagram}
\end{figure}

The main focus of this paper is the analysis of THCs via the
production of two SM-like Higgs bosons at $e^+e^-$ colliders.
For ILC energies, i.e.\ $\sqrt{s} \le 1000 \gev$,
the main di-Higgs production channel is 
double Higgs-strahlung \eeZhh.
This process is key to the future measurement of the THC of the SM-like Higgs boson, since this coupling 
enters at the tree level in the cross section prediction~\cite{Durig:2016jrs}.
In the 2HDM, the scalar coupling $\lahhH$ also enters in the cross section prediction at the tree level,
with a possible resonant intermediate heavy Higgs boson $H$.
The Feynman diagrams contributing to the process are depicted in \reffi{fig:diagram}.%
\footnote{These diagrams were plotted with {\tt tikz-feynman}~\cite{Ellis:2016jkw}.}
The calculation also includes the diagrams with the identical final state Higgs bosons $h$ exchanged 
in the second and third diagrams.
As sketched in the last diagram with a blob, we introduce the one-loop corrected THCs $\lahhhone$ and $\lahhHone$,
computed with the methods described in \refse{sec:1LTHCs}. 

The inclusion of $\lahhhone$ and $\lahhHone$ in the cross section prediction includes the full subset of the 
one-loop purely scalar corrections.
They are expected to be the most relevant ones in the 2HDM (or in other BSM models with extended Higgs sectors), 
because in general scalar couplings in the 2HDM can be much larger than the EW gauge coupling $g$.
Other one-loop corrections to the cross section with potentially large scalar couplings are expected to be subleading in comparison, since they would have at least a factor of $g$ coming from a coupling between Higgs bosons and the $Z$ boson.

As in the SM, the diagram with $\kala$ (the bottom one in \reffi{fig:diagram} mediated by $h$) has a constructive
interference with the rest of the SM-like contributions (see for example~\citeres{DiVita:2017vrr,Torndal:2023fky}).
Therefore, it is expected that the allowed deviations of $\kalaone > 1$ that can be realized in the 2HDM at the 
one-loop level (see \refse{sec:scan}) lead to an increase of the double Higgs-strahlung cross section.
Consequently, this would imply a higher accuracy in the experimental measurement of the cross section and 
also of the SM-like Higgs self-coupling, as discussed in \citeres{Durig:2016jrs,Torndal:2023fky,Torndal:2023mmr}.

For our calculation we use the analytic formulas for the unpolarized differential cross section 
from~\citeres{Djouadi:1999gv,Muhlleitner:2000jj}. They were derived in the MSSM, and we
adapted them to the 2HDM case.
The differential cross section is given in terms of the reduced energies of the Higgs bosons, defined as $x_{1,2}=2E_{1,2}/\sqrt{s}$.
To obtain the differential cross section with respect to the invariant mass $m_{hh}$ of the two Higgs bosons, we performed a change of integration variables. 

The polarization of the initial electron-positron beams plays an important role in the process, 
since it can significantly enhance the production cross section.
In the particular case of the process \eeZhh, the tree-level polarized cross section can be related to 
the unpolarized cross section $\sigma_{\rm unpol}$  as
\begin{equation}
    \sigma_{RL} = \frac{4 g_R^2}{g_L^2 + g_R^2}\, \sigma_{\rm unpol}\simeq 1.57\,\sigma_{\rm unpol} \,,\ \ \ \ \ \ \  
    \sigma_{LR} = \frac{4 g_L^2}{g_L^2 + g_R^2}\, \sigma_{\rm unpol}\simeq 2.43\, \sigma_{\rm unpol} \,,
    \label{eq:xspol}
\end{equation}
where $\sigma_{RL,LR}$ refers to 100\% right(left)-handed polarized electrons and 100\%
left(right)-handed polarized positrons, and $g_{L,R}$ are the left (right) couplings of the electrons to the $Z$ boson, 
\ie $g_L=-1/2+\SW^2$ and $g_R=\SW^2$.
The same-polarization cross sections $\sigma_{RR}$ and $\sigma_{LL}$ are zero because the spin of the $e^+e^-$ pair 
must add up to one to produce a $Z$ boson in the $s$-channel.
The derivation of the expressions in \refeq{eq:xspol} can be found in \refap{app:pol}.

In the case of partially-polarized $e^-e^+$ beams, the cross section is given by~\cite{Moortgat-Pick:2005jsx}
\begin{equation}
    \sigma\!\left(P_{e^-}, P_{e^+}\right) = \frac{1}{4}\Bigr[ \left(1+P_{e^-}\right)\left(1-P_{e^+}\right)\sigma_{RL} + \left(1-P_{e^-}\right)\left(1+P_{e^+}\right)\sigma_{LR} \Bigr]\,,
\end{equation}
where $P_{e^-,e^+}$ denotes the electron and positron polarization, respectively.
The baseline design for the ILC foresees a maximum polarization of $\left|P_{e^-}\right|=80\%$ for electrons and $\left|P_{e^+}\right|=30\%$ for positrons \cite{Bambade:2019fyw}.
The possible two beam polarizations with opposite sign yield the following cross sections,
\begin{align}
    \sigma\!\left(-80\%, +30\%\right) = 0.035 \sigma_{RL} + 0.585 \sigma_{LR} \simeq 1.476\, \sigma_{\rm unpol} \,, \\
    \sigma\!\left(+80\%, -30\%\right) = 0.585 \sigma_{RL} + 0.035 \sigma_{LR} \simeq 1.004\, \sigma_{\rm unpol} \,,
\end{align}
where we have used $\SW^2 = 1-\LP\MW/\MZ\RP^2\simeq0.223$, with the values of $\MW$ and $\MZ$ from~\citere{ParticleDataGroup:2022pth}.
The consideration of the beam polarization constitutes one key difference with respect to the tree-level analysis of \citere{Arco:2021bvf}, where only unpolarized cross sections for the di-Higgs production were considered.
The foreseen ILC operating stages relevant for our work, and the corresponding polarized and unpolarized cross section prediction in the SM, can be found in \refta{tab:ILC}.

The total widths for the $H$ and $A$ Higgs bosons also enter in the total cross section production. 
We compute them with the public code {\tt HDECAY}.
Furthermore, we have included the effect of one-loop corrected THCs in the allowed Higgs-to-Higgs decays, 
\ie for $h\to AA$, $H\to AA$ and $H\to hh$.
For instance, for a generic Higgs-to-Higgs decay $\phi\to\chi\chi$, we rescale the partial decay width as
\begin{equation}
    \Gamma^{\left(1\right)}\!\left(\phi\to\chi\chi\right) = \left(\frac{\lambda_{\phi\chi\chi}^{\left(1\right)}}{\lambda_{\phi\chi\chi}^{\left(0\right)}}\right)^{\!2} \Gamma^{\left(0\right)}\!\left(\phi\to\chi\chi\right)\,,
\end{equation}
where we take the one-loop THC $\lambda_{\phi\chi\chi}^{\left(1\right)}$ from the effective potential
approach (see \refse{sec:1LTHCs}).
The change of these partial widths implies also a change in the total width of the $\phi$ boson.
We refer to this ``corrected'' decay width as $\Gacorr_\phi$, and it can be derived from the original decay
width $\Gamma_\phi$ as
\begin{equation}
    \Gacorr_\phi=\Gamma_\phi + \sum_\chi \left[ \Gamma^{\left(1\right)}\!\left(\phi\to\chi\chi\right) - \Gamma^{\left(0\right)}\!\left(\phi\to\chi\chi\right) \right]\,.
    \label{eq:Gacorr}
\end{equation}
In this way, we include the main scalar corrections to double Higgs production cross section also in the decay
width of the Higgs bosons.

It is important to keep in mind
that in the alignment limit the only possible BSM effect that one can expect in our computation 
arises entirely from the one-loop contributions to the coupling $\lahhh$.
Even if the coupling $\lahhHone$ is different from zero in the alignment limit, which is possible in general, 
the coupling $HZZ$ is always zero (see \refeq{eq:xi-gauge}).
A  complete one-loop computation would also include a BSM effect from the $H$-mediated diagram 
involving a one-loop corrected $HZZ$ coupling. 
In consequence, only at the two-loop level a non-vanishing $H$-resonance production can be generated 
in the exact alignment limit, 
as it involves a one-loop correction to $\lahhH$ as well as the loop-corrected $HZZ$ coupling.

\begin{table}[tb!]
    \centering
    \begin{tabular}{ccccc} \hline 
        $\sqrt{s}$ [GeV]    &   ${\cal L}_{\rm int}$ [fb$^{-1}$] &   $\sigma_\SM\!\left(-80\%, +30\%\right)$ [fb]   & $\sigma_\SM\!\left(+80\%, -30\%\right)$ [fb] &   $Zhh$ events \\ \hline
        500     &   1600 $\times$ 2   &   0.232  &   0.158   &   371 + 253   \\
        1000    &   3200 $\times$ 2    &   0.177  &   0.121   &   566 + 387   \\ \hline
    \end{tabular}
    \caption{SM prediction for \eeZhh\ for the foreseen ILC operating phases with beam polarization 
    $P_{e^-} = \mp80\%$ and $P_{e^+} = \pm30\%$~\cite{Barklow:2015tja,Bambade:2019fyw}.
    The ``$\times 2$'' in the second column refers to the sum of the two polarizations, corresponding to the
    sum in the last column.
    }
    \label{tab:ILC}
\end{table}


\subsection{Benchmark Points}
\label{sec:BPs}

To study the effect of the one-loop corrected THCs we defined specific benchmark points (BPs) that exhibit an interesting phenomenology while being in agreement with the current constraints as described in \refse{sec:constraints}.
The input parameters for these BPs are summarized in~\refta{tab:BP}.
We also show their predicted tree-level and one-loop THCs, and the tree-level and ``corrected'' 
$H$ width $\Gacorr_H$ (see \refeq{eq:Gacorr}).%
\footnote{The $A$ resonance does not play an important role in any of the studied BPs, 
so for brevity we do not show the predictions for the total width of $A$.}

BPal (benchmark point alignment) is the only point valid in all four Yukawa types. It is defined
in the alignment limit ($\CBA=0$), and it predicts a large value of $\kala$ at one-loop level, due to the large 
splitting between $\MA=\MHp$ and $\MH=\mbar$ (see~\refse{sec:scan}).
This BP is chosen specifically to demonstrate the important effects of the one-loop corrections 
to $\kala$ alone on the di-Higgs production cross section.
Additionally, BPal constitutes a good reference point to compare results of the cross section predictions 
considering the one-loop corrected coupling $\kalaone$ computed by means of the effective potential against the full
diagramatic computation (see \refse{sec:1LTHCs} for more details
).
This allows us to estimate the importance of the finite-momentum effects in the one-loop corrections to $\kala$,
which are neglected in the effective potential approach.

The other points, named BP1, BP2, BP3, BPsign and BPext, are chosen to illustrate interesting phenomenology
involving the coupling $\lahhH$ at one-loop level and, consequently, the $H$ resonant peak.
Focusing on the ILC with a center-of-mass energy of $500\gev$ (ILC500),
we consider BPs with masses for the $H$ boson between $2\Mh\sim\  250 \gev$ and $\sqrt{s}-\MZ\sim409 \gev$, such that the $H$ boson 
can be produced on-shell. All these points have a value of $|\CBA| \geq 0.1$, to ensure a relatively large coupling $ZZH \propto \CBA$, to yield a relevant effect from the $H$ resonance.
Taking into account current experimental constraints points with such large values of $|\CBA|$ are only allowed in the 2HDM Yukawa type I.

Each of these points is chosen to illustrate different aspects of the phenomenology.
BP1, BP2 and BP3 are chosen such that their one-loop prediction $\lahhHone$ is substantially larger than 
the tree-level prediction $\lahhHzero$.
In particular, BP1 has a value for $\lahhHzero \sim 0.02$ (slightly outside the alignment limit), 
while the one-loop prediction is $\lahhHone\sim 0.2$.
For BP2, the one-loop prediction $\lahhHone$ is about twice as large as the tree-level prediction $\lahhHzero$.
This is also the case for BP3, but in addition the decay width of $H$ is significantly larger than for the other 
BPs, due to the kinematically allowed decay $H\to AA$, and therefore the $H$ resonant peak is expected 
to be broader.
The point BPsign is chosen because it exhibits a sign change in $\lahhH$: the one-loop corrected coupling
$\lahhHone$ has a positive sign, while the tree-level prediction $\lahhHzero$ is negative.
The final point BPext (benchmark point extreme) features a large enhancement 
for $\kala$ and $\lahhH$ at one loop level, together with a very light $H$ boson close to the production threshold.
Therefore, we expect BPext to have a very large cross section, close to the maximum cross section allowed by 
the current constraints.

It should be noticed that the values of $\lahhHone$ predicted by our BPs are far from the extremal 
values discussed in \refse{sec:scan}.
This is related to the fact that we are focusing on the low-$\MH$ region of the 2HDM, 
while the larger values for $\lahhH$ (at tree and one-loop level) are found for a heavy $H$ boson. 

\begin{table}[t!]
    \centering
    \begin{tabular}{ccccccccccccc} \hline
        Point   & $\MH$   & $\MA$   & $\MHp$ & $\tb$ & $\CBA$   & $\bar{m}$   & $\kalazero$   & $\kalaone$    & $\lahhHzero$    & $\lahhHone$ & $\Gamma_H$ & $\Gacorr_H$ \\
        \hline
        BPal*   & 400     & 800     & 800   & 3.0     & 0.00   & 400   & 1.00   & 5.75  & 0.00   & 0.01 & 0.484 & 0.485 \\
        BP1   & 300   & 650   & 650   & 12    & 0.12  & 300 & 0.95  & 4.69  & 0.02  & 0.21    & 0.120 & 0.319 \\
        BP2   & 350   & 600   & 350   & 5.0 & 0.12  & 330   & 0.87  & 1.33  & 0.18  & 0.33  & 0.362 & 0.739 \\
        BP3   & 300   & 100   & 300   & 2.5   & -0.18 & 300   & 1.06  & 1.40  & 0.24  & 0.44  & 15.7  & 16.3 \\
        BPsign  &   350 &   650 & 650   & 20    & 0.10   & 350   & 0.995 & 5.47  & -0.08 & 0.16  & 0.175 & 0.275 \\
        BPext   &   260 & 700   & 700   & 8.0 & 0.10   & 260   & 0.96  & 5.81   & 0.07  & 0.24  & 0.059 & 0.189  \\ \hline
        
    \end{tabular}
    \caption{
    Benchmark points studied in this work to illustrate relevant one-loop effects from triple Higgs couplings.
    The asterisk (*) indicates that BPal is allowed in the four 2HDM types, while the others are only allowed in the 
    2HDM type I. Mass-dimension parameters are given in GeV.
    }
    \label{tab:BP}
\end{table}


\subsection{Experimental Sources of Uncertainty for the Access to THCs}
\label{sec:exp}

\subsubsection{Detection of the Final \texorpdfstring{\boldmath{$Z+4b$}}{Z+4b} Events}
\label{sec:cuts}

In this work we focus on the main decay channel of the SM-like Higgs boson into bottom quarks, 
\ie we consider the process $\eeZhhbb$.
Therefore, the experimental signature consists of four $b$-flavored jets and a $Z$ boson.
We estimate the expected number of final $Z+4b$ events by considering the following reduction factors,
\begin{equation}
    \bar N_{Z4b} = N_{Zhh} \times \left(\br\!\left(h\to b\bar b\right)\right)^2 \times {\cal A} \times \epsilon_b\ \equiv N_{Z4b} \times {\cal A} \times \epsilon_b\,,
    \label{eq:cuts}
\end{equation}
where $N_{Zhh}$ is the number of $Zhh$ events predicted by the cross section computed as detailed in \refse{sec:eehhZ} 
and $\br\!\left(h\to b\bar b\right)$ is the branching ratio of the decay $h\to b\bar b$ predicted in the 2HDM.
To obtain the number of $Z + 4b$ events  we consider the luminosities in \refta{tab:ILC}.
The $b$-tagging efficiency of detecting the 4 final $b$-jets is denoted by $\epsilon_b$, where we use $\epsilon_b=0.85$,
following \citeres{Tian:2013qmi,Durig:2016jrs}. In these works it was shown that this choice
leads to an optimal Higgs mass resolution in the  $Zhh\to q\bar q b\bar bb\bar b$ channel.
The detector acceptance is denoted by $\cal A$, which we estimate by simulating the process $\eeZhh$ with the subsequent decay $h\to b\bar b$, and considering the following preselection cuts on the final particle states, 
\begin{itemize}
    \item $E_{b} > 20\gev$: 
    At future $e^+e^-$ colliders, the energy of hadronic jets will be reconstructed applying Particle Flow calorimetry
    techniques~\cite{Thomson:2009rp}. 
     Despite the improvement w.r.t.\ traditional calorimetry, the energy resolution diminishes rapidly for energies 
    below 20--30~GeV.  
    \item $\left| \eta_{b} \right| < 2.5$ and $\left| \eta_{Z} \right| < 2.5$, where $\eta$ is the pseudo-rapidity. 
    Final state particles which are very collimated with the $e^+e^-$ beams, would be impossible to detect. 
    It is expected that final states can be reconstructed with Particle Flow calorimetry techniques until polar angles 
    of around 10\textdegree~\cite{Thomson:2009rp}, which corresponds to a pseudo-rapidity of approximately~2.5.
    \item $y_{bb} > 0.0025$ at $\sqrt{s}=500\gev$ and $y_{bb} > 0.0010$ at $\sqrt{s}=1\tev$, with 
    ${y_{ij}=2\min\!\left(E_i^2,E_j^2\right)\left(1-\cos\theta_{ij}\right)/s}$, where $\theta_{ij}$ is the angle 
    between the momenta of the particles $i$ and $j$. 
    The variable $y_{ij}$ is a definition of the distance between particles, which is widely used at $e^+e^-$ 
    colliders to perform the jet clustering procedure via the Durham algorithm~\cite{Catani:1991hj}.
    In our analysis, we impose a minimum separation between the final state $b$ jets, in order to be correctly identified 
    as jets.
    Similar cuts on $y_{bb}$ have been considered in experimental analyses at the ILC with 4$b$ jets as final state
    at $500\gev$~\cite{Durig:2014lfa,Tian:2010np,Yonamine:2010su}. 
\end{itemize}
The acceptance is estimated as the ratio of events with and without the above cuts, 
\begin{equation}
    {\cal A} = \frac{N^{\rm w/\; cuts}}{N^{\rm w/o\; cuts}}\,.
\end{equation}
We computed the number of final events with and without cuts for the process $\eeZhhbb$ with  {\tt MadGraph5\textunderscore aMC v2.9.17}~\cite{Alwall:2014hca} at the parton level and hence did not consider any hadronization of the $b$-jets.
In an experimental analysis, further cuts on the invariant mass of $b$ jet pairs would be considered to reconstruct the Higgs boson $h$, which can be challenging due to the finite detector resolution.
However, we  did not consider such cuts, since our simulations are only at the parton level, and thus the vast majority of simulated events shows a pronounced $h$ peak at 125 GeV, unaffected by the detector resolution.

The acceptances for the considered BPs and for the SM case (for comparison) are shown in \refta{tab:Acc} for center-of-mass energies of $\sqrt{s}=500\gev$ and $\sqrt{s}=1\tev$.
In the 500~GeV case we obtain acceptances around 73\% for all considered BPs, which is very close to the acceptance obtained in the SM.
In the case of a 1~TeV collider, the acceptances of the studied BPs are between 64\% and 73\%, which are slightly
worse compared to the acceptance obtained in the SM of 76\%. 
We found that the cut on the ``distance'' between $b$-jets, $y_{bb}$, is the cut that reduces the most the
number of events yielding $\bar N_{Z4b}$.

The fraction of events obtained after the $b$-tagging reconstruction and the preselection cuts, \ie $\bar N_{Z4b}/N_{Z4b} = {\cal A}\times \epsilon_b$, is about 62\% at 500~GeV and  between 54\% and 62\% at 1~TeV.
The 500~GeV case is in good agreement with the results of the experimental analysis in \citere{Durig:2016jrs}.%
\footnote{To our knowledge, there is no analysis at $\sqrt{s} = 1 \tev$ for the double Higgs-strahlung channel which we could compare our results to.}
In that work, after their preselection cuts they retain 61.6\% of the total predicted $Zb\bar bb\bar b$ events, after studying the $Z$ decay channels to $e^+e^-$, $\mu^+\mu^-$, $\nu\bar\nu$ and $q\bar q$ separately.%
\footnote{There is currently intense progress to update the experimental projections of~\citere{Durig:2016jrs}.
See for instance~\citeres{Munch:effic,Bliewert:2024hed}, where they expect sizable improvements in the projected accuracy in the measurement of $\sigma\!\LP Zhh\RP$ and $\lahhh$ due to better
jet tagging, particle identification and the usage kinematic fitting.}
In addition, they consider more cuts to further suppress the SM background (where $ZZZ$ and $ZZh$ are the most challenging ones) with respect to the $Zhh$ signal. 
These additional cuts are more severe and only 17.0\% of the $Zb\bar bb\bar b$ events survive them.
However, we only consider the events after the simple preselection cuts and $b$-tagging identification, since a realistic experimental analysis including backgrounds is beyond the scope of this work.  

\begin{table}[t!]
    \centering
    \begin{tabular}{cccccccc}
        \hline
        & SM & BPal & BP1 & BP2 & BP3 & BPsign & BPext \\
        \hline
        $\sqrt{s}=500\gev$ & 0.736 & 0.743 & 0.739 & 0.733 & 0.738 & 0.741 & 0.745 \\
        $\sqrt{s}=1000\gev$ & 0.758 & 0.652 & 0.645 & 0.689 & 0.726 & 0.657 & 0.641 \\
        \hline
    \end{tabular}
    \caption{Acceptances for the considered benchmark points (BPs) after the preselection cuts.}
    \label{tab:Acc}
\end{table}


\subsubsection{Smearing and Binning of the Cross Section Distributions}
\label{sec:smearing}

The experimental measurements of the invariant mass distributions are affected by the finite resolution of the 
hadronic calorimeters. 
Additionally, pairing the final state four $b$-jets to reconstruct the two SM-like Higgs bosons introduces an extra source 
of experimental uncertainty. 
To account for this, we apply an artificial smearing to the theoretical predictions for $\minv$, 
assuming Gaussian uncertainties. 
The smearing is characterized by a percentage, $p\%$, such that each value of $\minv$ in the distribution has an associated Gaussian uncertainty with a full-width at half maximum (FWHM) defined as $p\%$ of $\minv$.
In other words, the FWHM of the Gaussian distributions is given by $\mathrm{FWHM} = 2\sqrt{2\log2}\, \sigma = \minv\times p\% $, where $\sigma$ is the standard deviation of the distribution.

Currently, the expected experimental resolution on the invariant mass of the final-state Higgs pair 
at the ILC is unclear.
In consequence, we will consider different values for the smearing parameter, namely 0\% (no smearing), 2\%, 5\% and 10\% 
in order to give a notion of the required experimental resolution to access possible BSM signals from THCs at future
high-energy $e^+e^-$ colliders.
Very recently, there has been progress in the achievable $\minv$ resolution at a high-energy $e^+e^-$ collider 
operating at 500 GeV~\cite{Munch:mhhres,LinearColliderVision:2025hlt}. In these preliminary results, the authors find a $1\sigma$
uncertainty on $\minv$ of 2.1\% in the $Zhh\to\mu^+\mu^-b\bar bb\bar b$ channel and
2.6\% in the $Zhh\to\nu\bar\nu b\bar bb\bar b$ channel,
while they remark that there is still room for further improvements.
These values would correspond to a smearing of 4.9\% and 6.1\%, respectively, thus our 5\% smearing scenario would
represent the most realistic assumption given these results. 

Another important aspect in the reconstruction of the differential cross section in terms of $\minv$ 
is the determination of the bin size in the distribution. 
A smaller bin size is desirable, as it allows for a more detailed reconstruction of the differential distributions. 
However, the bin size strongly depends on the number of final events detected at the collider.
To determine the bin size after smearing, we require that each bin in the kinematically allowed region 
($2\Mh < \minv < \sqrt{s} - \MZ$) contains at least two reconstructed events, \ie~$\bar N_{Z4b} \geq 2$.
This approach may be somewhat conservative, as a larger number of events is expected somewhat above
the production threshold ($\minv \gtrsim 2\Mh$) compared to the distribution tails ($\minv \sim 2 \Mh$ and
$\minv \lesssim \sqrt{s} - \MZ$). 
As a result, the resolution could potentially be higher in the more relevant region, particularly
around $\mhh \sim \MH$.


\section{Sensitivity to One-Loop THCs at  \texorpdfstring{\boldmath{$e^+e^-$}}{e+e-} Colliders}
\label{sec:diffxs}

In this section we analyze the potential sensitivity to the (one-loop corrected) THCs at $e^+e^-$ colliders.
We employ the differential  distributions of the invariant mass of the final-state di-Higgs
pair, $\minv$, which are shown to be sensitive to the effects of THCs (see \citere{Arco:2021bvf} and references therein).
As discussed in \refse{sec:eehhZ}, the effects of $\kalaone$ enter via a non-resonant diagram mediated by
$h$-exchange, while the effects of $\lahhHone$ enter via a resonant diagram mediated by $H$-exchange.
Furthermore, in the analysis presented in this section, we consider the relevant experimental factors that can
potentially degrade the experimental measurement of the $Zhh$ distributions, as discussed in
\refse{sec:exp}, and consequently reduce the projected sensitivity to the THCs.
To explore all these effects, in this section we show the 2HDM prediction of the  $m_{hh}$ distributions for the
BPs defined in \refta{tab:BP}, which are chosen to exhibit a variety of phenomenological effects
that can potentially be expected at a high-energy $e^+e^-$ collider, such as the ILC.


\subsection{Sensitivity to the SM-like THC \texorpdfstring{\boldmath{$\lahhh$}}{lahhh}}
\label{sec:lahhh}

\subsubsection{General Impact on the Invariant Mass Distributions}


\begin{figure}[t!]
\centering
    \includegraphics[width=0.49\textwidth]{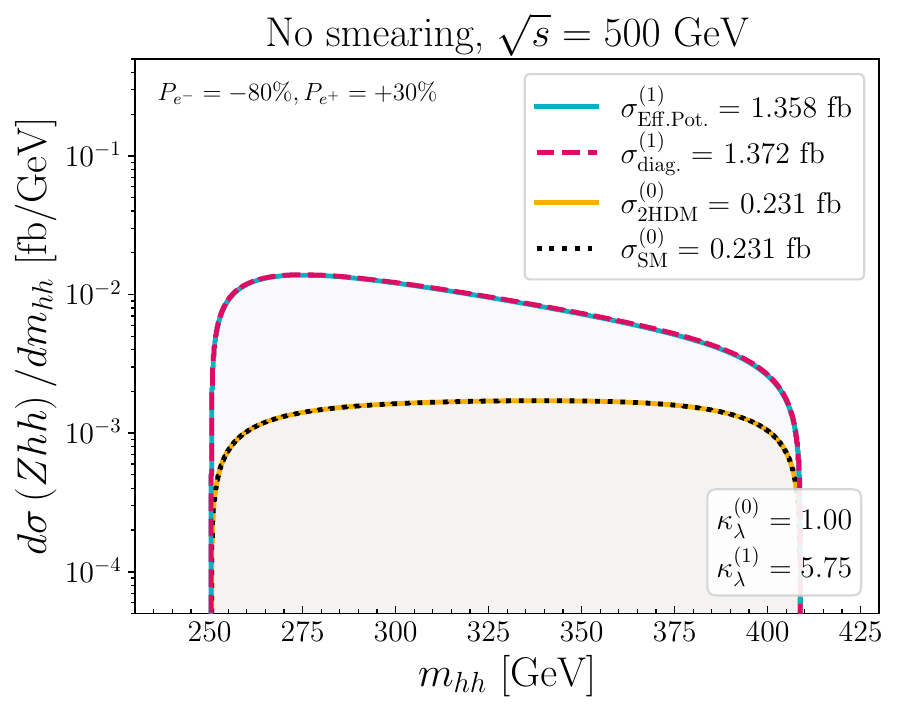}
    \includegraphics[width=0.49\textwidth]{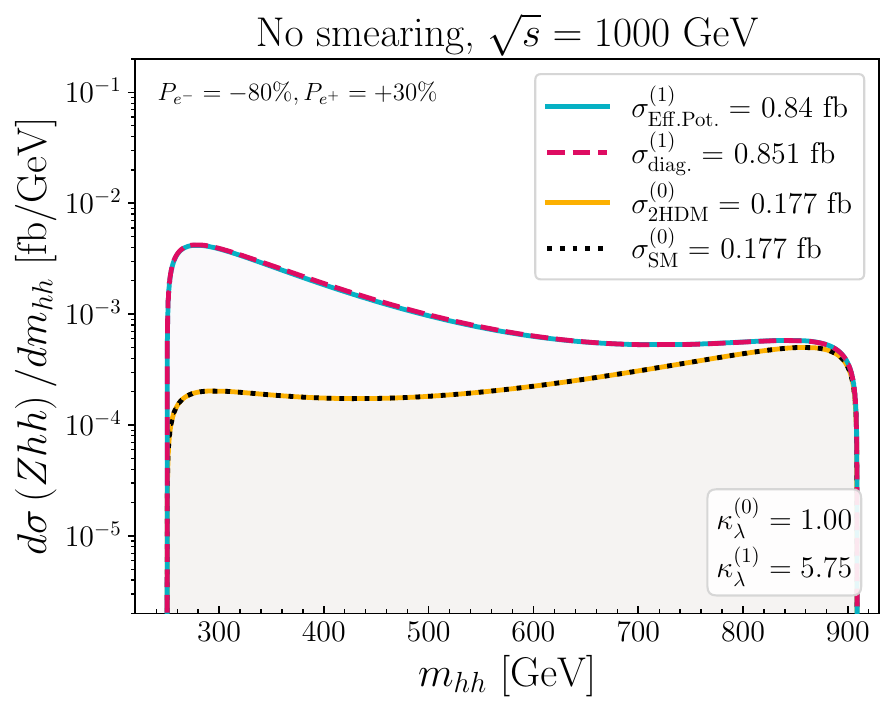}
    \caption{Differential distribution \wrt $\minv$ for BPal at $\sqrt{s}=500\gev$ (left) and $\sqrt{s}=1\tev$ (right) for $P_{e^-}=-80\%$ and $P_{e^+}=+30\%$. 
    The blue lines include the one-loop values of $\kalaone$
    from the effective potential, the dashed red lines include the full diagrammatic prediction for $\kalaone$, the yellow lines show the 2HDM tree-level prediction and the dotted black lines show the SM tree-level prediction.}
     \label{fig:BPal}
\end{figure}

In \reffi{fig:BPal} we display the differential distribution of the cross section with respect to the invariant di-Higgs mass 
$\minv$ for center-of-mass energies of 500~GeV (left) and 1~TeV (right) for BPal, see \refta{tab:BP}.
The polarization has been chosen as $P_{e^-}=-80\%$ and $P_{e^+}=+30\%$.
The solid blue lines show the cross section $\sigeff$ including the one-loop corrected value of $\kala$ from the effective potential. 
The dashed red lines show the cross section $\sigmom$, which includes the fully diagrammatic one-loop result for $\kala$
(which will be discussed in the next subsection).
For comparison, the tree-level cross section for the 2HDM, $\sigtree$, is plotted with solid yellow lines, while the tree-level result for 
the SM, $\sigSM$, is plotted with black dotted lines.
This benchmark point has been chosen specifically to demonstrate the effects that the one-loop corrections to
\lahhh\ alone can have on the di-Higgs production cross section.
Since the alignment limit ($\CBA=0$) is assumed for this benchmark point, BSM physics can only enter via the one-loop corrections  to
$\kalaone$ (see also the discussion in \refse{sec:eehhZ}).

BPal predicts a one-loop corrected triple Higgs coupling $\kalaone=5.75$ as given by the one-loop effective potential, 
which is close to the maximum allowed values for $\kalaone$ shown in \refta{tab:THCranges}. 
This  means that we have a large deviation from the tree-level prediction $\kalazero=1$ as given in the alignment limit.
This change in the value of $\kala$ is reflected in a strong enhancement of the di-Higgs production cross section, 
as can be seen in \reffi{fig:BPal}.
The cross section including the one-loop corrected $\kalaone$ is 5.9 (4.8) times larger than the tree-level prediction at $\sqrt{s}=500\ (1000)\gev$.
This enhancement is most pronounced in the low $\minv$ region for both center-of-mass energies, which is known to be the most sensitive region to the variation of $\kala$ (see \citere{Arco:2021bvf} and references therein).
These one-loop corrections to the $Zhh$ cross section induced by $\kalaone$ are considerably larger than those expected in the SM, which are not larger than 10\% for 500--1000~GeV center-of-mass energies~\cite{Belanger:2003ya,Zhang:2003jy}.

It should be noted that this large increase in the $Zhh$ production cross section happens in the alignment limit (\ie $\CBA=0$).
In the alignment limit the Higgs boson $h$ has production rates and decay branching ratios very similar to those predicted in the SM,  
therefore a BSM signal for this BP is unlikely to be detectable in single Higgs production at the HL-LHC \cite{Cepeda:2019klc,CMS:2025hfp}.
This makes di-Higgs production a key process to investigate possible BSM effects and it is crucial
to fully test the nature of the Higgs potential.

Regarding the potential experimental sensitivity to $\kala$, the analyses of \citeres{Durig:2016jrs,Torndal:2023fky,Torndal:2023mmr,List:2024ukv,LinearColliderVision:2025hlt} 
show the projections of the expected precision of the measurement of $\kala$ at $e^+e^-$ colliders in the case that the SM prediction is 
not realized, \ie $\kala\neq1$. 
These results are obtained after extrapolating the full simulation results from the SM case, which leads to a determination of $\kala$ 
with an accuracy of 15\% at the ILC operating at 550~GeV with an integrated luminosity of 4.4~\iab~\cite{List:2024ukv,LinearColliderVision:2025hlt}.
Considering a possible luminosity upgrade of 8~\iab, the accuracy for $\kala=1$ drops to 11\%.
Assuming that integrated luminosity, \citere{LinearColliderVision:2025hlt} gives a very similar projected precision of about 10\% in the case that $\kala>1$ for the $Zhh$ production channel.
The combination with the $\nu\bar\nu hh$ channel would result in a precision of about 8\% for $\kala\geq1$ at 550 GeV and 8~\iab. 
Here it is important to note that the analysis in this subsection only takes into account the theoretical differential distributions, 
i.e.\ we do not consider the smearing or binning as discussed in \refse{sec:exp}.
However, the sensitivity to $\kala$ discussed above already takes such effects into account, as it is based on ILC experimental analyses.
This will be different in \refse{sec:lahhH}, where smearing and binning have a crucial effect on the potential experimental 
sensitivity on $\lahhH$.

On more general grounds, since in the 2HDM we mainly find values of $\kalaone\gtrsim1$, we conclude that the experimental determination 
of this coupling via di-Higgs production at $e^+e^-$ colliders would be of about 10\% independently of the value realized for $\kalaone$ (if other BSM effects are negligible), 
which presents this channel as a great opportunity to determine the Higgs boson self-coupling.
It should be mentioned that if an $H$~resonance peak is realized in the $\minv$ distributions, especially close to the 
production threshold, this could potentially degrade the experimental sensitivity to $\kala$ discussed above.
In such a case, it would be necessary to efficiently disentangle the contributions of the $H$~resonance from the 
non-resonant contribution mediated by the  $h$~exchange.
In the next sections we discuss the potential sensitivity to the $H$~peak at $e^+e^-$ colliders, but we do not analyze this scenario 
where the effects of $\lahhH$ and $\kala$ are mixed (which are beyond the scope of our paper). 
However, in the case of the discovery of a $\CP$-even resonance around $300\gev$, a detailed experimental analysis  would be required.


\subsubsection{Finite Momentum Effects from \texorpdfstring{\boldmath{$\lahhh$}}{lahhh}}
\label{sec:finitemom} 

\begin{figure}[t!]
    \centering
    \includegraphics[width=0.7\textwidth]{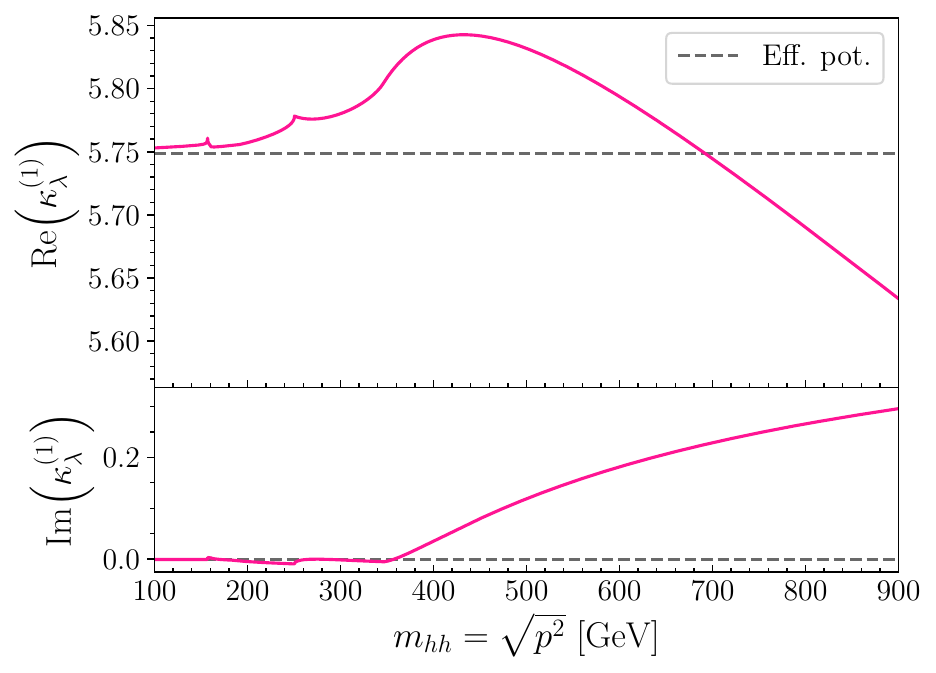}
    \caption{Real (upper) and imaginary (lower) part of the fully diagrammatic one-loop prediction $\kalaone$ for BPal as a function of the invariant mass of the final two Higgs bosons. 
    The gray dashed line shows the respective prediction from the effective potential. }
    \label{fig:momkala}
\end{figure}

In this subsection we analyze the finite-momentum effects in the loop-corrected trilinear Higgs self-coupling by comparing the effects when using the diagrammatic calculation of $\kappa_\lambda^{(1)}$ to those when using $\kappa_\lambda^{(1)}$ obtained from the effective potential (see the discussion in \refse{sec:1LTHCs}).
In \reffi{fig:BPal} it can be seen that the predictions for $\sigeff$ and 
$\sigmom$ (\ie the solid blue \vs the dashed red lines) are very close to each other. This implies
that the the finite-momentum effects are very small.
To test this further, we show in \reffi{fig:momkala} the result of the fully diagrammatic computation of $\kalaone$ for  BPal 
as a function of the loop momentum $\sqrt{p^2}$.
In the case of di-Higgs production, it corresponds to the invariant mass of the final Higgs pair $\minv$. 
The upper (lower) plot shows the real (imaginary) part of the one-loop corrected THC, $\kalaone$.
We also show the prediction from the effective potential with horizontal dashed gray lines.
As plot range we have chosen to start at~$100 \gev$, going up to~$900 \gev$, \ie covering the full range that is relevant for
$\sqrt{s} = 1000 \gev$.
Starting with the real part, for small values of $\minv=\sqrt{p^2}$, the predictions from both approaches are very close to each other.
Above $\minv\gtrsim 200\gev$ the real part of $\kalaone$ obtained with the diagrammatic computation starts deviating more significantly from the prediction of the 
effective potential, reaching the maximum value of 5.84 at $\minv\sim400\gev$.
For larger values of $\minv$, the real part of $\kalaone$ goes down to $\sim 5.6$ for $\minv=1000 \gev$.
Concerning the imaginary part of $\kala$, it is negligible below the di-top threshold, and reaches a maximum value of less than 0.3 for the highest values of $\sqrt{p^2} = \minv$.
One can see the $WW$, $hh$ and $tt$ threshold effects at $p=2m_W,\ 2\Mh,\ 2m_t$, respectively, both in the real and the imaginary 
parts of $\kalaone$.%
\footnote{The $ZZ$ threshold is also present at $2m_Z$, but it is not visible in the plot as its numerical effect is very small.}
The $HH$ threshold is not visible because  for this benchmark point $\lahHH$ is proportional to $\Mh^2/v^2$ (see \refeq{eq:THC3}) and hence very small. 
The $AA$ and $H^+H^-$ thresholds are present and very prominent as expected given the large values of the $hAA$ and $hH^+H^-$ THCs, 
but they are at $p=2\MA=2\MHp=1600\gev$, outside the range of the kinematically allowed region for $\minv$.
Our results agree  with the corresponding findings in \citere{Bahl:2023eau}.

The disagreement between the two computations of $\kalaone$ shown in \reffi{fig:momkala} is not large enough to have a phenomenologically 
relevant impact on the final prediction for the cross section, nor on the $\minv$ predictions, as we have seen also in \reffi{fig:BPal}.
In fact, even though it is not visible by eye, the difference between the $\sigeff$ and 
$\sigmom$ has a very similar dependence on $\minv$ as that of $\kalaone$ in \reffi{fig:momkala}. 
We can therefore conclude that for the current experimental sensitivities expected at $e^+e^-$ colliders, the simpler $\kalaone$ 
calculation using the effective potential is sufficient to capture the relevant one-loop corrections to the total di-Higgs production cross section.
Consequently, we will stick to the effective potential calculation in the discussions of the other benchmark points below.


\subsection{Sensitivity to the THC \texorpdfstring{\boldmath{$\lahhH$}}{lahhH}}
\label{sec:lahhH}

The remaining BPs shown in \refta{tab:BP} are specifically chosen to exhibit relevant BSM effects related to the one-loop corrected 
BSM THC $\lahhHone$.
To assess the potential experimental sensitivity to the triple Higgs coupling $\lahhHone$ at $e^+e^-$ colliders, we evaluate the statistical 
significance of the $H$ Higgs boson resonance peak for center-of-mass energies of 500~GeV and 1~TeV.
To do this, we perform a profile likelihood analysis exploiting the information of the distributions \wrt the invariant mass $
\minv$, following the procedure of \citere{Cowan:2010js} (as detailed below).
We use the number of $Z+4b$ signal events after the preselection cuts and the $b$-tagging efficiency discussed in \refse{sec:cuts}.
We also analyze how some sources of uncertainty related to the experimental resolution of the invariant mass distributions
affect the statistical significance of the $H$~peaks for our BPs.


\subsubsection{Estimation of the Statistical Significance of the \texorpdfstring{\boldmath{$H$}}{H} Resonance Peak}
\label{sec:sig}

In our statistical analysis the signal and background events  in the $i$th bin are given, respectively, by
\begin{align}
    s_i &= \bar N_{i,4bZ} - \bar N^C_{i,4bZ} \,,\label{eq:si}\\
    b_i &= \bar N^C_{i,4bZ}\,, \label{eq:bi}
\end{align}
where $\bar N_{i,4bZ}$ are the number of $Z+4b$ events, as discussed in \refse{sec:cuts}, and $\bar N^C_{i,4bZ}$ are the predicted number of $Z+4b$ events from the same parameter point but with the THC $\lambda_{hhH}$ artificially set to zero, corresponding to the events from the ``continuum'', \ie without resonance.
Consequently, the absolute number of events per bin $n_i$ is given by
\begin{equation}
    n_i = s_i + b_i = \bar N_{i,4bZ}\,.
\end{equation}

We test two hypotheses where the expected number of events in the $i$th bin is given by $E(n_i) = \mu s_i + b_i$, where $\mu$ is 
known as the strength parameter.
The value $\mu=0$ corresponds to the no $H$ resonance hypothesis, while $\mu=1$ is the nominal signal hypothesis.
To test a given value of $\mu$, one can construct the profile likelihood ratio, defined by
\begin{equation}
    \lambda\!\left(\mu\right) = \frac{L\!\left(\mu\right)}{L\!\left(\hat\mu\right)}\,,
    \label{eq:PLR}
\end{equation}
where $L\!\left(\mu\right)$ is the likelihood function, which is given by the product of the Poisson probabilities of all bins as
\begin{equation}
    L\!\left(\mu\right) = \prod_{i} \frac{\left(\mu s_i + b_i\right)^{n_i}}{n_i!} e^{-\left(\mu s_i + b_i\right)} \,,
    \label{eq:LF}
\end{equation}
\ie the product of the probabilities of measuring $n_i$ events in the $i$th bin when $\mu s_i + b_i$ events are expected.
The parameter $\hat \mu$ in \refeq{eq:PLR} is the value for the strength parameter that maximizes the likelihood, also known as unconditional 
maximum-likelihood estimator (MLE).
In our case, we consider the simplest case where $\hat\mu=1$, meaning that the measured number of events corresponds exactly to the 
nominal signal case.
This is usually known as a ``Asimov'' data set, and is typically used to calculate expected sensitivities.
Therefore, to compute the significance of the $H$ resonance peak signal, we need the profile likelihood ratio when $\mu=0$, that is
\begin{equation}
    \lambda\!\left(0\right) = \frac{L\!\left(0\right)}{L\!\left(1\right)}\,,
\end{equation}
such that we test how likely it would be to not measure the signal of $H$ against the scenario with no resonance ($\mu=0$).
To obtain the statistical significance $Z$ we assume that the likelihood in \refeq{eq:PLR} can be approximated by a Gaussian distribution,
\begin{equation}
    Z = \sqrt{-2\log\!\left(\lambda(0)\right)}\,,
\end{equation}
which is true for a large data sample~\cite{Cowan:2010js}.
With all these expressions we can derive the following formula for the statistical significance:
\begin{equation}
     Z = \sqrt{\sum_i \left(Z_i\right)^2}\,,
     \label{eq:signif}
\end{equation}
where $Z_i$ could be understood as the separate significance of each bin $i$, given by
\begin{equation}
    Z_i = \sqrt{2\left( \left(s_i+b_i\right)\log\!\left(1+\frac{s_i}{b_i}\right)-s_i\right)}\,.
\end{equation}
The expression above reduces to the well-known expression for the statistical significance $Z_i\simeq s_{i}/\sqrt{b_{i}}$ 
in the limit $b_i\gg s_i$. It should be noted
that with this estimation of statistical significance by means of the profile likelihood ratio, we fully exploit the information of the differential shapes of our invariant mass distributions.
More specifically, the $H$ peak shape, which can be realized as a peak-dip or a dip-peak structure, is captured by the fact that the variable $s_i$ in \refeq{eq:si} can be negative.

In this work we consider two scenarios for the initial polarization of the incoming $e^+e^-$ pair, as shown in \refta{tab:ILC}.
Therefore, we can obtain a statistical significance for each initial polarization states $P_{e^-}=\mp80\%$ and $P_{e^+}=\pm30\%$, namely $Z_{-+}$ and $Z_{+-}$ respectively.
In the following, we will refer to the {\it combined} statistical significance given by
\begin{equation}
    Z = \sqrt{\left( Z_{-+}\right)^2 + \left( Z_{+-}\right)^2}\,.
    \label{eq:signiftot}
\end{equation}
To compute $Z_{-+}$ and $Z_{+-}$ we consider the different bin size expected for each polarization scenarios following the same procedure
as discussed in \refse{sec:smearing}.

One should keep in mind that a rigorous experimental analysis would be much more complex.
For example, a Monte Carlo simulation of the signal and background, together with a full reconstruction of the detector signal, 
would be required, as well as the consideration of other sources of background (with their corresponding uncertainties), nuisance parameters 
in the likelihood function, bin-by-bin correlations, etc.
Nevertheless, our simplified analysis will shed light on the potential sensitivity to $\lahhHone$ and to the $H$ resonance peak that can be 
achieved at $e^+e^-$ colliders under the assumption of negligible background.
We hope that it serves as a starting point for future, more complete, experimental analyses.


\subsubsection{Access to \texorpdfstring{\boldmath{$\lahhH$}}{lahhH} via the \texorpdfstring{\boldmath{$H$}}{H} Resonance Peak}

\begin{figure}[p]
    \includegraphics[width=\textwidth]{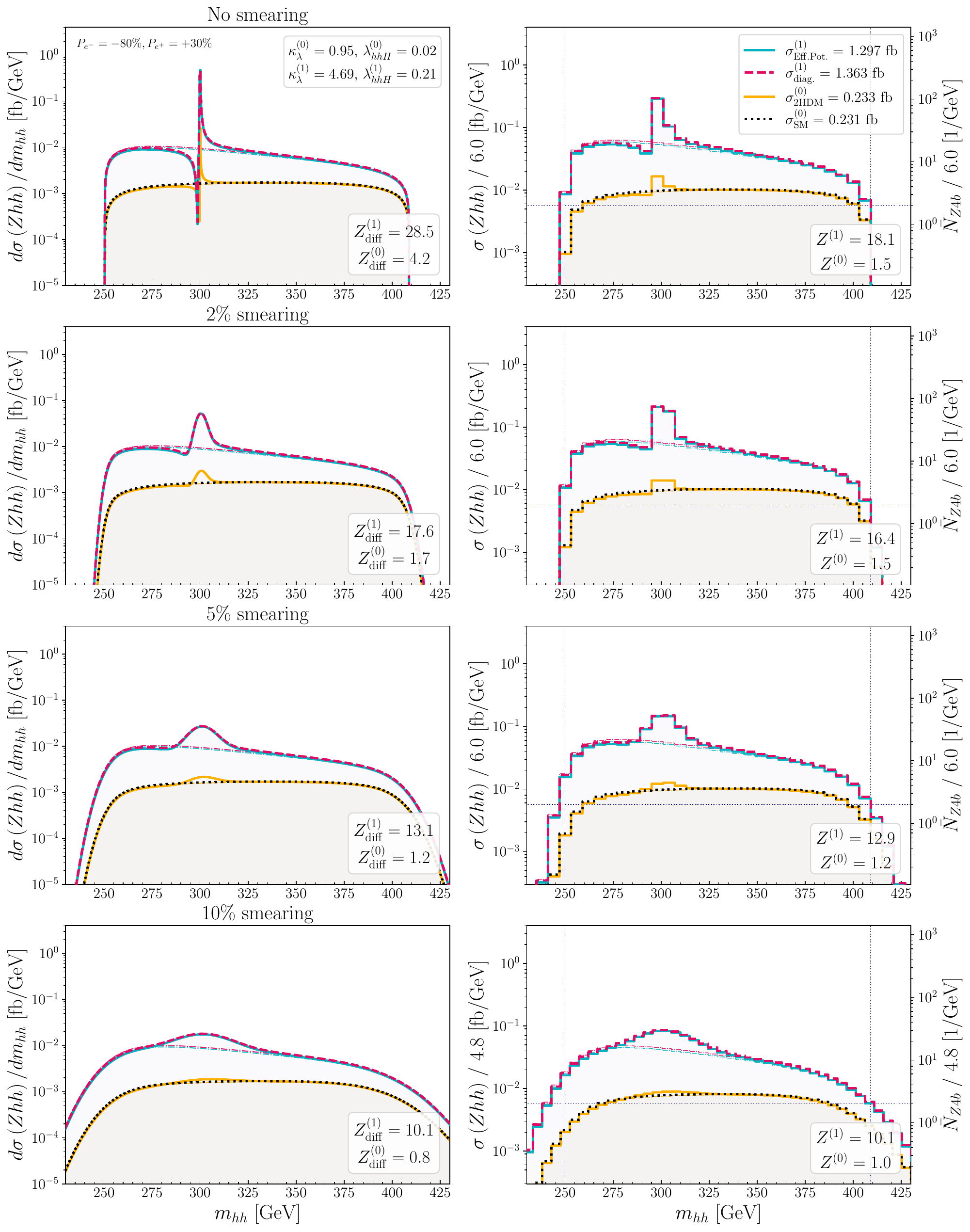}
    \caption{Differential distribution \wrt $\minv$ for BP1 at $\sqrt{s}=500$ GeV for $P_{e^-}=-80\%$ and $P_{e^+}=+30\%$. 
    The color coding is the same as in \reffi{fig:BPal}, and the red and blue lines also include $\lahhHone$ as predicted by the effective potential.
    From top to bottom, the distributions have a smearing of 0\% (no smearing), 2\%, 5\% and 10\%.
    The left plots show the theoretical differential distributions, and the right ones show the distributions with a bin size such that all bins inside the kinematically allowed region have $\bar N_{i,Z4b}\geq2$ (dotted navy lines). }
    \label{fig:BP1-500}
\end{figure}

\begin{figure}[p]
    \includegraphics[width=\textwidth]{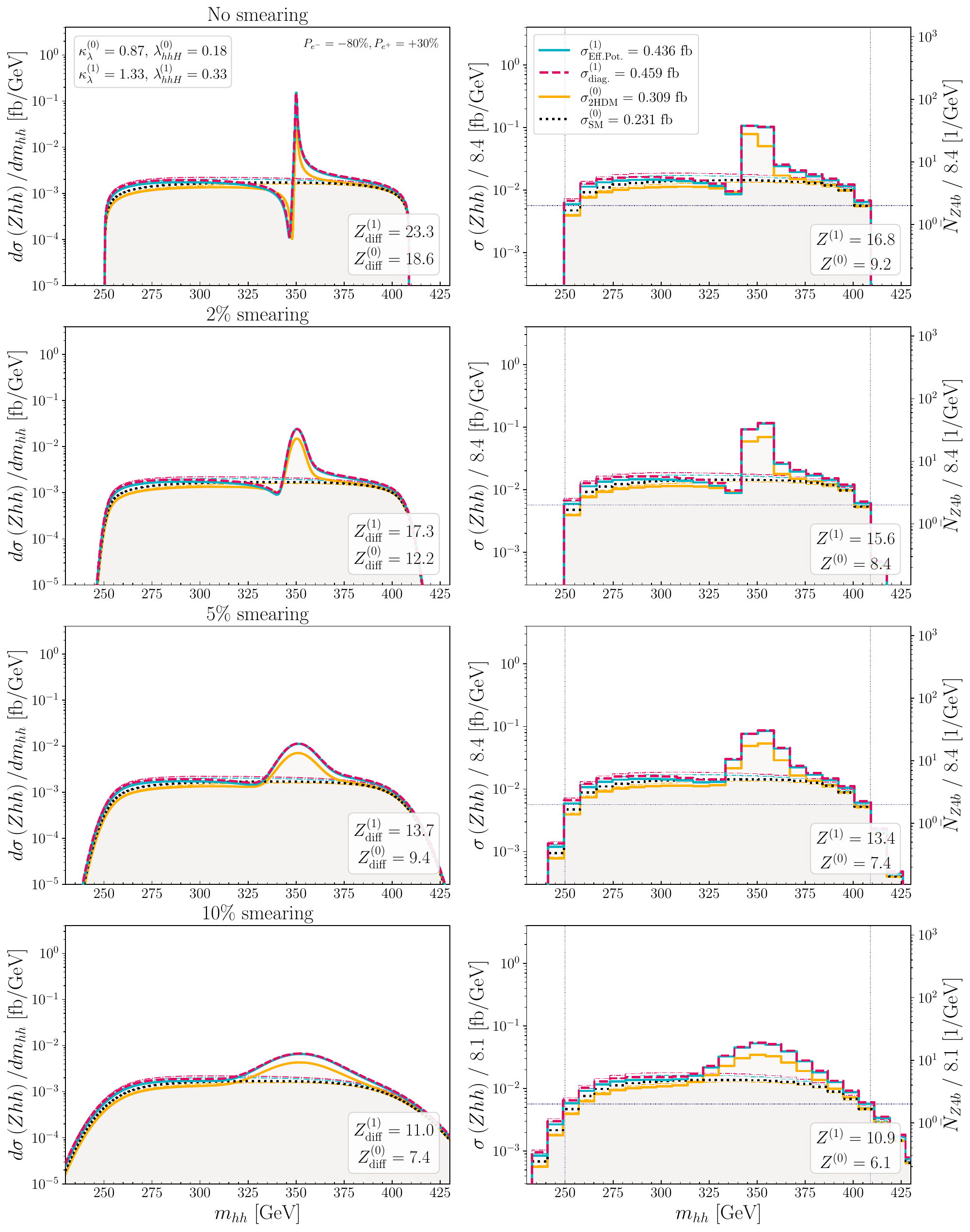}
    \caption{Differential distribution as a function of $\minv$ for BP2 at $\sqrt{s}=500$ GeV for 
    $P_{e^-}=-80\%$ and $P_{e^+}=+30\%$. The color coding is the same as in \protect\reffi{fig:BP1-500}.}
    \label{fig:BP2-500}
\end{figure}

\begin{figure}[p]
    \includegraphics[width=\textwidth]{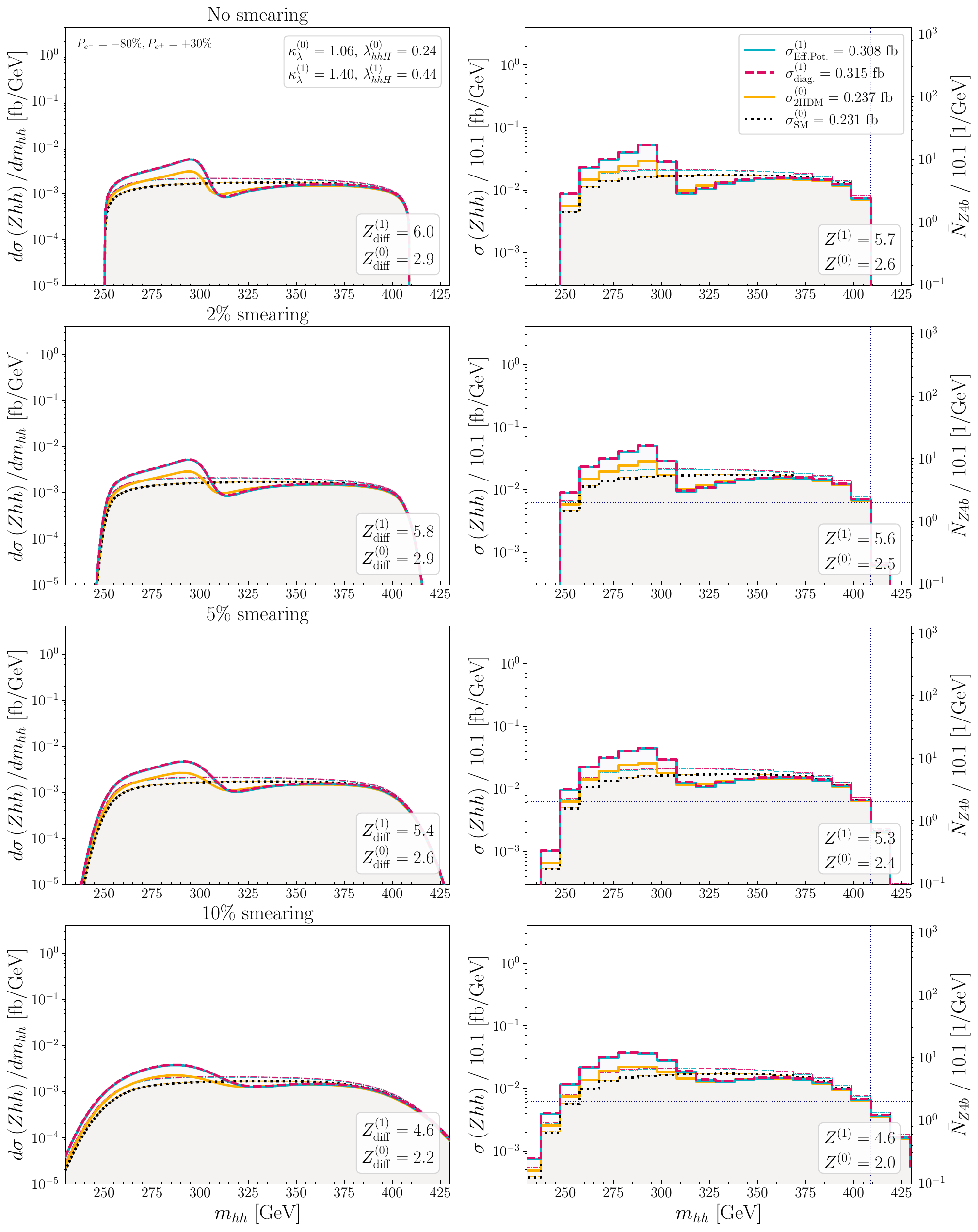}
    \caption{Differential distribution as a function of $\minv$ for BP3 at $\sqrt{s}=500$ GeV  
    for $P_{e^-}=-80\%$ and $P_{e^+}=+30\%$. The color coding is the same as in \protect\reffi{fig:BP1-500}.}
    \label{fig:BP3-500}
\end{figure}

\begin{figure}[p]
    \includegraphics[width=\textwidth]{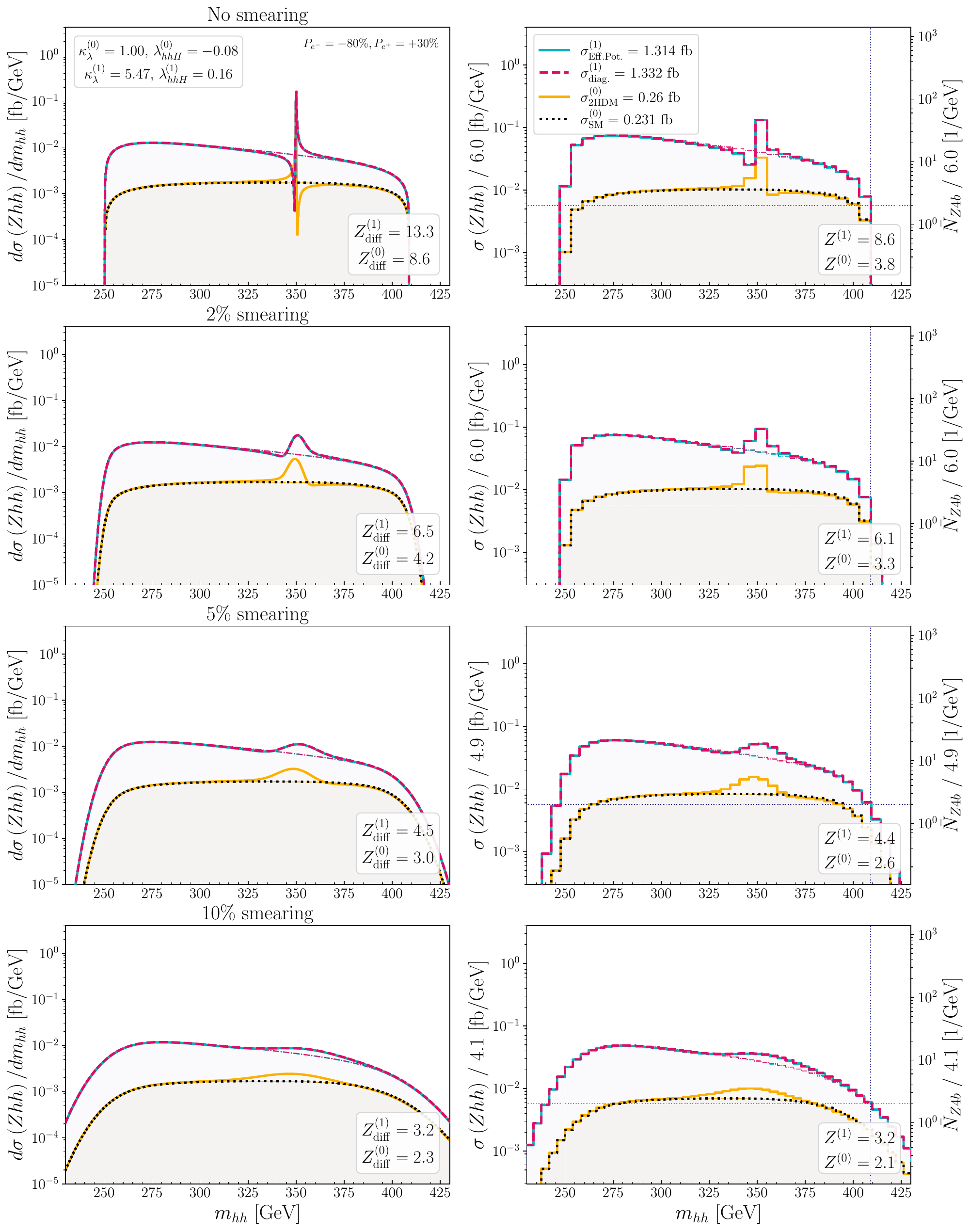}
    \caption{Differential distribution as a function of $\minv$ for BPsign at $\sqrt{s}=500$ GeV for 
    $P_{e^-}=-80\%$ and $P_{e^+}=+30\%$ The color coding is   the same as in \protect\reffi{fig:BP1-500}.}
    \label{fig:BPsign-500}
\end{figure}

\begin{figure}[p]
    \includegraphics[width=\textwidth]{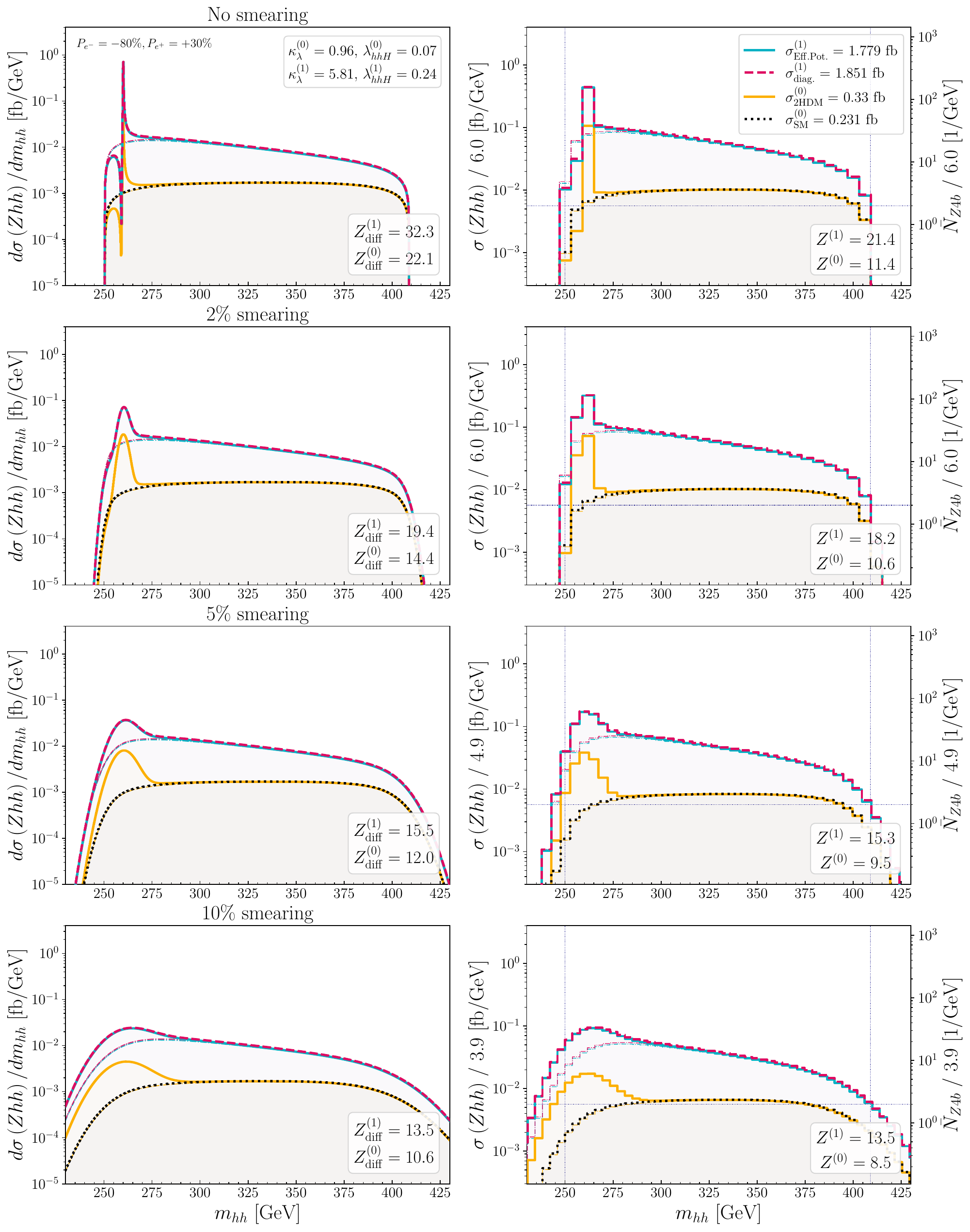}
    \caption{Differential distribution as a function of $\minv$ for BPext at $\sqrt{s}=500$ GeV  for 
    $P_{e^-}=-80\%$ and $P_{e^+}=+30\%$. The color coding is the same as in \protect\reffi{fig:BP1-500}.}
    \label{fig:BPext-500}
\end{figure}

In \reffis{fig:BP1-500} to \ref{fig:BPext-500} we show the differential cross sections for 
the remaining five BPs displayed in \refta{tab:BP} for $\sqrt{s}=500\gev$ and $P_{e^-}=-80\%$ and $P_{e^+}=+30\%$.
For each benchmark point we present eight plots in two columns.
Each row of plots shows the differential cross sections \wrt to $\minv$ for different assumptions of smearing, 
namely a smearing of 0\% (no smeared distributions), 2\%, 5\% and 10\%, from top to bottom.
The left plots present the theoretical differential  distributions%
\footnote{In practice, the differential distributions were obtained by computing the cross section for points separated by 0.2~GeV 
in the invariant mass.},
while the right plots show the distributions binned such that all bins within the kinematically allowed region have 
$\bar N_{Z4b} \geq 2$.
For more details on the smearing and binning of the cross sections see \refse{sec:smearing}.
Furthermore, in these binned distributions we show $\bar N_{Z4b}$ on the right vertical axis.
We indicate the minimum required number of events per bin $\bar N_{i,Z4b}=2$ with a horizontal dotted navy line and the kinematically allowed region with two vertical dotted navy lines.
In short, the left plots show the theoretical smeared differential cross section of the $Zhh$ process, while the right plots attempt to replicate the binned distributions that could potentially be measured at an $e^+e^-$ collider.
The color labeling and notation for the cross section predictions are the same as in \reffi{fig:BPal}.
We additionally include thin dash-dotted lines corresponding to the cross section distributions with the coupling $\lahhH$ 
artificially set to zero, such that there is no $H$ resonance peak.
For those lines the color coding is the same as for the complete distributions.
These distributions without the $H$~resonance contribution serve to determine
the ``background'' events $b_i$ discussed in \refse{sec:sig}.

In addition, the plots indicate the statistical significance of the $H$~resonance peak for each assumption of 
smearing as given by \refeq{eq:signiftot}.
We remark here that we display the statistical significance after the combination of the two polarization runs considered, 
as explained in \refse{sec:sig}, even though we only show the cross section distributions for $P_{e^-}=-80\%$ and $P_{e^+}=+30\%$.
The upper labels ${}^{\LP0\RP}$ and ${}^{\LP1\RP}$ denote the significance for the $H$~peak for the tree-level and one-loop 
distributions, respectively.
In the case of the tree-level significance, we obtain them with the corresponding size of the bins such that $\bar N_{Z4b} \geq 2$, 
although in the plots, for simplicity, we show the distributions of the tree-level prediction with the bin size determined from the prediction including one-loop THCs.
In the plots in the left columns we also provide the significance values for the differential cross section distributions without 
binning, which we denote as $Z_{\rm diff}$.
With these significance values, we can analyze the impact that the resolution on the invariant mass distributions can have on the potential access
to the $H$~resonance peak, and thus the sensitivity to the $\lahhHone$ coupling, at $e^+e^-$ colliders.
We gather all these significance values after the smearing and binning of the distributions in \refta{tab:Z500}, 
where we also include the size of the bins given by our prescription for the one-loop and tree-level distributions.

As stated above, here we concentrate on the results for $\sqrt{s} = 500 \gev$. The
results for a center-of-mass energy of 1~TeV (\ie the cross section distributions for the considered BPs and their respective $H$ resonance significance) can be found in \refap{app:1TeV}. Comparing the $500 \gev$ and the $1000 \gev$ results, one can see that the latter
do not provide any further qualitative information \wrt the 500~GeV case. 
This is to be expected, since our BPs have been chosen such that an $H$ resonance peak can be observed at $\sqrt{s}=500\gev$
and the $Zhh$ cross section decreases with $1/s$. 
Nevertheless, it should be mentioned that a collider operating at $\sqrt{s}=1\tev$  would be of great importance in both cases where 
we observe an $H$ resonance or not.
In the former case, the 1~TeV machine can help to reduce the statistical and systematic uncertainties in the measured properties of the 
new scalar, and in the latter case higher center-of-mass energies provide potential access to heavier states.
Another advantages of a 1~TeV collider is given by the fact that it would  have access to the vector boson fusion production 
channel, which is not considered in the present work
(for more details on this see \eg \citere{Arco:2021bvf} and references therein).

We start the discussion of our results with the distributions for BP1 shown in \reffi{fig:BP1-500}.
We can see an enhancement by a factor of 5.6 in the cross section $\sigeff$ with one-loop corrected THCs compared to the tree-level prediction $\sigtree$.
This is partly due to the large value of $\kalaone=4.7$ at one loop, which increases the non-resonant contributions to the cross section, similar to the case of BPal as described in \refse{sec:lahhh}. 
We now turn to the effect of the one-loop corrections of $\lahhHone$, which enters the cross section through a resonant diagram mediated by the heavy Higgs boson $H$.
In BP1, the resonance peak is expected at $\minv = \MH = 300\gev$, as it can be seen in all plots in \reffi{fig:BP1-500}.
Furthermore, it can be seen in all plots that the $H$~resonance peak is more prominent, \ie it is more separated from the non-resonant ``continuum'' contributions, in the distributions with one-loop THCs compared to ones with the tree-level couplings.
This is due to the fact that for our benchmark point the tree-level prediction for $\lahhH$ is very close to zero, while the one-loop correction increases the value of this coupling to a much larger value of $\lahhHone=0.21$.
This effect can be also quantified by the statistical significance of the differential distributions, $\Zdiffone=28.5$ and $\Zdiffzero=4.2$
in the case of no smearing.
For all considered smearing values and bin sizes, we find that the significance with one-loop couplings is roughly one order of 
magnitude larger than the significance with tree-level couplings.
This implies that the $\lahhH$ resonant peak is potentially accessible after considering the one-loop corrections to $\lahhH$, whereas the 
tree-level prediction would naively suggest that it is inaccessible.  

We can also discuss the effect of the binning size and the smearing on the cross section distributions, and in particular on the significance of the $H$ resonance peak.
We find that the smearing has the greater effect on $\Zdiffone$. 
The significance decreases from $\Zdiffone=28.5$ with no smearing, to $\Zdiffone=17.6$ with a 2\% smearing, to $\Zdiffone=13.1$ with 
a 5\% smearing, to $\Zdiffone=10.1$ with a 10\% smearing.
This loss of sensitivity is also clearly visible in the plots, as the resonant peak becomes less prominent and broader as the smearing
percentage increases.
Binning the unsmeared distribution also reduces the sensitivity to the $H$ resonance,  namely from
$\Zdiffone=28.5$ down to $\Zone=18.1$.
However, the significance is not affected in a relevant way by the binning of the distributions when smearing is considered.
The reason  is that the resonance peak in BP1 is very sharp and narrow, which can only be reconstructed with extremely good 
experimental resolution.
But as soon as the distribution is smeared and the resonance becomes broader, the binning does not reduce the sensitivity to $H$  significantly.

We also find that the smearing and binning of the distributions reduces the potential sensitivity to the sign of $\lahhH$.
The $H$ mediated diagram changes sign exactly at $\minv=\MH$, which results in a change of the interference with the rest of the non-resonant
diagrams as well (see for instance \cite{Heinemeyer:2024hxa,Arco:2022lai}).
Therefore, one can find the so-called dip-peak or peak-dip structures around the resonant peak depending on the sign of the $\lahhH$ coupling.
For BP1, we have a dip-peak structure due to the fact that $\lahhH>0$, as can be clearly seen in the unsmeared differential distribution 
(top left plot in \reffi{fig:BP1-500}).
However, compared to the binned distribution without smearing (top right plot), an increased smearing results in ``losing'' the
dip structure to the left of the resonance, and only an enhancement to the right is visible.
For the 2\% smeared distributions, the differential distribution barely shows the dip-peak structure.
For larger percentages of smearing there are no traces of any complex structure around the $H$~resonance peak.

We turn to the results obtained for BP2, shown in \reffi{fig:BP2-500}. 
For this benchmark point, the  tree-level and one-loop predictions for $\kala$ are both rather close to 1, and thus the non-resonant 
contributions in $\sigeff$, $\sigmom$ and $\sigtree$ are rather close to the SM prediction $\sigSM$. 
On the other hand, the one-loop corrections to $\lahhH$ are sizable for this point.
In BP2 the one-loop corrected coupling is $\lahhHone=0.33$, which is almost twice the tree-level prediction $\lahhHzero=0.18$.
The effect of this change in the value of $\lahhH$ is visible in the resonance peaks of all the plots in \reffi{fig:BP2-500}, 
which are found around $\minv=\MH=350\gev$.
The $H$ resonance peaks are higher in the cross section predictions with the one-loop corrected THCs compared to the tree-level predictions.
This effect can be quantified with the change in the statistical significance.
In the unsmeared differential distributions, we find $\Zdiffone=23.3$, while $\Zdiffzero=18.6$.
As in BP1, the statistical significance of the $H$~resonance peak decreases when we consider smearing in the differentia distributions.
For the largest smearing percentage considered, 10\%, but no binning, we find $\Zdiffone=11.0$ and $\Zdiffzero=7.4$. 
The effect of the binning of the distributions on the significance is similar to that found for BP1.
Binning the unsmeared distributions reduces the sensitivity to the $H$ resonance peak, from $\Zdiffzero=18.6$ to $\Zzero=9.2$ and from $\Zdiffone=23.3$ to $\Zone=16.8$.
On the contrary, when the distributions are smeared, the subsequent binning does not significantly worsen the statistical significance 
of the $H$~peak.
Consequently, in BP2 the one-loop corrections to $\lahhH$ lead to a modest improvement for the potential sensitivity to the 
$H$~resonance peak, and hence to the value of $\lahhH$. 
The unsmeared and unbinned distributions also exhibit a very clear dip-peak structure, 
which depends on the sign of $\lahhH$, as discussed above.
However, as with the distributions for BP1, the smearing of the cross section distributions, and to a lesser extent the binning of them, 
erases any trace of this dip-peak structure, making the experimental access to the sign and size of the $\lahhH$ coupling more challenging.

The  differential distributions for BP3 are shown in \reffi{fig:BP3-500}.
For this point the pseudoscalar boson $A$ is light enough that the decay channel $H\to AA$ is kinematically allowed.
Consequently, the total decay width of $H$ is relatively large, amounting to $\Gacorr_H=16.3\gev$ after considering the one-loop corrected 
coupling $\lambda_{HAA}$ (see \refeq{eq:Gacorr}).
For comparison, in all other BPs $\Gacorr_H$ is always less than 1 GeV.
This large value of $\Gacorr_H$ is the reason why there is no narrow resonant peak-dip structure around  $\minv=\MH=300\gev$, 
in contrast to the other studied points.
The large width of $H$ yields a very broad peak-dip structure that extends approximately from the threshold 
at $\minv=250\gev$ to $\minv\simeq350\gev$. It is furthermore interesting to note that the
distributions in the plots exhibit a peak-dip structure, instead of a dip-peak structure as in the previously discussed BP1 and BP2.
The reason lies in the choice of $\CBA < 0$ in BP3, which changes the sign of the $ZHH$ coupling. Despite $\lahhH>0$ in all BPs, 
the product $g_{HZZ} \times \lahhH$ changes sign \wrt the other BPs, and thus changes the global sign of the $H$-resonance diagram.
The absence of a narrow $H$~peak results in a lower statistical significance compared to the other BPs.
Specifically, in the unsmeared differential distributions we find $\Zdiffone=6.0$ and $\Zdiffzero=2.9$.
The significance is larger after including the one-loop corrected THCs since for this point $\lahhHone=0.44$, while $\lahhHzero=0.24$
(the effect of $\kala$ is negligible as it  it increases solely by roughly 30\% when including the NLO corrections in contrast to an 80\% increase in case of the $\lambda_{hhH}$ coupling), but this enhancement is not sufficient
to reach large values of the statistical significance.
Smearing and binning of the distributions does not play an important role for BP3, since the differential distributions are already very smooth.
For instance, the binning of the unsmeared distribution yields a significance of $\Zone=5.7$, and a smearing of 10\% yields $\Zone=4.6$,
very close to the values of the unbinned distribution. 
For all these reasons, BP3 is a more challenging point to probe at future $e^+e^-$ colliders.

We continue our discussion of the sensitivity to $\lahhH$ via the $H$ resonace diagram with {BPsign}, 
whose predicted invariant mass distributions are shown in \reffi{fig:BPsign-500}.
BPsign was specifically chosen because the one-loop corrections to $\lahhH$ change the sign of this THC.
Specifically, the tree-level prediction is $\lahhHzero=-0.08$, while we at the one-loop level we find $\lahhHone=0.16$.
As discussed earlier, this changes the shape of the cross section around the resonance peak, which is located at $\minv=\MH=350\gev$ for this point.
The tree-level differential distribution without smearing exhibits a peak-dip structure, which changes to a dip-peak structure
once the one-loop corrected THCs are taken into account.
The resonance peak also becomes more prominent as the absolute value of $\lahhHone$ is increased relative to $\lahhHzero$.
This is reflected in the statistical significance in both cases: the distribution with one-loop corrections yields a value of  
$\Zdiffone=13.3$, while the tree-level prediction gives $\Zdiffzero=8.6$.
The dip-peak and peak-dip structures of the $H$~peaks are still visible in the binned distributions without smearing, 
although the significances are reduced to $\Zone=8.6$ and $\Zzero=3.8$.
However, any visible hint from the sign of $\lahhH$ via the shape of the $H$ resonance is lost when we consider the smeared 
distributions, even with the smallest smearing percentage considered of a 2\%. 
Furthermore, the significance is also reduced to roughly half in the differential distributions.
As in the previous points analyzed, once the smearing is included in the distributions, the binning of the distribution does not 
further reduce the statistical significance of the $H$~resonance peak.
This implies that the experimental access to $\lahhH$ will require a high resolution in $\minv$, 
similar to our findings for the previously discussed BPs.
For this point, the one-loop corrections to $\kala$ are also considerable, going from $\kalazero = 1$ to $\kalaone = 5.47$.
Consequently, we observe an overall enhancement in the differential cross sections with one-loop corrected THCs for values of $\minv$ away from the resonance peak around $\minv = \MH = 350 \gev$.
This overall enhancement due to a large value of $\kalaone$ is another factor that can facilitate the experimental access to the 
$H$ resonance peak, and thus to $\lahhH$, since it implies more final $Z+4b$ events.

The  differential distributions in $\minv$ for the last considered point BPext (``BP extreme'') are shown in \reffi{fig:BPext-500}.
The total cross section with one-loop corrected THCs is with 1.8~fb much larger than the tree-level cross section of 0.3~fb. 
Given that in BPext the loop-corrected THCs are $\kalaone=5.81$ and $\lahhHone=0.24$, together with the fact that $\MH=260\gev$ is very close to the threshold,
we find that BPext predicts a total cross section close to the maximum that can be found in the 2HDM for an $e^+e^-$ collider at 500~GeV.
This very light heavy $\cp$-even $H$~boson and the large value of $\lahhHone$ imply a very prominent and large resonance peak at $\minv=\MH$.
Such a large $H$~peak results in a large statistical significance of $\Zdiffone=32.3$ from the differential cross section 
taking into account the THCs at the one-loop level.
Even after binning the distributions one finds a large significance of $\Zone=21.4$. Conversely,
the smearing of the distributions has a very large impact on the statistical significance, since the $H$~resonance peak in this point is very narrow.
For a smearing of 2\%, the significance drops to $\Zdiffone=19.4$, for 5\% to $\Zdiffone=15.5$, and for 10\% to $\Zdiffone=13.5$.
Similar to the other points, the binning of the cross section distributions after the smearing does not have a relevant impact on the values 
obtained for the statistical significance, especially when the smearing is large.
Although the significance is reduced by the smearing, the resonance is so pronounced that the statistical significance is $\Zone=13.5$ 
even in the worst case analyzed~(10\%~smearing).

\begin{table}[t!]
    \centering
    \begin{tabular}{cccccccccc} \hline
        Point & Smearing & Bin$^{\LP1\RP}_{-+}$ & Bin$^{\LP1\RP}_{+-}$ & $\Zdiffone$ & $\Zone$ & Bin$^{\LP0\RP}_{+-}$ & Bin$^{\LP0\RP}_{-+}$ & $\Zdiffzero$ & $\Zzero$ \\ \hline
        \multirow{4}{*}{BP1} & 0\% & 6.0 & 7.6 & 28.5 & 18.1 & 11.4 & 13.3 & 4.2 & 1.4 \\ 
         & 2\% & 6.0 & 7.6 & 17.6 & 16.4 & 11.4 & 13.3 & 1.7 & 1.4 \\ 
         & 5\% & 6.0 & 7.2 & 13.1 & 12.9 & 11.4 & 13.3 & 1.2 & 1.1 \\ 
         & 10\% & 4.8 & 6.4 & 10.1 & 10.1 & 10.8 & 13.3 & 0.8 & 0.8 \\ \hline
        \multirow{4}{*}{BP2} & 0\% & 8.4 & 12.3 & 23.3 & 16.8 & 11.4 & 14.9 & 18.6 & 11.1 \\ 
         & 2\% & 8.4 & 12.3 & 17.3 & 15.6 & 11.4 & 14.9 & 12.2 & 10.1 \\ 
         & 5\% & 8.4 & 12.3 & 13.7 & 13.4 & 11.4 & 14.9 & 9.4 & 9.0 \\ 
         & 10\% & 8.1 & 11.6 & 11.0 & 10.9 & 10.9 & 14.9 & 7.4 & 7.3 \\ \hline
        \multirow{4}{*}{BP3} & 0\% & 10.1 & 15.1 & 6.0 & 5.7 & 10.8 & 13.3 & 2.9 & 2.9 \\ 
         & 2\% & 10.1 & 15.1 & 5.8 & 5.6 & 10.8 & 13.3 & 2.9 & 2.8 \\ 
         & 5\% & 10.1 & 15.1 & 5.4 & 5.3 & 10.8 & 13.3 & 2.6 & 2.6 \\ 
         & 10\% & 10.1 & 14.0 & 4.6 & 4.6 & 10.8 & 14.8 & 2.2 & 2.2 \\ \hline
        \multirow{4}{*}{BPsign} & 0\% & 6.0 & 7.6 & 13.3 & 8.6 & 11.4 & 13.3 & 8.6 & 4.3 \\ 
         & 2\% & 6.0 & 7.6 & 6.5 & 6.1 & 11.4 & 13.3 & 4.2 & 3.8 \\ 
         & 5\% & 4.9 & 6.5 & 4.5 & 4.4 & 10.8 & 13.3 & 3.0 & 2.9 \\ 
         & 10\% & 4.1 & 6.0 & 3.2 & 3.2 & 10.8 & 13.3 & 2.3 & 2.2 \\ \hline
        \multirow{4}{*}{BPext} & 0\% & 6.0 & 7.6 & 32.3 & 21.4 & 11.4 & 11.4 & 22.1 & 14.3 \\ 
         & 2\% & 6.0 & 7.6 & 19.4 & 18.2 & 9.5 & 11.4 & 14.4 & 12.8 \\ 
         & 5\% & 4.9 & 6.0 & 15.5 & 15.3 & 9.5 & 12.1 & 12.0 & 11.8 \\ 
         & 10\% & 3.9 & 5.6 & 13.5 & 13.5 & 9.3 & 12.1 & 10.6 & 10.5 \\ \hline
    \end{tabular}
    \caption{Statistical significance $Z$ for all the benchmark points (BPs) for a center-of-mass energy of $\sqrt{s}=500\gev$. 
    $Z_{\rm diff}$ corresponds to the significance obtained from the differential distributions, while $Z$ is the significance
    considering the corresponding bin size (chosen such that the expected events for the bins within the kinematically allowed 
    area are larger than 2).
    The upper labels ${}^{\LP0\RP}$ and ${}^{\LP1\RP}$ refer to the significance and bin size including  tree-level or one-loop THCs, respectively. 
    The lower labels $_{-+}$ and $_{+-}$ refer to the polarization scenarios with $P_{e^-}=\mp80\%$ and  $P_{e^+}=\pm30\%$, respectively.}
    \label{tab:Z500}
\end{table}

\medskip

The statistical significances of the $H$~resonance peaks for the studied BPs for all considered values of smearing, 
with and without binning, are summarized in \refta{tab:Z500}.
This table also includes the bin size (labeled as ``Bin'')
obtained with the method described in \refse{sec:smearing}, both for the tree-level and 
the one-loop corrected  differential distributions, and for the two considered polarization running scenarios.
For all points considered, the significance obtained from the distributions with one-loop THCs is always larger than that obtained 
with the tree-level predictions.
This is due to two facts: first, our points were chosen such that $|\lahhHone| > |\lahhHzero|$, 
resulting in a more pronounced 
$H$~resonance peak, and second for some points $\kalaone>\kalazero$, which increases the predicted number of final $Z+4b$ events.
Thus, we demonstrated that one-loop corrections to THCs can have a large impact on a phenomenological analyses 
in di-Higgs production.

The values for the statistical significances allow us to also quantify the potential relative sensitivity to the $H$~resonance
peak, which is the main experimental access to the THC $\lahhH$ at $e^+e^-$ colliders.
For all points, the larger significance is always given by the unsmeared and unbinned distributions $\Zdiffone$.
This is to be expected, since a differential distributions without smearing yields the purely theoretical prediction 
for the cross section. In the case that the $H$~peak is narrow, it
can only be observed in these theoretical distributions.
Correspondingly, after binning, the significance always worsens, especially in the unsmeared case and in the 2\% smearing case, 
or at best remains the same, especially when the smearing is already large (5\% or 10\%).
Smearing also degrades the sensitivity to the $H$~resonance peak, because the larger the smearing, the smaller the significance obtained. 
However, it should be noted that once the distributions are smeared, the subsequent binning does not have a large impact on the significance.
Therefore, from our analysis we  conclude that smearing is the limiting factor in the sensitivity to the $H$~resonance peak, 
and thus to the value of $\lahhH$. Binning is only important in the case of unsmeared distributions.

It should be noted that we do not consider any experimental backgrounds in our analysis of the significance to the $H$~resonance.
Therefore, the significance values given in this section are only accurate in the case that it is possible to efficiently subtract 
all the background events from the di-Higgs signal. 
Consequently, our values for the experimental significance to the $H$~peak should be considered as optimistic.
Nevertheless, we have decided to not consider any further experimental cuts for our estimation of the final accessible events 
apart from those discussed in \refse{sec:cuts}, since they are based on a non-resonant search for a $\kala$ signal.
In the case of a resonant search, as in the present study, it is likely that an experimental study follows a different strategy, 
similar to the di-Higgs resonant and non-resonant searches performed by the ATLAS and CMS collaborations at the LHC.
For reference, we show in \refap{app:DuerigCuts} our prediction for the statistical significance for our BPs, but with only 17\% of the 
theoretical $Zb\bar bb\bar b$ events, which corresponds to the result of \citere{Durig:2016jrs} to suppress the signal \vs background in the 
$Zhh$ channel in the SM (but based on non-resonant di-Higgs production).


\section{Conclusions}
\label{sec:conclusions}

In this work we studied the impact of one-loop corrections to triple Higgs couplings (THCs) on di-Higgs production in BSM models 
with extended Higgs sectors at high-energy $e^+e^-$~colliders.
We furthermore explored the experimental sensitivity to THCs and how to access them at $e^+e^-$ colliders.
In particular, we focused on the di-Higgs-strahlung process, which is the dominant production channel of two SM-like Higgs bosons 
for center-of-mass energies between roughly 500~GeV and 1~TeV.
In extended BSM Higgs sectors, the main one-loop corrections to this process are expected to be induced by the involved THCs at the one-loop 
level, due to the large scalar couplings that can be realized in such models~\cite{Kanemura:2002vm,Kanemura:2004mg,Bahl:2023eau}. 

As a theoretical framework to explore the effect of one-loop THCs on di-Higgs production, we used the two-Higgs doublet model (2HDM).
In our study, we identified the lighter of the two CP-even Higgs bosons, $h$,  with the SM-like Higgs boson observed at the LHC with a mass of $\sim 125 \gev$.  
In the 2HDM, the THCs  $\lahhh$ and $\lahhH$ enter in the prediction of the cross section via diagrams mediated by the $h$~and 
the $H$~Higgs boson, respectively.
The size of most of the 2HDM scalar couplings involved in these THCs at the one-loop level, namely $\lahhhone$ and $\lahhHone$, 
are constrained only by the unitarity requirement of the model, which allows potentially large one-loop corrections from the scalar sector. 
We computed the one-loop corrections to these THCs using the Coleman-Weinberg effective potential, 
and we furthermore compared our results to those of a fully diagrammatic computation of the one-loop corrections to $\lahhhone$.
This allowed us to estimate the importance of the finite-momentum effects in the one-loop corrections to $\lahhhone$.

The first part of our analysis evaluated the currently allowed ranges for $\lahhh$ (or alternatively 
$\kala=\lahhh/\lambda_{\mathrm{SM}}^{\LP0\RP}$) and $\lahhH$, both at the tree and at the one-loop level, 
considering all relevant current experimental and theoretical constraints.
We found that one-loop corrections can significantly impact the allowed values of $\kalaone$ and $\lahhHone$ within the 
viable 2HDM parameter space, as summarized in \refta{tab:THCranges}.
Specifically, $\kalazero$ is tightly constrained to be close to~1 (\ie the SM value), due to the necessity to be close 
to the alignment limit. However, one-loop corrections can enhance its prediction up to $\sim 6$ in all 2HDM types, even in the 
alignment limit. For $\lahhHzero$, values around $\pm 1.5$ are allowed at tree level, extending to approximately $\pm 2$ 
including the one-loop corrections. These large corrections arise from strong couplings of $h$ and $H$ to heavy Higgs bosons, 
such as $A$, $H^\pm$, or $H$ itself.

Next, we investigated the potential sensitivity to one-loop corrected THCs at high-energy $e^+e^-$~collider via the double Higgs-strahlung
process, \ie \eeZhh.
We defined six benchmark points (see \refta{tab:BP}) allowed by all current constraints, chosen to exhibit
a variety of interesting phenomenology related to THCs. In particular, the benchmark points illustrate how one-loop corrections 
to THCs can affect the final (absolute and differential) di-Higgs production cross section.
We focused on an $e^+e^-$ collider operating at 500~GeV, close to where the maximum production cross section is found in the SM 
(but we also evaluated results for a center-of-mass energies of 1 TeV in some cases). 
We also took into account the beam polarization at an $e^+e^-$ collider, with $P_{e^-}=\mp 80\%$ and $P_{e^+}=\pm 30\%$, 
which enhances the di-Higgs production cross section \wrt unpolarized beams. 
To disentangle the effects of the THCs involved in the process, we studied the differential  distributions  \wrt the invariant mass of the final pair of Higgs bosons, $\minv$.

The effect of $\kalaone$ enters via a non-resonant diagram mediated by $h$, with maximum sensitivity near the threshold at 
$\minv=2\Mh \simeq 250\gev$. 
We found that large values of $\kalaone$ strongly enhance the production cross section, even in the alignment limit. 
For our benchmark point BPal, which predicts $\kalaone=5.75$, the di-Higgs production cross-section is enhanced by factors of 5.9 (4.8)
at $\sqrt{s}=500$ $(1000) \gev$ relative to the SM.
This is due to the positive interference between the diagram with $\kala$ and the other non-resonant contributions.
Therefore, as $\kalaone\gtrsim1$ in most cases, the di-Higgs cross section in the 2HDM is likely to be enhanced when the 
one-loop corrected THCs are taken into account. 
Current projections for measuring deviations in $\kala$ via di-Higgs production at $e^+e^-$ colliders 
give an accuracy of about 10\% for the values of $\kalaone$ found in our paper~\cite{LinearColliderVision:2025hlt}.
This reinforces di-Higgs production 
as a promising probe of BSM physics, even in the absence of significant deviations in other Higgs-boson couplings.
We furthermore found that the two methods of determining $\kalaone$ (with the effective potential and with a fully diagrammatic approach)
yield similar di-Higgs cross section predictions, both absolute and differential, with a relative difference within 1-5\%.
This confirms the effective potential approach as a good approximation to account for the one-loop corrections to $\kala$ in the $Zhh$ 
cross section.

We also examined the impact of the one-loop corrected THC $\lahhHone$ on the production cross section.
The effect of this coupling enters through a resonant diagram mediated by $H$, which can potentially produce a resonant peak at $\minv=\MH$.
To estimate the potential sensitivity to the $\lahhHone$ via the $H$~resonance peak we considered the main Higgs decay channel to $b\bar b$,
giving a final state $Zhh\to Zb\bar bb\bar b$, incorporating acceptance cuts inspired by $e^+e^-$ experimental analyses and considering 
$b$-tagging efficiencies.
To model experimental uncertainties, we applied Gaussian smearing to theoretical cross section distributions with smearing values 
of~0\%, 2\%, 5\%, and 10\%, where 5\% reflects current expectations for $\minv$ resolution~\cite{Munch:mhhres,LinearColliderVision:2025hlt} (which could potentially improve in the future). 
Additionally, we set bin sizes ensuring at least two events in each bin in the kinematically allowed region ($2\Mh<\minv<\sqrt{s}-\MZ$).
We quantified the sensitivity to the $H$~resonance peak using a likelihood ratio statistical test, which gives the statistical significance 
of the $H$~resonance peak against the no-resonance hypothesis, denoted by $Z$.
For all our benchmark points (except for BPal), the significance of the $H$~resonance peak is enhanced when the one-loop corrections 
to the THCs are considered, since the BPs were chosen such that $\lahhHone>\lahhHzero$.
We found promising significance values for all studied benchmark points.
For BP1, BP2, and BPext, we obtained $\Zone>10$ even under pessimistic smearing conditions (10\%).
For BP3 and BPsign, $\Zone\sim5$ was achieved in all considered scenarios. 
These promising results suggest that high-energy $e^+e^-$ colliders could provide a unique opportunity to probe a BSM $H$ boson 
in the mass range $250 \gev \lsim \MH \lsim 400 \gev$ and its triple coupling to two SM-like Higgs bosons, \lahhH.

Finally, we analyzed the degrading effect that smearing and binning of the cross section distributions have on the sensitivity to the $H$~resonance 
and hence to $\lahhHone$.
Without smearing (which is an unrealistic scenario), the statistical significance of the $H$~resonance peak is significantly reduced 
after the binning of the cross section, \ie compared to the significance values obtained from the differential cross section.  
However, even for small smearing values of 2\%, the degradation of the significance is dominated by the detector resolution to 
determine $\minv$, since the subsequent binning has a minimal effect on the significance.
For larger smearing values of 5\% and 10\%, the detector resolution completely dominates the degradation of the $H$~resonance peak, 
and binning effects become negligible in comparison.
We concluded that the primary limiting factor in accessing $\lahhH$ via the $H$~resonance peak is the finite experimental resolution
in $\minv$.
These experimental uncertainties, but especially smearing, dilute the resonant peak, making it more difficult to distinguish it from the 
continuum background. This effect is especially important for narrow resonances.
In particular, the sensitivity to the sign of the THC $\lahhH$ is also affected by these experimental uncertainties, as we observed in BPsign.
Therefore, at future $e^+e^-$ colliders, a high detector resolution will be crucial to probe $H$~resonances in the di-Higgs channel 
and thus to access the THC $\lahhH$. 

In summary, our work emphasizes the fact that higher-order corrected THCs can significantly modify the di-Higgs production cross section 
in BSM Higgs models, and in many cases they can enhance it even when the alignment limit is imposed.
Therefore, with this work we emphasize that it is crucial to include these higher-order corrections in any phenomenological analysis 
of di-Higgs production.
Moreover, our study highlights the challenge of accessing the $H$~resonance peak and its associated THC $\lahhH$ at $e^+e^-$ colliders. 
We provided an estimate of the experimental precision required to achieve this, underlining the importance of high-resolution detectors in 
future collider experiments.


\subsection*{Acknowledgments}
\begingroup
We would like to thank the authors of \citere{Bahl:2023eau} for their help in using and installing {\tt anyH3/anyBSM} and for fruitful discussions. 
We also thank Jenny List for valuable discussions.
F.A.~acknowledge support by the Deutsche
Forschungsgemeinschaft (DFG, German Research
Foundation) under Germany's Excellence
Strategy -- EXC 2121 ``Quantum Universe'' --
390833306.
The work of F.A.~has also been partially funded
by the Deutsche Forschungsgemeinschaft 
(DFG, German Research Foundation) -- 491245950.
The work of S.H.\ has received financial support from the
PID2022-142545NB-C21 funded by MCIN/AEI/10.13039/501100011033/ FEDER, UE
and in part by the grant IFT Centro de Excelencia Severo Ochoa CEX2020-001007-S
funded by MCIN/AEI/ 10.13039/501100011033. 
The work of M.M.\ is supported by the BMBF-Project 05H24VKB.

\endgroup



\appendix

\section{Enhancement Factors for Polarized Cross Section}
\label{app:pol}

The amplitude of the di-Higgs process $\eeZhh$ can be generically written as
\begin{equation}
    \mathcal{M} = \bar v \gamma_\mu (g_L P_L + g_R P_R) u X^{\mu\nu}\epsilon^\ast_\nu\,,
\end{equation}
where $X^{\mu\nu}$
denotes the part of the diagram attached to the right end of the $Z$ boson propagators in \reffi{fig:diagram}.
If one considers initially polarized electron-positron pairs, the polarized amplitudes can be written as
\begin{equation}
     \mathcal{M}_{LR} = \bar {v}_R \gamma_\mu \left(g_L P_L + g_R P_R\right) u_L X^{\mu\nu}\epsilon^\ast_\nu 
     = g_L \bar {v}_R \gamma_\mu  P_L  u_L X^{\mu\nu}\epsilon^\ast_\nu \, ,
\end{equation}
\begin{equation}
    \mathcal{M}_{RL} = \bar {v}_L \gamma_\mu \left(g_L P_L + g_R P_R\right) u_R X^{\mu\nu}\epsilon^\ast_\nu 
     = g_R \bar {v}_L \gamma_\mu  P_R  u_R X^{\mu\nu}\epsilon^\ast_\nu \,.
\end{equation}
To compute the unpolarized amplitude squared we have
\begin{equation}
    \begin{split}
    |\bar {\mathcal{M}}|^2&=\frac{1}{4}\sum_{\rm spin} {\mathcal{M}} \\ &= \frac{1}{4} \frac{1}{2} \left[ \left(g_L^2+g_R^2\right) {\rm{tr}}\!\left(p_1\gamma_\mu p_2\gamma_\alpha\right) + \left(g_L^2-g_R^2\right) {\rm{tr}}\!\left(p_1\gamma_\mu p_2\gamma_\alpha\gamma^5\right) \right] X^{\mu\nu} \left({X^{\alpha\beta}}\right)^\dagger \epsilon^\ast_\nu \epsilon_\beta\,,
    \end{split}
\end{equation}
where $p_{1,2}$ is the momentum of the positron/electron.
Since $X^{\mu\nu}$ is real and $\sum \epsilon^\ast_\nu \epsilon_\beta$ is symmetric, only the trace without $\gamma^5$ contributes to the amplitude squared.
This can be seen explicitly in the expression for $X^{\mu\nu}$. 
For all diagrams one has
\begin{equation}
    X^{\mu\nu}\propto g^{\mu\nu}\,,
\end{equation}
except for the $A$-mediated ones, where
\begin{equation}
    X^{\mu\nu}\propto \left(p_{h_1}-p_A\right)^\mu \left(p_{h_2}+p_A\right)^\nu + \left(p_{h_1}\leftrightarrow p_{h_2} \right)\,
\end{equation}
so that the combination $X^{\mu\nu} \left({X}^{\alpha\beta}\right)^\dagger \sum_{\rm spin}\epsilon^\ast_\nu \epsilon_\beta $ gives a symmetric tensor under the $\mu$ and $\alpha$ indices.

In the case of the $LR$ polarized amplitude one has
\begin{equation}
    \begin{split}
     \mathcal{M}_{LR} &= \bar {v}_R \gamma_\mu \left(g_L P_L + g_R P_R\right) u_L X^{\mu\nu}\epsilon^\ast_\nu 
     = g_L \bar {v}_R \gamma_\mu  P_L  u_L X^{\mu\nu}\epsilon^\ast_\nu \\  & = g_L \left(\bar {v}_R + \bar {v}_L\right) \gamma_\mu  P_L  \left(u_L + u_R\right) X^{\mu\nu}\epsilon^\ast_\nu \, ,
     \end{split}
\end{equation}
where we used the relation of the chiral projectors. To compute the amplitude squared, 
\begin{equation}
    | \mathcal{M}_{LR} |^2 = \frac{1}{2} g_L^2 \left[ {\rm tr}\!\left(p_1\gamma_\mu p_2\gamma_\alpha\right) + {\rm tr}\!\left(p_1\gamma_\mu p_2\gamma_\alpha\gamma^5\right) \right] X^{\mu\nu} \left({X^{\alpha\beta}}\right)^\dagger \epsilon^\ast_\nu \epsilon_\beta\,, 
\end{equation}
where again only the traces without $\gamma^5$ contribute. 
The result for the $RL$ amplitude is the same but interchanging $g_L$ for $g_R$.
Comparing the expressions for polarized and unpolarized squared amplitudes, one arrives at the expressions in \refeq{eq:xspol}.

\section{Results for \texorpdfstring{\boldmath{$\sqrt{s}=1\tev$}}{sqrt(s) = 1 TeV}}
\label{app:1TeV}

\begin{figure}[p!]
    \includegraphics[width=\textwidth]{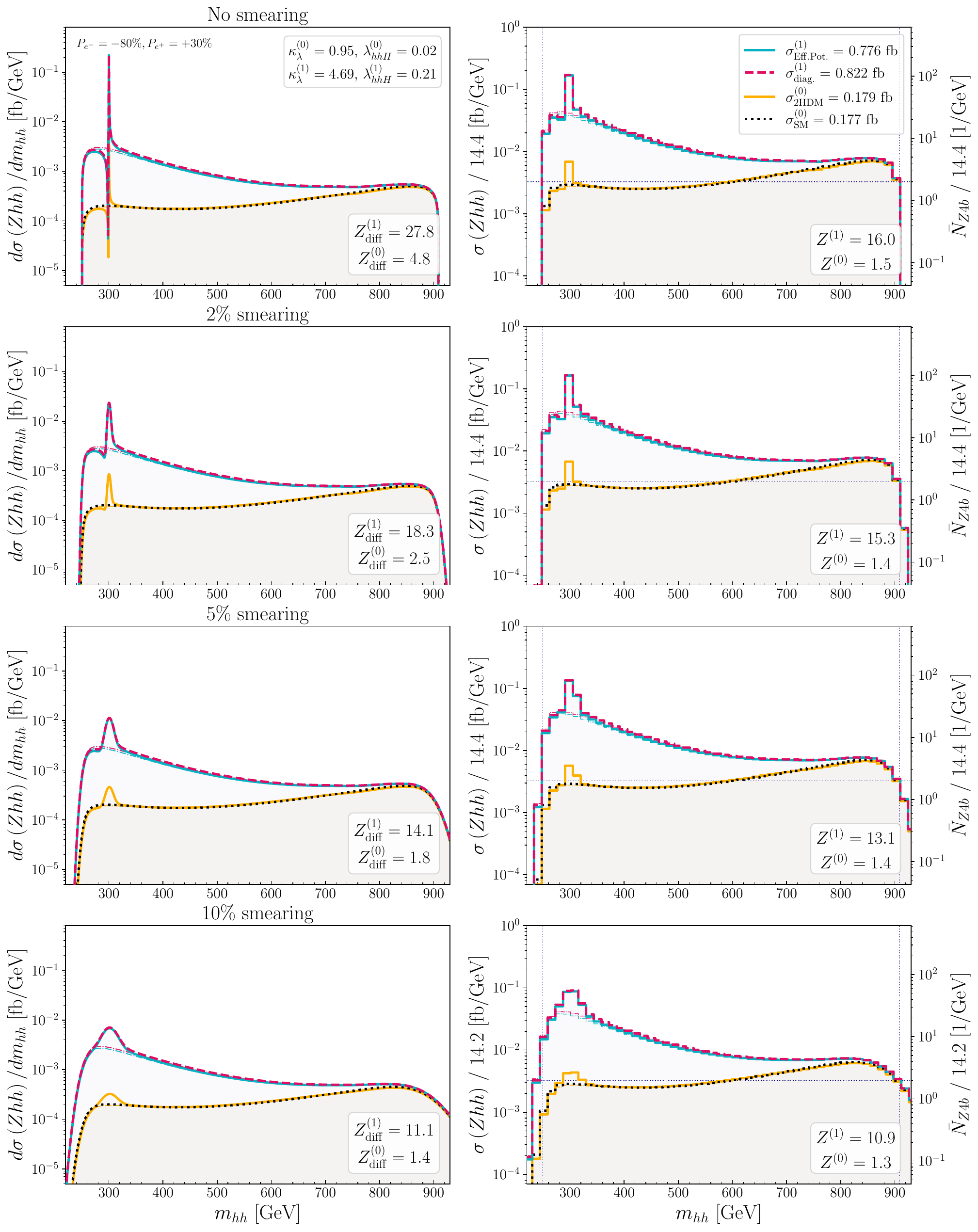}
    \caption{Differential distribution as a function of $\minv$ for BP1 at $\sqrt{s}=1\tev$  
    for $P_{e^-}=-80\%$ and $P_{e^+}=+30\%$. The color coding is the same as in \protect\reffi{fig:BP1-500}.}
    \label{fig:BP1-1000}
\end{figure}

\begin{figure}[p!]
    \includegraphics[width=\textwidth]{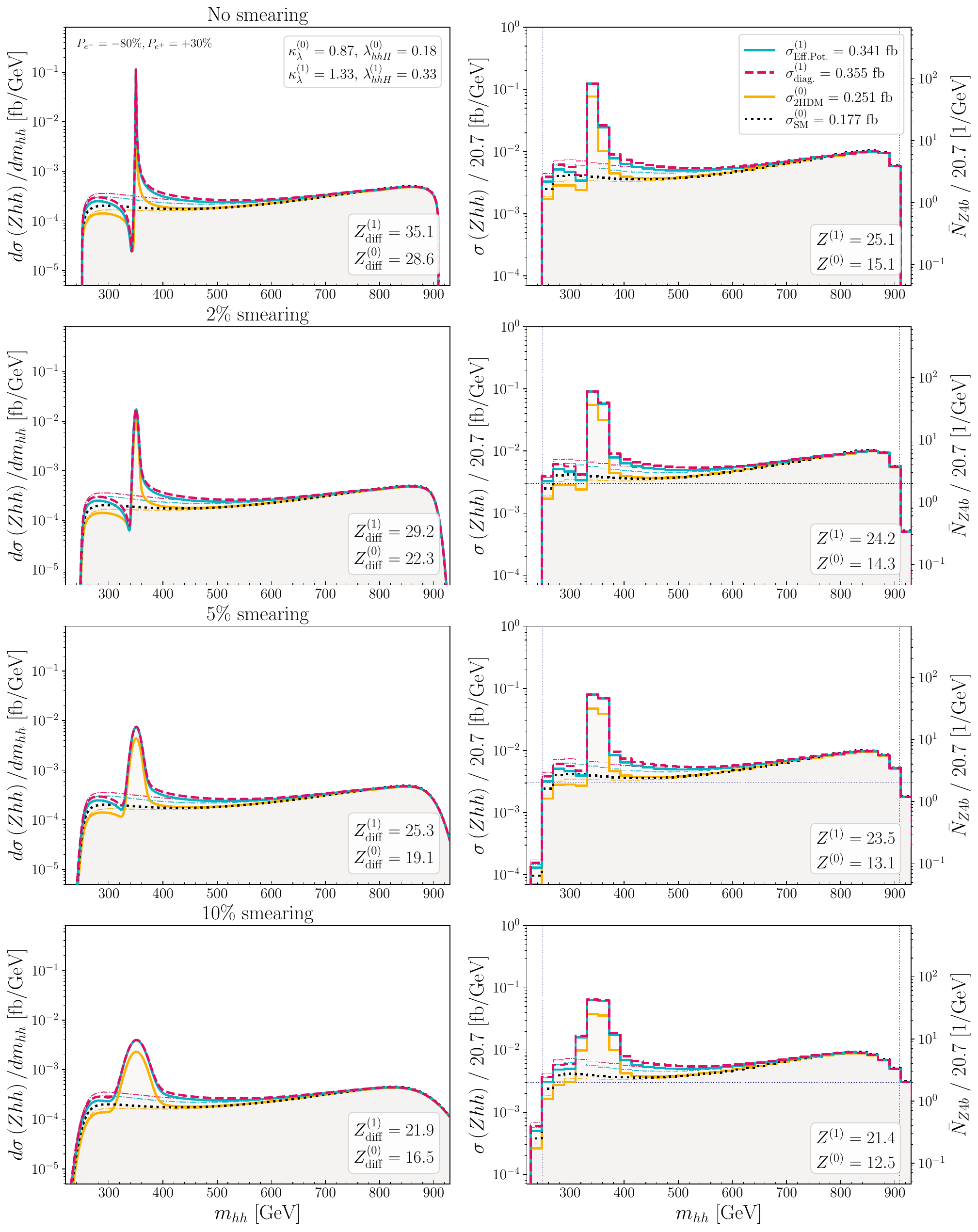}
    \caption{Differential distribution as a function of $\minv$ for BP2 at $\sqrt{s}=1\tev$  
    for $P_{e^-}=-80\%$ and $P_{e^+}=+30\%$. The color coding is the same as in \protect\reffi{fig:BP1-500}.}
    \label{fig:BP2-1000}
\end{figure}

\begin{figure}[p!]
    \includegraphics[width=\textwidth]{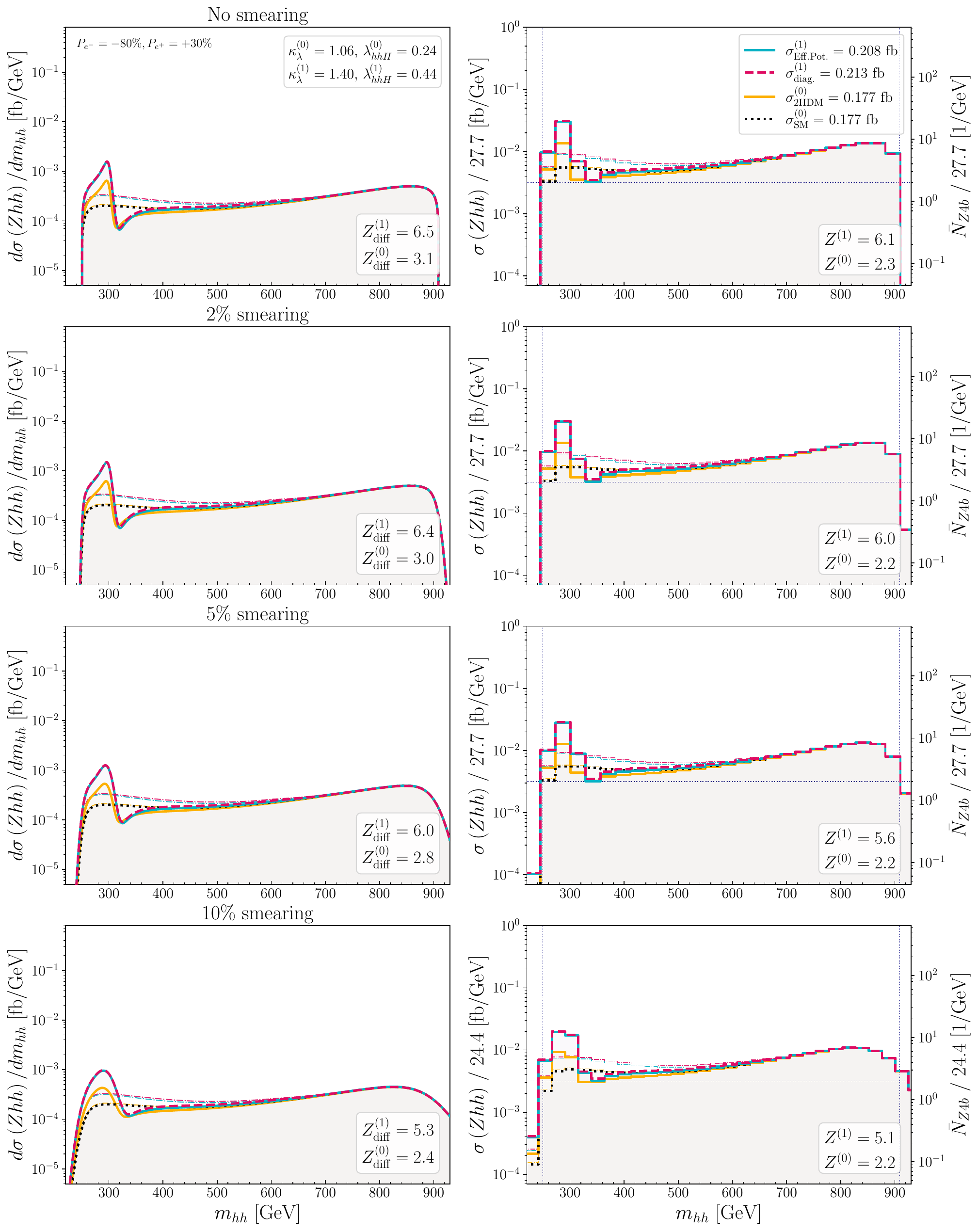}
    \caption{Differential  distribution as a function of $\minv$ for BP3 at $\sqrt{s}=1\tev$  
    for $P_{e^-}=-80\%$ and $P_{e^+}=+30\%$. The color coding is the same as in \protect\reffi{fig:BP1-500}.}
    \label{fig:BP3-1000}
\end{figure}

\begin{figure}[p!]
    \includegraphics[width=\textwidth]{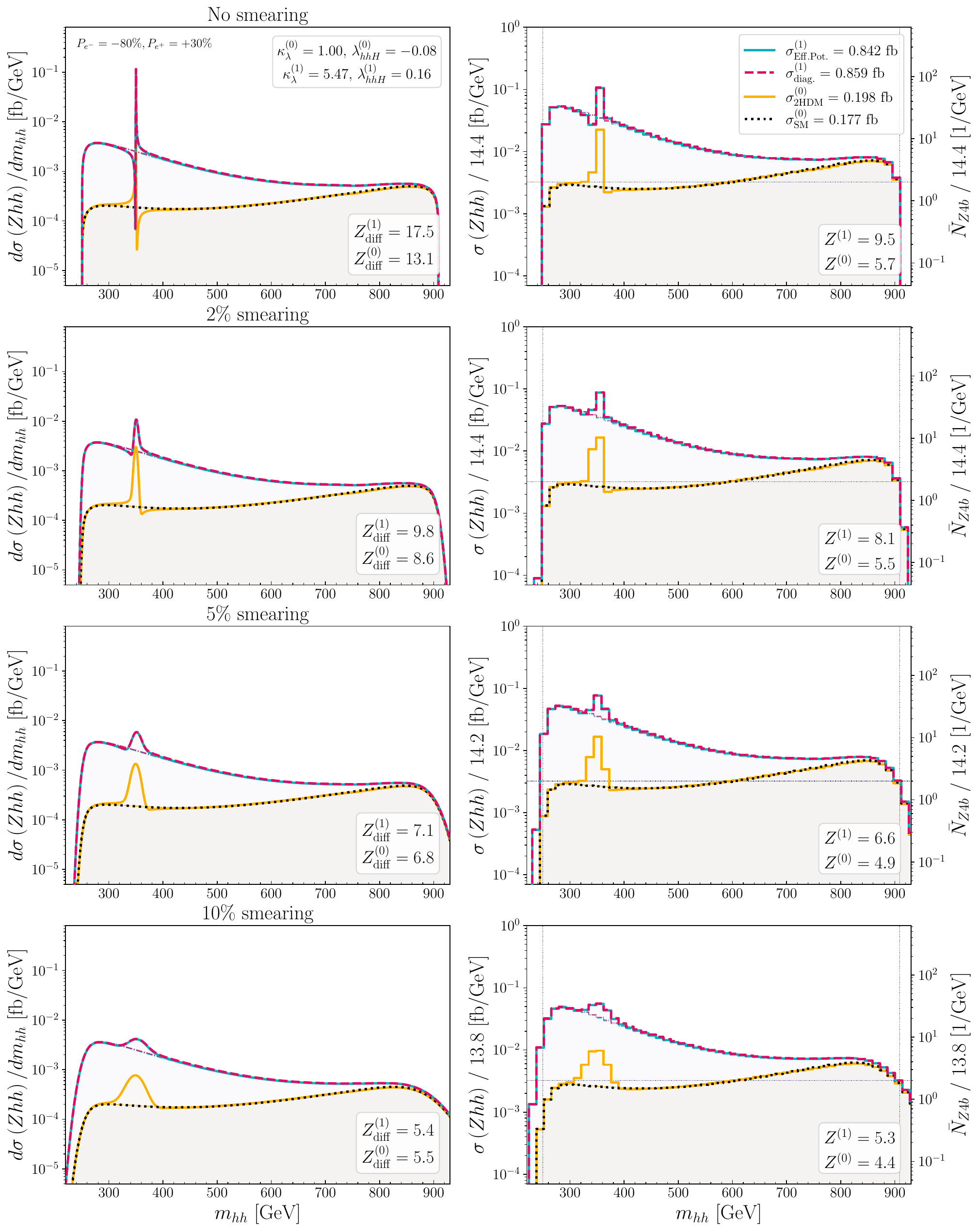}
    \caption{Differential distribution as a function of $\minv$ for BPsign at $\sqrt{s}=1\tev$  f
    or $P_{e^-}=-80\%$ and $P_{e^+}=+30\%$. The color coding is the same as in \protect\reffi{fig:BP1-500}}
    \label{fig:BPsign-1000}
\end{figure}

Here we briefly summarize our results of the sensitivity to $\lahhHone$ from the double Higgs-strahlung process at an $e^+e^-$ collider
operating at a center-of-mass energy of $\sqrt{s}=1\tev$.
The differential distributions for the BPs defined in \refta{tab:BP} can be found in \reffis{fig:BP1-1000} to \ref{fig:BPsign-1000},
where we use the same notation as in the $500 \gev$ analysis, see \refse{sec:lahhH}.
Overall, the obtained values of the total cross sections are smaller than in the  $\sqrt{s}=500\gev$ case, discussed in \refse{sec:lahhH}, 
as expected in the Higgs-strahlung channel.

The effects induced by $\kalaone$ are very similar to the $\sqrt{s}=500\gev$ case.
For the points where $\kalaone$ is large (namely BP1, BP2, BPsign and BPext), an enhancement of the non-resonant contributions 
is found in the differential distributions.
More specifically, this cross section enhancement is more important close to the threshold production.

Regarding the sensitivity to $\lahhHone$, it enters again via the $H$~resonance production.
For the considered BPs, we get smaller statistical significances $Z$ from all the $H$ resonances compared to the 
$\sqrt{s}=500\gev$ case, except for BP2 and BPsign.
The reason for this is that $\MH$ is relatively high, and therefore a larger center-of-mass energy favors the production the $H$~resonance.
Similar to the 500~GeV case, the enhancement from $\kalaone$
implies also more events in the $H$~resonance peak, 
which could overcome the problem of having a small number of events like in the SM.
However, as stated in \refse{sec:lahhH}, a complete analysis for an $e^+e^-$ collider at $\sqrt{s}=1\tev$ should include 
the $WW$ fusion di-Higgs production, $e^+e^- \to \nu\bar\nu \, hh$, which is beyond the scope of this paper.

\begin{figure}[p!]
    \includegraphics[width=\textwidth]{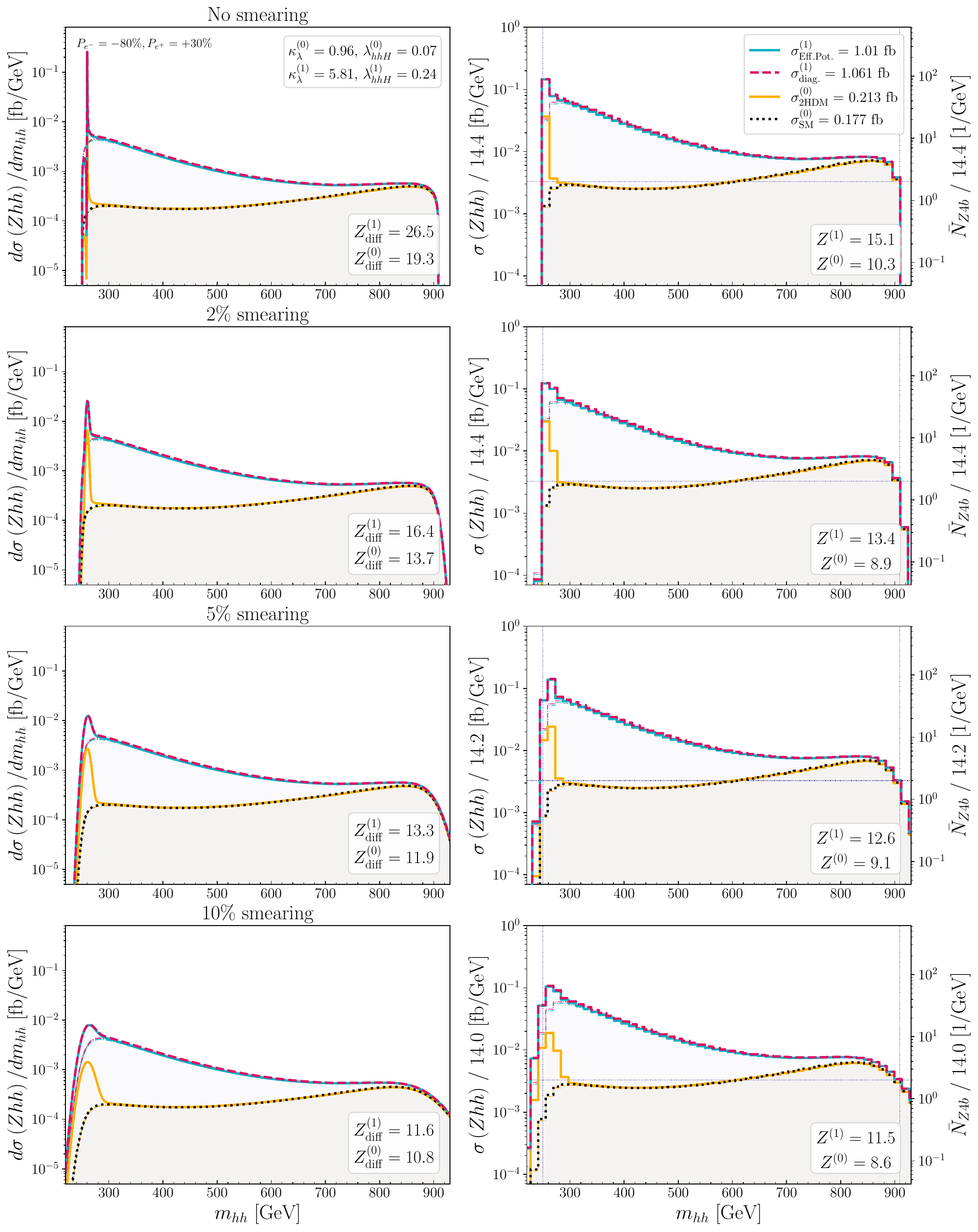}
    \caption{Differential distribution as a function of $\minv$ for BPext at $\sqrt{s}=1\tev$  
    for $P_{e^-}=-80\%$ and $P_{e^+}=+30\%$. The color coding is the same as in \protect\reffi{fig:BP1-500}.}
    \label{fig:BPext-1000}
\end{figure}

Smearing and binning of the distributions has similar effects as in the 500~GeV case.
Smearing is the limiting experimental effect to detect the $H$~resonance peak, as it can be seen in the values of the 
statistical significances~$Z$, shown in the \reffis{fig:BP1-1000} to \ref{fig:BPext-1000}, and also summarized in \refta{tab:Z1000}.
Binning is only important when considering unsmeared distributions, which are not realistic from an experimental point of view.
When smearing is considered, the posterior binning of the distributions does not have a big effect on the obtained values for~$Z$.

\begin{table}[t!]
    \centering
    \begin{tabular}{cccccccccc} \hline
        Point & Smearing & Bin$^{\LP1\RP}_{-+}$ & Bin$^{\LP1\RP}_{+-}$ & $\Zdiffone$ & $\Zone$ & Bin$^{\LP0\RP}_{+-}$ & Bin$^{\LP0\RP}_{-+}$ & $\Zdiffzero$ & $\Zzero$ \\ \hline
        \multirow{4}{*}{BP1} & 0\% & 14.4 & 16.2 & 27.8 & 16.0 & 26.6 & 39.2 & 4.8 & 1.5 \\ 
         & 2\% & 14.4 & 19.8 & 18.3 & 15.3 & 26.6 & 39.2 & 2.5 & 1.4 \\ 
         & 5\% & 14.4 & 19.8 & 14.1 & 13.1 & 26.6 & 39.2 & 1.8 & 1.4 \\ 
         & 10\% & 14.2 & 19.4 & 11.1 & 10.9 & 29.5 & 34.2 & 1.4 & 1.3 \\ \hline
        \multirow{4}{*}{BP2} & 0\% & 20.7 & 27.8 & 35.1 & 25.1 & 33.8 & 39.8 & 28.6 & 18.6 \\ 
         & 2\% & 20.7 & 27.8 & 29.2 & 24.2 & 33.6 & 39.8 & 22.3 & 17.9 \\ 
         & 5\% & 20.7 & 27.8 & 25.3 & 23.5 & 33.6 & 39.9 & 19.1 & 17.0 \\ 
         & 10\% & 20.7 & 27.9 & 21.9 & 21.4 & 33.7 & 46.5 & 16.5 & 15.6 \\ \hline
        \multirow{4}{*}{BP3} & 0\% & 27.7 & 34.5 & 6.5 & 6.1 & 28.6 & 39.3 & 3.1 & 2.3 \\ 
         & 2\% & 27.7 & 34.5 & 6.4 & 6.0 & 28.6 & 39.3 & 3.0 & 2.3 \\ 
         & 5\% & 27.7 & 34.5 & 6.0 & 5.6 & 28.5 & 39.3 & 2.8 & 2.2 \\ 
         & 10\% & 24.4 & 34.6 & 5.3 & 5.1 & 26.6 & 37.1 & 2.4 & 2.3 \\ \hline
        \multirow{4}{*}{BPsign} & 0\% & 14.4 & 16.2 & 17.5 & 9.5 & 23.9 & 33.8 & 13.1 & 6.6 \\ 
         & 2\% & 14.4 & 18.2 & 9.8 & 8.1 & 23.9 & 33.8 & 8.6 & 6.2 \\ 
         & 5\% & 14.2 & 19.4 & 7.1 & 6.6 & 26.5 & 33.8 & 6.8 & 5.6 \\ 
         & 10\% & 13.8 & 18.2 & 5.4 & 5.3 & 26.4 & 33.9 & 5.5 & 5.2 \\ \hline
        \multirow{4}{*}{BPext} & 0\% & 14.4 & 16.2 & 26.5 & 15.1 & 19.6 & 28.1 & 19.3 & 13.0 \\ 
         & 2\% & 14.4 & 18.2 & 16.4 & 13.4 & 19.2 & 27.5 & 13.7 & 10.7 \\ 
         & 5\% & 14.2 & 19.4 & 13.3 & 12.6 & 18.8 & 27.4 & 11.9 & 11.0 \\ 
         & 10\% & 14.0 & 18.2 & 11.6 & 11.5 & 18.6 & 27.3 & 10.8 & 10.5 \\ \hline
    \end{tabular}
    \caption{Statistical significance $Z$ for all the benchmark points (BPs) for a center-of-mass energy of $\sqrt{s}=1\tev$. 
    We use the same notation as in \protect\refta{tab:Z500}.}
    \label{tab:Z1000}
\end{table}



\section{Significance of the  \texorpdfstring{\boldmath{$H$}}{H} Resonance with More Stringent Event Cuts}
\label{app:DuerigCuts}

In this appendix we consider the same selection cuts as applied in \citere{Durig:2016jrs}.
This study, compared to ours, takes into account further cuts in the event selection to further suppress the $Zhh\to Z b\bar bb\bar b$ 
signal versus the SM background.
These extra cuts reduce the final detected number of events to 17\% of the inclusive theoretical prediction, 
compared to the $\sim60\%$ obtained when applying only the preselection cuts as we considered in the analysis in the main text.
However, it should be kept in mind that these cuts were optimized for non-resonant di-Higgs production.

\begin{table}[htb!]
    \centering
    \begin{tabular}{cccccccccc} \hline
        Point & Smearing & Bin$^{\LP1\RP}_{-+}$ & Bin$^{\LP1\RP}_{+-}$ & $\Zdiffone$ & $\Zone$ & Bin$^{\LP0\RP}_{+-}$ & Bin$^{\LP0\RP}_{-+}$ & $\Zdiffzero$ & $\Zzero$ \\ \hline
        \multirow{4}{*}{BP1} & 0\% & 15.1 & 18.6 & 14.8 & 7.8 & 26.6 & 39.9 & 2.2 & 0.5 \\ 
         & 2\% & 15.1 & 18.6 & 9.2 & 7.0 & 26.6 & 39.9 & 0.9 & 0.4 \\ 
         & 5\% & 14.0 & 18.6 & 6.8 & 6.2 & 26.6 & 39.9 & 0.6 & 0.4 \\ 
         & 10\% & 13.1 & 18.6 & 5.3 & 5.1 & 26.6 & 39.9 & 0.4 & 0.4 \\ \hline
        \multirow{4}{*}{BP2} & 0\% & 22.8 & 35.3 & 12.2 & 7.6 & 26.6 & 39.9 & 9.7 & 4.5 \\ 
         & 2\% & 22.8 & 35.3 & 9.0 & 7.4 & 26.6 & 39.9 & 6.4 & 4.4 \\ 
         & 5\% & 22.8 & 35.3 & 7.2 & 6.5 & 26.6 & 39.9 & 4.9 & 4.2 \\ 
         & 10\% & 22.8 & 35.2 & 5.7 & 5.5 & 26.6 & 39.9 & 3.9 & 3.7 \\ \hline
        \multirow{4}{*}{BP3} & 0\% & 22.8 & 34.7 & 3.1 & 2.7 & 26.6 & 33.3 & 1.5 & 1.4 \\ 
         & 2\% & 27.6 & 34.7 & 3.0 & 2.9 & 26.6 & 33.3 & 1.5 & 1.3 \\ 
         & 5\% & 27.6 & 34.7 & 2.8 & 2.7 & 26.6 & 33.3 & 1.4 & 1.3 \\ 
         & 10\% & 27.6 & 34.7 & 2.4 & 2.4 & 26.6 & 39.9 & 1.1 & 1.1 \\ \hline
        \multirow{4}{*}{BPsign} & 0\% & 12.3 & 16.3 & 6.9 & 3.4 & 26.6 & 33.6 & 4.5 & 1.5 \\ 
         & 2\% & 12.3 & 16.3 & 3.4 & 2.9 & 26.6 & 33.6 & 2.2 & 1.5 \\ 
         & 5\% & 12.7 & 16.3 & 2.3 & 2.2 & 26.6 & 33.6 & 1.6 & 1.3 \\ 
         & 10\% & 12.3 & 15.7 & 1.7 & 1.7 & 26.6 & 39.9 & 1.2 & 1.1 \\ \hline
        \multirow{4}{*}{BPext} & 0\% & 12.3 & 16.3 & 16.7 & 9.1 & 23.5 & 28.5 & 11.5 & 6.5 \\ 
         & 2\% & 12.3 & 16.3 & 10.1 & 8.3 & 23.5 & 28.5 & 7.4 & 6.5 \\ 
         & 5\% & 12.3 & 16.3 & 8.0 & 7.6 & 23.5 & 33.3 & 6.2 & 5.9 \\ 
         & 10\% & 11.6 & 15.7 & 7.0 & 6.9 & 25.0 & 33.3 & 5.5 & 5.2 \\ \hline
    \end{tabular}
    \caption{Statistical significance $Z$ for all the benchmark points (BPs) for a center-of-mass energy of $\sqrt{s}=500\gev$. Here we consider
    that only 17\% of the theoretically calculated $Zb\bar bb\bar b$ events enter our evaluation. This corresponds to the result of \citere{Durig:2016jrs} 
    to suppress the signal versus background in the $Zhh$ channel.
    We use the same notation as in \protect\refta{tab:Z500}.} 
    \label{tab:Z500-durig}
\end{table}

The results for the significances for the $500 \gev$ case are summarized in \refta{tab:Z500-durig}.
Due to the smaller numbers of events that we can reconstruct compared to the previous analysis, 
now binning plays also a dominant role in the degradation of the obtained values of~$Z$.
With these extra cuts, the bin size gets very large compared to the values shown in \ref{tab:Z500}.
For the considered points, the bin sizes obtained with these more stringent cuts are over a factor of two larger 
compared with the analysis presented in the main text.
Therefore, as expected, the sensitivity to $\lahhH$ would worsen with these more stringent cuts.
However, it should be noted that even with this low number of events, after considering the one-loop corrections to $\lahhH$ 
the values for the significance $Z$ are above 5 for the points BP1, BP2 and BPext for smearing values smaller or equal to 5\%.
This emphasizes again the relevance that one-loop corrections can have in a phenomenological analysis of the $Zhh$ signal in BSM models.

\begin{table}[ht!]
    \centering
    \begin{tabular}{cccccccccc} \hline
        Point & Smearing & Bin$^{\LP1\RP}_{-+}$ & Bin$^{\LP1\RP}_{+-}$ & $\Zdiffone$ & $\Zone$ & Bin$^{\LP0\RP}_{+-}$ & Bin$^{\LP0\RP}_{-+}$ & $\Zdiffzero$ & $\Zzero$ \\ \hline
        \multirow{4}{*}{BP1} & 0\% & 29.5 & 40.0 & 15.5 & 7.0 & 75.0 & 102.5 & 2.7 & 0.7 \\ 
         & 2\% & 30.5 & 40.0 & 10.2 & 6.0 & 75.0 & 102.5 & 1.4 & 0.7 \\ 
         & 5\% & 34.0 & 46.1 & 7.8 & 5.6 & 75.0 & 102.5 & 1.0 & 0.6 \\ 
         & 10\% & 35.3 & 48.6 & 6.2 & 5.7 & 73.1 & 103.0 & 0.8 & 0.6 \\ \hline
        \multirow{4}{*}{BP2} & 0\% & 52.4 & 89.1 & 18.9 & 10.6 & 112.4 & 113.2 & 15.4 & 7.5 \\ 
         & 2\% & 52.4 & 89.2 & 15.7 & 10.4 & 112.4 & 112.4 & 12.0 & 6.9 \\ 
         & 5\% & 52.4 & 89.3 & 13.6 & 10.5 & 107.3 & 112.4 & 10.3 & 7.2 \\ 
         & 10\% & 52.7 & 68.2 & 11.8 & 10.6 & 77.7 & 106.2 & 8.9 & 7.8 \\ \hline
        \multirow{4}{*}{BP3} & 0\% & 78.2 & 109.1 & 3.4 & 1.7 & 92.8 & 114.1 & 1.6 & 0.8 \\ 
         & 2\% & 78.2 & 109.1 & 3.3 & 1.7 & 92.8 & 114.2 & 1.6 & 0.8 \\ 
         & 5\% & 78.2 & 108.8 & 3.1 & 1.7 & 92.6 & 114.2 & 1.5 & 0.8 \\ 
         & 10\% & 75.2 & 108.1 & 2.8 & 1.8 & 92.4 & 114.2 & 1.3 & 0.8 \\ \hline
        \multirow{4}{*}{BPsign} & 0\% & 29.5 & 38.3 & 9.7 & 4.3 & 75.0 & 106.7 & 7.2 & 2.6 \\ 
         & 2\% & 29.5 & 40.0 & 5.4 & 3.1 & 75.0 & 106.7 & 4.7 & 2.6 \\ 
         & 5\% & 32.7 & 46.1 & 3.9 & 2.5 & 75.0 & 106.5 & 3.7 & 2.6 \\ 
         & 10\% & 35.3 & 46.1 & 3.0 & 2.5 & 75.0 & 105.0 & 3.0 & 2.5 \\ \hline
        \multirow{4}{*}{BPext} & 0\% & 29.5 & 40.0 & 14.8 & 8.0 & 64.3 & 89.9 & 10.8 & 6.5 \\ 
         & 2\% & 29.5 & 40.0 & 9.2 & 6.9 & 64.3 & 89.9 & 7.6 & 6.5 \\ 
         & 5\% & 32.7 & 46.1 & 7.4 & 6.2 & 61.7 & 89.9 & 6.7 & 5.4 \\ 
         & 10\% & 35.3 & 46.1 & 6.5 & 6.2 & 60.9 & 86.0 & 6.0 & 5.3 \\ \hline
    \end{tabular}
    \caption{Statistical significance $Z$ for all the benchmark points (BPs) for a center-of-mass energy of $\sqrt{s}=1\tev$. Here we consider
    that only 17\% of the theoretically calculated $Zb\bar bb\bar b$ events enter our evaluation. This corresponds to the result of \citere{Durig:2016jrs} 
    to suppress the signal versus background in the $Zhh$ channel.
    We use the same notation as in \protect\refta{tab:Z500}.}
    \label{tab:Z1000-durig}
\end{table}

\printbibliography

@article{ATLAS:2012yve,
    author = "Aad, Georges and others",
    collaboration = "ATLAS",
    title = "{Observation of a new particle in the search for the Standard Model Higgs boson with the ATLAS detector at the LHC}",
    eprint = "1207.7214",
    archivePrefix = "arXiv",
    primaryClass = "hep-ex",
    reportNumber = "CERN-PH-EP-2012-218",
    doi = "10.1016/j.physletb.2012.08.020",
    journal = "Phys. Lett. B",
    volume = "716",
    pages = "1--29",
    year = "2012"
}

@article{CMS:2012qbp,
    author = "Chatrchyan, Serguei and others",
    collaboration = "CMS",
    title = "{Observation of a New Boson at a Mass of 125 GeV with the CMS Experiment at the LHC}",
    eprint = "1207.7235",
    archivePrefix = "arXiv",
    primaryClass = "hep-ex",
    reportNumber = "CMS-HIG-12-028, CERN-PH-EP-2012-220",
    doi = "10.1016/j.physletb.2012.08.021",
    journal = "Phys. Lett. B",
    volume = "716",
    pages = "30--61",
    year = "2012"
}

@book{Gunion:1989we,
    author = "Gunion, John F. and Haber, Howard E. and Kane, Gordon L. and Dawson, Sally",
    title = "{The Higgs Hunter's Guide}",
    reportNumber = "SCIPP-89/13, UCD-89-4, BNL-41644",
    volume = "80",
    year = "2000"
}

@article{Aoki:2009ha,
    author = "Aoki, Mayumi and Kanemura, Shinya and Tsumura, Koji and Yagyu, Kei",
    title = "{Models of Yukawa interaction in the two Higgs doublet model, and their collider phenomenology}",
    eprint = "0902.4665",
    archivePrefix = "arXiv",
    primaryClass = "hep-ph",
    reportNumber = "TU-839, UT-HET-022, IC-2009-007",
    doi = "10.1103/PhysRevD.80.015017",
    journal = "Phys. Rev. D",
    volume = "80",
    pages = "015017",
    year = "2009"
}

@article{Branco:2011iw,
    author = "Branco, G. C. and Ferreira, P. M. and Lavoura, L. and Rebelo, M. N. and Sher, Marc and Silva, Joao P.",
    title = "{Theory and phenomenology of two-Higgs-doublet models}",
    eprint = "1106.0034",
    archivePrefix = "arXiv",
    primaryClass = "hep-ph",
    doi = "10.1016/j.physrep.2012.02.002",
    journal = "Phys. Rept.",
    volume = "516",
    pages = "1--102",
    year = "2012"
}

@article{Arco:2021bvf,
    author = "Arco, F. and Heinemeyer, S. and Herrero, M. J.",
    title = "{Sensitivity to triple Higgs couplings via di-Higgs production in the 2HDM at $e^+e^-$ colliders}",
    eprint = "2106.11105",
    archivePrefix = "arXiv",
    primaryClass = "hep-ph",
    reportNumber = "IFT-UAM/CSIC-21-020",
    doi = "10.1140/epjc/s10052-021-09665-w",
    journal = "Eur. Phys. J. C",
    volume = "81",
    number = "10",
    pages = "913",
    year = "2021"
}

@article{Basler:2018cwe,
    author = {Basler, Philipp and M\"uhlleitner, Margarete},
    title = "{BSMPT (Beyond the Standard Model Phase Transitions): A tool for the electroweak phase transition in extended Higgs sectors}",
    eprint = "1803.02846",
    archivePrefix = "arXiv",
    primaryClass = "hep-ph",
    doi = "10.1016/j.cpc.2018.11.006",
    journal = "Comput. Phys. Commun.",
    volume = "237",
    pages = "62--85",
    year = "2019"
}

@article{Basler:2020nrq,
    author = {Basler, Philipp and M\"uhlleitner, Margarete and M\"uller, Jonas},
    title = "{BSMPT v2 a tool for the electroweak phase transition and the baryon asymmetry of the universe in extended Higgs Sectors}",
    eprint = "2007.01725",
    archivePrefix = "arXiv",
    primaryClass = "hep-ph",
    doi = "10.1016/j.cpc.2021.108124",
    journal = "Comput. Phys. Commun.",
    volume = "269",
    pages = "108124",
    year = "2021"
}

@article{Basler:2024aaf,
    author = {Basler, Philipp and Biermann, Lisa and M\"uhlleitner, Margarete and M\"uller, Jonas and Santos, Rui and Viana, Jo\~ao},
    title = "{BSMPT v3 A Tool for Phase Transitions and Primordial Gravitational Waves in Extended Higgs Sectors}",
    eprint = "2404.19037",
    archivePrefix = "arXiv",
    primaryClass = "hep-ph",
    reportNumber = "KA-TP-08-2024",
    month = "4",
    year = "2024"
}

@article{Alwall:2014hca,
    author = "Alwall, J. and Frederix, R. and Frixione, S. and Hirschi, V. and Maltoni, F. and Mattelaer, O. and Shao, H. -S. and Stelzer, T. and Torrielli, P. and Zaro, M.",
    title = "{The automated computation of tree-level and next-to-leading order differential cross sections, and their matching to parton shower simulations}",
    eprint = "1405.0301",
    archivePrefix = "arXiv",
    primaryClass = "hep-ph",
    reportNumber = "CERN-PH-TH-2014-064, CP3-14-18, LPN14-066, MCNET-14-09, ZU-TH-14-14",
    doi = "10.1007/JHEP07(2014)079",
    journal = "JHEP",
    volume = "07",
    pages = "079",
    year = "2014"
}

@article{Bahl:2023eau,
    author = "Bahl, Henning and Braathen, Johannes and Gabelmann, Martin and Weiglein, Georg",
    title = "{anyH3: precise predictions for the trilinear Higgs coupling in the Standard Model and beyond}",
    eprint = "2305.03015",
    archivePrefix = "arXiv",
    primaryClass = "hep-ph",
    doi = "10.1140/epjc/s10052-023-12173-8",
    journal = "Eur. Phys. J. C",
    volume = "83",
    number = "12",
    pages = "1156",
    year = "2023"
}

@article{Bambade:2019fyw,
    author = "Bambade, Philip and others",
    title = "{The International Linear Collider: A Global Project}",
    eprint = "1903.01629",
    archivePrefix = "arXiv",
    primaryClass = "hep-ex",
    reportNumber = "DESY 19-037, DESY-19-037, FERMILAB-FN-1067-PPD, IFIC/19-10, IRFU-19-10,
  JLAB-PHY-19-2854, KEK Preprint 2018-92, JLAB-PHY-19-2854, KEK
  Preprint 2018-92, LAL/RT 19-001, PNNL-SA-142168,
  SLAC-PUB-17412, SLAC-PUB-17412",
    month = "3",
    year = "2019"
}

@article{CLICdp:2018cto,
    author = "Charles, T. K. and others",
    editor = "Burrows, P. N. and Catalan Lasheras, N. and Linssen, L. and Petri\v{c}, M. and Robson, A. and Schulte, D. and Sicking, E. and Stapnes, S.",
    collaboration = "CLICdp, CLIC",
    title = "{The Compact Linear Collider (CLIC) - 2018 Summary Report}",
    eprint = "1812.06018",
    archivePrefix = "arXiv",
    primaryClass = "physics.acc-ph",
    reportNumber = "CERN-2018-005-M, CERN-2018-005",
    doi = "10.23731/CYRM-2018-002",
    volume = "2/2018",
    month = "12",
    year = "2018",
}

@inproceedings{Bai:2021rdg,
    author = "Bai, Mei and others",
    title = "{C$^3$: A ''Cool'' Route to the Higgs Boson and Beyond}",
    booktitle = "{Snowmass 2021}",
    eprint = "2110.15800",
    archivePrefix = "arXiv",
    primaryClass = "hep-ex",
    reportNumber = "SLAC-PUB-17629",
    month = "10",
    year = "2021"
}

@article{Djouadi:2018xqq,
    author = {Djouadi, Abdelhak and Kalinowski, Jan and M\"uhlleitner, Margarete and Spira, Michael},
    collaboration = "HDECAY",
    title = "{HDECAY: Twenty$_{++}$ years after}",
    eprint = "1801.09506",
    archivePrefix = "arXiv",
    primaryClass = "hep-ph",
    reportNumber = "LPT-ORSAY-18-04, CERN-TH-2017-262, LPT-Orsay-18-04, KA-TP-03-2018, PSI-PR-18-02",
    doi = "10.1016/j.cpc.2018.12.010",
    journal = "Comput. Phys. Commun.",
    volume = "238",
    pages = "214--231",
    year = "2019"
}

@article{Djouadi:1997yw,
    author = "Djouadi, A. and Kalinowski, J. and Spira, M.",
    title = "{HDECAY: A Program for Higgs boson decays in the standard model and its supersymmetric extension}",
    eprint = "hep-ph/9704448",
    archivePrefix = "arXiv",
    reportNumber = "DESY-97-079, IFT-96-29, PM-97-04",
    doi = "10.1016/S0010-4655(97)00123-9",
    journal = "Comput. Phys. Commun.",
    volume = "108",
    pages = "56--74",
    year = "1998"
}

@article{ATLAS:2024ish,
    author = "Aad, Georges and others",
    collaboration = "ATLAS",
    title = "{Combination of Searches for Higgs Boson Pair Production in pp Collisions at s=13\,\,TeV with the ATLAS Detector}",
    eprint = "2406.09971",
    archivePrefix = "arXiv",
    primaryClass = "hep-ex",
    reportNumber = "CERN-EP-2024-160",
    doi = "10.1103/PhysRevLett.133.101801",
    journal = "Phys. Rev. Lett.",
    volume = "133",
    number = "10",
    pages = "101801",
    year = "2024"
}

@article{ATLAS:2022vkf,
    author = "Aad, Georges and others",
    collaboration = "ATLAS",
    title = "{A detailed map of Higgs boson interactions by the ATLAS experiment ten years after the discovery}",
    eprint = "2207.00092",
    archivePrefix = "arXiv",
    primaryClass = "hep-ex",
    reportNumber = "CERN-EP-2022-057",
    doi = "10.1038/s41586-022-04893-w",
    journal = "Nature",
    volume = "607",
    number = "7917",
    pages = "52--59",
    year = "2022",
    note = "[Erratum: Nature 612, E24 (2022)]"
}

@article{CMS:2022dwd,
    author = "Tumasyan, Armen and others",
    collaboration = "CMS",
    title = "{A portrait of the Higgs boson by the CMS experiment ten years after the discovery.}",
    eprint = "2207.00043",
    archivePrefix = "arXiv",
    primaryClass = "hep-ex",
    reportNumber = "CMS-HIG-22-001, CERN-EP-2022-039",
    doi = "10.1038/s41586-022-04892-x",
    journal = "Nature",
    volume = "607",
    number = "7917",
    pages = "60--68",
    year = "2022",
    note = "[Erratum: Nature 623, (2023)]"
}

@article{Heinemeyer:2024vqw,
    author = {Heinemeyer, Sven and M\"uhlleitner, Margarete and Radchenko, Kateryna and Weiglein, Georg},
    title = "{Higgs Pair Production and Triple Higgs Couplings at the LHC in the 2HDM framework}",
    reportNumber = "DESY-23-165, IFT\textendash{}UAM/CSIC-23-141, KA-TP-23-2023",
    doi = "10.22323/1.449.0411",
    journal = "PoS",
    volume = "EPS-HEP2023",
    pages = "411",
    year = "2024"
}

@article{Heinemeyer:2024hxa,
    author = {Heinemeyer, S. and M\"uhlleitner, M. and Radchenko, K. and Weiglein, G.},
    title = "{Higgs pair production in the 2HDM: impact of loop corrections to the trilinear Higgs couplings and interference effects on experimental limits}",
    eprint = "2403.14776",
    archivePrefix = "arXiv",
    primaryClass = "hep-ph",
    doi = "10.1140/epjc/s10052-025-14124-x",
    journal = "Eur. Phys. J. C",
    volume = "85",
    number = "4",
    pages = "437",
    year = "2025"
}

@article{Bahl:2022jnx,
    author = "Bahl, Henning and Braathen, Johannes and Weiglein, Georg",
    title = "{New Constraints on Extended Higgs Sectors from the Trilinear Higgs Coupling}",
    eprint = "2202.03453",
    archivePrefix = "arXiv",
    primaryClass = "hep-ph",
    reportNumber = "DESY-22-018, EFI-22-2",
    doi = "10.1103/PhysRevLett.129.231802",
    journal = "Phys. Rev. Lett.",
    volume = "129",
    number = "23",
    pages = "231802",
    year = "2022"
}

@article{Kanemura:2002vm,
    author = "Kanemura, Shinya and Kiyoura, Shingo and Okada, Yasuhiro and Senaha, Eibun and Yuan, C. P.",
    title = "{New physics effect on the Higgs selfcoupling}",
    eprint = "hep-ph/0211308",
    archivePrefix = "arXiv",
    reportNumber = "KEK-TH-856, MSUHEP-21119",
    doi = "10.1016/S0370-2693(03)00268-5",
    journal = "Phys. Lett. B",
    volume = "558",
    pages = "157--164",
    year = "2003"
}

@article{Kanemura:2004mg,
    author = "Kanemura, Shinya and Okada, Yasuhiro and Senaha, Eibun and Yuan, C. -P.",
    title = "{Higgs coupling constants as a probe of new physics}",
    eprint = "hep-ph/0408364",
    archivePrefix = "arXiv",
    doi = "10.1103/PhysRevD.70.115002",
    journal = "Phys. Rev. D",
    volume = "70",
    pages = "115002",
    year = "2004"
}

@article{Braathen:2019pxr,
    author = "Braathen, Johannes and Kanemura, Shinya",
    title = "{On two-loop corrections to the Higgs trilinear coupling in models with extended scalar sectors}",
    eprint = "1903.05417",
    archivePrefix = "arXiv",
    primaryClass = "hep-ph",
    reportNumber = "OU-HET-1001",
    doi = "10.1016/j.physletb.2019.07.021",
    journal = "Phys. Lett. B",
    volume = "796",
    pages = "38--46",
    year = "2019"
}

@article{Braathen:2019zoh,
    author = "Braathen, Johannes and Kanemura, Shinya",
    title = "{Leading two-loop corrections to the Higgs boson self-couplings in models with extended scalar sectors}",
    eprint = "1911.11507",
    archivePrefix = "arXiv",
    primaryClass = "hep-ph",
    reportNumber = "OU-HET-1030",
    doi = "10.1140/epjc/s10052-020-7723-2",
    journal = "Eur. Phys. J. C",
    volume = "80",
    number = "3",
    pages = "227",
    year = "2020"
}

@article{Djouadi:1999gv,
    author = {Djouadi, A. and Kilian, W. and M\"uhlleitner, M. and Zerwas, P. M.},
    title = "{Testing Higgs selfcouplings at e+ e- linear colliders}",
    eprint = "hep-ph/9903229",
    archivePrefix = "arXiv",
    reportNumber = "DESY-99-001, TTP-99-02, PM-99-01",
    doi = "10.1007/s100529900082",
    journal = "Eur. Phys. J. C",
    volume = "10",
    pages = "27--43",
    year = "1999"
}

@phdthesis{Muhlleitner:2000jj,
    author = {M\"uhlleitner, Milada Margarete},
    title = "{Higgs particles in the standard model and supersymmetric theories}",
    eprint = "hep-ph/0008127",
    archivePrefix = "arXiv",
    reportNumber = "DESY-THESIS-2000-033",
    school = "Hamburg U.",
    year = "2000"
}

@article{Arco:2020ucn,
    author = "Arco, F. and Heinemeyer, S. and Herrero, M. J.",
    title = "{Exploring sizable triple Higgs couplings in the 2HDM}",
    eprint = "2005.10576",
    archivePrefix = "arXiv",
    primaryClass = "hep-ph",
    reportNumber = "IFT-UAM/CSIC-20-30, FTUAM-20-3",
    doi = "10.1140/epjc/s10052-020-8406-8",
    journal = "Eur. Phys. J. C",
    volume = "80",
    number = "9",
    pages = "884",
    year = "2020"
}

@article{Arco:2022xum,
    author = "Arco, F. and Heinemeyer, S. and Herrero, M. J.",
    title = "{Triple Higgs couplings in the 2HDM: the complete picture}",
    eprint = "2203.12684",
    archivePrefix = "arXiv",
    primaryClass = "hep-ph",
    reportNumber = "IFT--UAM/CSIC-22-032",
    doi = "10.1140/epjc/s10052-022-10485-9",
    journal = "Eur. Phys. J. C",
    volume = "82",
    number = "6",
    pages = "536",
    year = "2022"
}

@article{deBlas:2019rxi,
    author = "de Blas, J. and others",
    title = "{Higgs Boson Studies at Future Particle Colliders}",
    eprint = "1905.03764",
    archivePrefix = "arXiv",
    primaryClass = "hep-ph",
    reportNumber = "DESY-19-079",
    doi = "10.1007/JHEP01(2020)139",
    journal = "JHEP",
    volume = "01",
    pages = "139",
    year = "2020"
}

@article{DiMicco:2019ngk,
    author = "Alison, J. and others",
    editor = "Di Micco, Biagio and Gouzevitch, Maxime and Mazzitelli, Javier and Vernieri, Caterina",
    title = "{Higgs boson potential at colliders: Status and perspectives}",
    eprint = "1910.00012",
    archivePrefix = "arXiv",
    primaryClass = "hep-ph",
    reportNumber = "FERMILAB-CONF-19-468-E-T, LHCXSWG-2019-005",
    doi = "10.1016/j.revip.2020.100045",
    journal = "Rev. Phys.",
    volume = "5",
    pages = "100045",
    year = "2020"
}

@phdthesis{Durig:2016jrs,
    author = {D\"urig, Claude Fabienne},
    title = "{Measuring the Higgs Self-coupling at the International Linear Collider}",
    reportNumber = "DESY-THESIS-2016-027",
    doi = "10.3204/PUBDB-2016-04283",
    school = "Hamburg U.",
    address = "Hamburg",
    year = "2016",
    url = {https://bib-pubdb1.desy.de/record/310520},
}

@article{Roloff:2019crr,
    author = "Roloff, Philipp and Schnoor, Ulrike and Simoniello, Rosa and Xu, Boruo",
    collaboration = "CLICdp",
    title = "{Double Higgs boson production and Higgs self-coupling extraction at CLIC}",
    eprint = "1901.05897",
    archivePrefix = "arXiv",
    primaryClass = "hep-ex",
    reportNumber = "v1: CLICdp-Note-2018-006, v2: CLICdp-Pub-2019-007, CLICdp-Note-2018-006",
    doi = "10.1140/epjc/s10052-020-08567-7",
    journal = "Eur. Phys. J. C",
    volume = "80",
    number = "11",
    pages = "1010",
    year = "2020"
}

@inproceedings{Torndal:2023fky,
    author = "Torndal, Julie Munch and List, Jenny",
    title = "{Higgs self-coupling measurement at the International Linear Collider}",
    booktitle = "{International Workshop on Future Linear Colliders}",
    eprint = "2307.16515",
    archivePrefix = "arXiv",
    primaryClass = "hep-ph",
    month = "7",
    year = "2023"
}

@phdthesis{Weinberg:1973am,
    author = "Weinberg, Erick J.",
    title = "{Radiative corrections as the origin of spontaneous symmetry breaking}",
    eprint = "hep-th/0507214",
    archivePrefix = "arXiv",
    school = "Harvard U.",
    year = "1973"
}

@article{Coleman:1973jx,
    author = "Coleman, Sidney R. and Weinberg, Erick J.",
    title = "{Radiative Corrections as the Origin of Spontaneous Symmetry Breaking}",
    doi = "10.1103/PhysRevD.7.1888",
    journal = "Phys. Rev. D",
    volume = "7",
    pages = "1888--1910",
    year = "1973"
}

@phdthesis{ArcoGarcia:2023zjz,
    author = "Arco, Francisco",
    title = "{Searching for Triple Higgs Couplings: a phenomenological analysis in the Two Higgs Doublet Model}",
    school = "Madrid, Autonoma U.",
    year = "2023",
    url = "http://hdl.handle.net/10486/712636"
}

@article{Muhlleitner:2020wwk,
    author = {M\"uhlleitner, Margarete and Sampaio, Marco O. P. and Santos, Rui and Wittbrodt, Jonas},
    title = "{ScannerS: parameter scans in extended scalar sectors}",
    eprint = "2007.02985",
    archivePrefix = "arXiv",
    primaryClass = "hep-ph",
    reportNumber = "KA-TP-05-2020, LU TP 20-38",
    doi = "10.1140/epjc/s10052-022-10139-w",
    journal = "Eur. Phys. J. C",
    volume = "82",
    number = "3",
    pages = "198",
    year = "2022"
}

@article{Bahl:2022igd,
    author = {Bahl, Henning and Biek\"otter, Thomas and Heinemeyer, Sven and Li, Cheng and Paasch, Steven and Weiglein, Georg and Wittbrodt, Jonas},
    title = "{HiggsTools: BSM scalar phenomenology with new versions of HiggsBounds and HiggsSignals}",
    eprint = "2210.09332",
    archivePrefix = "arXiv",
    primaryClass = "hep-ph",
    doi = "10.1016/j.cpc.2023.108803",
    journal = "Comput. Phys. Commun.",
    volume = "291",
    pages = "108803",
    year = "2023"
}

@article{Cowan:2010js,
    author = "Cowan, Glen and Cranmer, Kyle and Gross, Eilam and Vitells, Ofer",
    title = "{Asymptotic formulae for likelihood-based tests of new physics}",
    eprint = "1007.1727",
    archivePrefix = "arXiv",
    primaryClass = "physics.data-an",
    doi = "10.1140/epjc/s10052-011-1554-0",
    journal = "Eur. Phys. J. C",
    volume = "71",
    pages = "1554",
    year = "2011",
    note = "[Erratum: Eur.Phys.J.C 73, 2501 (2013)]"
}

@article{Moortgat-Pick:2005jsx,
    author = "Moortgat-Pick, G. and others",
    title = "{The Role of polarized positrons and electrons in revealing fundamental interactions at the linear collider}",
    eprint = "hep-ph/0507011",
    archivePrefix = "arXiv",
    reportNumber = "CERN-PH-TH-2005-036, DCPT-04-100, DESY-05-059, FERMILAB-PUB-05-060-T, IPPP-04-50, KEK-2005-16, PRL-TH-05-01, SHEP-05-03, SLAC-PUB-11087",
    doi = "10.1016/j.physrep.2007.12.003",
    journal = "Phys. Rept.",
    volume = "460",
    pages = "131--243",
    year = "2008"
}

@article{Ellis:2016jkw,
    author = "Ellis, Joshua",
    title = "{TikZ-Feynman: Feynman diagrams with TikZ}",
    eprint = "1601.05437",
    archivePrefix = "arXiv",
    primaryClass = "hep-ph",
    doi = "10.1016/j.cpc.2016.08.019",
    journal = "Comput. Phys. Commun.",
    volume = "210",
    pages = "103--123",
    year = "2017"
}

@article{Barklow:2015tja,
    author = "Barklow, T. and Brau, J. and Fujii, K. and Gao, J. and List, J. and Walker, N. and Yokoya, K.",
    title = "{ILC Operating Scenarios}",
    eprint = "1506.07830",
    archivePrefix = "arXiv",
    primaryClass = "hep-ex",
    reportNumber = "ILC-NOTE-2015-068, DESY-15-102, IHEP-AC-2015-002, KEK-PREPRINT --2015-17, SLAC-PUB-16309",
    month = "6",
    year = "2015"
}

@article{Glashow:1976nt,
    author = "Glashow, Sheldon L. and Weinberg, Steven",
    title = "{Natural Conservation Laws for Neutral Currents}",
    reportNumber = "HUTP-76-A158",
    doi = "10.1103/PhysRevD.15.1958",
    journal = "Phys. Rev. D",
    volume = "15",
    pages = "1958",
    year = "1977"
}

@article{Paschos:1976ay,
    author = "Paschos, E. A.",
    title = "{Diagonal Neutral Currents}",
    reportNumber = "BNL-21870",
    doi = "10.1103/PhysRevD.15.1966",
    journal = "Phys. Rev. D",
    volume = "15",
    pages = "1966",
    year = "1977"
}

@article{Gunion:2002zf,
    author = "Gunion, John F. and Haber, Howard E.",
    title = "{The CP conserving two Higgs doublet model: The Approach to the decoupling limit}",
    eprint = "hep-ph/0207010",
    archivePrefix = "arXiv",
    reportNumber = "SCIPP-02-10",
    doi = "10.1103/PhysRevD.67.075019",
    journal = "Phys. Rev. D",
    volume = "67",
    pages = "075019",
    year = "2003"
}

@article{Biekotter:2022kgf,
    author = {Biek\"otter, Thomas and Heinemeyer, Sven and No, Jos\'e Miguel and Olea-Romacho, Mar\'\i{}a Olalla and Weiglein, Georg},
    title = "{The trap in the early Universe: impact on the interplay between gravitational waves and LHC physics in the 2HDM}",
    eprint = "2208.14466",
    archivePrefix = "arXiv",
    primaryClass = "hep-ph",
    reportNumber = "DESY-22-127, IFT--UAM/CSIC--22-015",
    doi = "10.1088/1475-7516/2023/03/031",
    journal = "JCAP",
    volume = "03",
    pages = "031",
    year = "2023"
}

@article{Bertolini:1985ia,
    author = "Bertolini, Stefano",
    title = "{Quantum Effects in a Two Higgs Doublet Model of the Electroweak Interactions}",
    reportNumber = "NYU/TR10/85",
    doi = "10.1016/0550-3213(86)90341-X",
    journal = "Nucl. Phys. B",
    volume = "272",
    pages = "77--98",
    year = "1986"
}

@article{Akeroyd:2000wc,
    author = "Akeroyd, Andrew G. and Arhrib, Abdesslam and Naimi, El-Mokhtar",
    title = "{Note on tree level unitarity in the general two Higgs doublet model}",
    eprint = "hep-ph/0006035",
    archivePrefix = "arXiv",
    reportNumber = "UFR-HEP-00-06, KEK-TH-00-699, KEK-TH-699",
    doi = "10.1016/S0370-2693(00)00962-X",
    journal = "Phys. Lett. B",
    volume = "490",
    pages = "119--124",
    year = "2000"
}

@article{Ginzburg:2005dt,
    author = "Ginzburg, I. F. and Ivanov, I. P.",
    title = "{Tree-level unitarity constraints in the most general 2HDM}",
    eprint = "hep-ph/0508020",
    archivePrefix = "arXiv",
    doi = "10.1103/PhysRevD.72.115010",
    journal = "Phys. Rev. D",
    volume = "72",
    pages = "115010",
    year = "2005"
}

@article{Deshpande:1977rw,
    author = "Deshpande, Nilendra G. and Ma, Ernest",
    title = "{Pattern of Symmetry Breaking with Two Higgs Doublets}",
    reportNumber = "OITS-81",
    doi = "10.1103/PhysRevD.18.2574",
    journal = "Phys. Rev. D",
    volume = "18",
    pages = "2574",
    year = "1978"
}

@article{Barroso:2013awa,
    author = "Barroso, A. and Ferreira, P. M. and Ivanov, I. P. and Santos, Rui",
    title = "{Metastability bounds on the two Higgs doublet model}",
    eprint = "1303.5098",
    archivePrefix = "arXiv",
    primaryClass = "hep-ph",
    doi = "10.1007/JHEP06(2013)045",
    journal = "JHEP",
    volume = "06",
    pages = "045",
    year = "2013"
}

@article{Haller:2018nnx,
    author = {Haller, Johannes and Hoecker, Andreas and Kogler, Roman and M\"onig, Klaus and Peiffer, Thomas and Stelzer, J\"org},
    title = "{Update of the global electroweak fit and constraints on two-Higgs-doublet models}",
    eprint = "1803.01853",
    archivePrefix = "arXiv",
    primaryClass = "hep-ph",
    doi = "10.1140/epjc/s10052-018-6131-3",
    journal = "Eur. Phys. J. C",
    volume = "78",
    number = "8",
    pages = "675",
    year = "2018"
}

@article{Abouabid:2021yvw,
    author = {Abouabid, Hamza and Arhrib, Abdesslam and Azevedo, Duarte and Falaki, Jaouad El and Ferreira, Pedro. M. and M\"uhlleitner, Margarete and Santos, Rui},
    title = "{Benchmarking di-Higgs production in various extended Higgs sector models}",
    eprint = "2112.12515",
    archivePrefix = "arXiv",
    primaryClass = "hep-ph",
    doi = "10.1007/JHEP09(2022)011",
    journal = "JHEP",
    volume = "09",
    pages = "011",
    year = "2022"
}

@article{ParticleDataGroup:2022pth,
    author = "Workman, R. L. and others",
    collaboration = "Particle Data Group",
    title = "{Review of Particle Physics}",
    doi = "10.1093/ptep/ptac097",
    journal = "PTEP",
    volume = "2022",
    pages = "083C01",
    year = "2022"
}

@article{Basler:2017uxn,
    author = {Basler, Philipp and M\"uhlleitner, Margarete and Wittbrodt, Jonas},
    title = "{The CP-Violating 2HDM in Light of a Strong First Order Electroweak Phase Transition and Implications for Higgs Pair Production}",
    eprint = "1711.04097",
    archivePrefix = "arXiv",
    primaryClass = "hep-ph",
    reportNumber = "DESY-17-174, KA-TP-39-2017",
    doi = "10.1007/JHEP03(2018)061",
    journal = "JHEP",
    volume = "03",
    pages = "061",
    year = "2018"
}

@inproceedings{Tian:2013qmi,
    author = "Tian, Junping",
    title = "{Study of Higgs self-coupling at the ILC based on the full detector simulation at \ensuremath{\sqrt{}} s = 500 GeV and \ensuremath{\sqrt{}} s = 1 TeV}",
    booktitle = "{3rd Linear Collider Forum}",
    publisher = "DESY",
    address = "Hamburg",
    pages = "224--247",
    year = "2013"
}

@article{Thomson:2009rp,
    author = "Thomson, M. A.",
    title = "{Particle Flow Calorimetry and the PandoraPFA Algorithm}",
    eprint = "0907.3577",
    archivePrefix = "arXiv",
    primaryClass = "physics.ins-det",
    reportNumber = "CU-HEP-09-11",
    doi = "10.1016/j.nima.2009.09.009",
    journal = "Nucl. Instrum. Meth. A",
    volume = "611",
    pages = "25--40",
    year = "2009"
}

@inproceedings{Durig:2014lfa,
    author = {D\"urig, Claude and Fujii, Keisuke and List, Jenny and Tian, Junping},
    title = "{Model Independent Determination of $HWW$ coupling and Higgs total width at ILC}",
    booktitle = "{International Workshop on Future Linear Colliders}",
    eprint = "1403.7734",
    archivePrefix = "arXiv",
    primaryClass = "hep-ex",
    month = "3",
    year = "2014"
}

@article{Tian:2010np,
    author = "Tian, Junping and Fujii, Keisuke and Gao, Yuanning",
    title = "{Study of Higgs Self-coupling at ILC}",
    eprint = "1008.0921",
    archivePrefix = "arXiv",
    primaryClass = "hep-ex",
    month = "8",
    year = "2010"
}

@article{Yonamine:2010su,
    author = "Yonamine, Ryo and Ikematsu, Katsumasa and Uozumi, Satoru and Fujii, Keisuke",
    title = "{A Study of top-quark Yukawa coupling measurement in $e^+e^- -> t \bar{t} H$ at sqrt(s) = 500 GeV}",
    eprint = "1008.1110",
    archivePrefix = "arXiv",
    primaryClass = "hep-ex",
    month = "8",
    year = "2010"
}

@article{Catani:1991hj,
    author = "Catani, S. and Dokshitzer, Yuri L. and Olsson, M. and Turnock, G. and Webber, B. R.",
    title = "{New clustering algorithm for multi - jet cross-sections in e+ e- annihilation}",
    reportNumber = "CAVENDISH-HEP-91-5",
    doi = "10.1016/0370-2693(91)90196-W",
    journal = "Phys. Lett. B",
    volume = "269",
    pages = "432--438",
    year = "1991"
}

@article{Belanger:2003ya,
    author = "Belanger, G. and Boudjema, F. and Fujimoto, J. and Ishikawa, T. and Kaneko, T. and Kurihara, Y. and Kato, K. and Shimizu, Y.",
    title = "{Full 0(alpha) electroweak corrections to double Higgs strahlung at the linear collider}",
    eprint = "hep-ph/0309010",
    archivePrefix = "arXiv",
    reportNumber = "LAPTH-994, KEK-CP-142",
    doi = "10.1016/j.physletb.2003.09.080",
    journal = "Phys. Lett. B",
    volume = "576",
    pages = "152--164",
    year = "2003"
}

@article{Zhang:2003jy,
    author = "Zhang, Ren-You and Ma, Wen-Gan and Chen, Hui and Sun, Yan-Bin and Hou, Hong-Sheng",
    title = "{Full O(alpha(ew)) electroweak corrections to e+ e- ---\ensuremath{>} H H Z}",
    eprint = "hep-ph/0308203",
    archivePrefix = "arXiv",
    doi = "10.1016/j.physletb.2003.10.040",
    journal = "Phys. Lett. B",
    volume = "578",
    pages = "349--358",
    year = "2004"
}

@article{Abramowicz:2016zbo,
    author = "Abramowicz, H. and others",
    title = "{Higgs physics at the CLIC electron\textendash{}positron linear collider}",
    eprint = "1608.07538",
    archivePrefix = "arXiv",
    primaryClass = "hep-ex",
    reportNumber = "CLICDP-PUB-2016-001",
    doi = "10.1140/epjc/s10052-017-4968-5",
    journal = "Eur. Phys. J. C",
    volume = "77",
    number = "7",
    pages = "475",
    year = "2017"
}

@article{Torndal:2023mmr,
    author = "Torndal, Julie Munch and List, Jenny and Ntounis, Dimitrios and Vernieri, Caterina",
    title = "{Higgs self-coupling measurement at future~$e^+e^-$ colliders}",
    eprint = "2311.16774",
    archivePrefix = "arXiv",
    primaryClass = "hep-ex",
    reportNumber = "DESY-23-168, DESY-23-186",
    doi = "10.22323/1.449.0406",
    journal = "PoS",
    volume = "EPS-HEP2023",
    pages = "406",
    year = "2024"
}

@article{DiVita:2017vrr,
    author = "Di Vita, Stefano and Durieux, Gauthier and Grojean, Christophe and Gu, Jiayin and Liu, Zhen and Panico, Giuliano and Riembau, Marc and Vantalon, Thibaud",
    title = "{A global view on the Higgs self-coupling at lepton colliders}",
    eprint = "1711.03978",
    archivePrefix = "arXiv",
    primaryClass = "hep-ph",
    reportNumber = "DESY-17-131, FERMILAB-PUB-17-462-T",
    doi = "10.1007/JHEP02(2018)178",
    journal = "JHEP",
    volume = "02",
    pages = "178",
    year = "2018"
}

@article{Arco:2022lai,
    author = {Arco, F. and Heinemeyer, S. and M\"uhlleitner, M. and Radchenko, K.},
    title = "{Sensitivity to triple Higgs couplings via di-Higgs production in the 2HDM at the (HL-)LHC}",
    eprint = "2212.11242",
    archivePrefix = "arXiv",
    primaryClass = "hep-ph",
    reportNumber = "DESY-22-203, IFT-UAM/CSIC-22-073, KA-TP-30-2022",
    doi = "10.1140/epjc/s10052-023-12193-4",
    journal = "Eur. Phys. J. C",
    volume = "83",
    number = "11",
    pages = "1019",
    year = "2023"
}

@inproceedings{Munch:mhhres,
  author    = {Munch Torndal, Julie},
  title     = "{Neutrino Correction in ZHH events}",
  booktitle = "ILD Analysis/Software Meeting",
  year      = 2024,
  url       = "https://agenda.linearcollider.org/event/10545/contributions/55901/attachments/40107/63560/ILDmeeting_NuCorrection.pdf"
}

@inproceedings{Munch:effic,
  author    = {Munch Torndal, Julie},
  title     = "{ILC capabilities for the measurement of double Higgs production at 500 GeV}",
  booktitle = "IDT-WG3-Phys Open Meeting",
  year      = 2023,
  url       = "https://agenda.linearcollider.org/event/9881/contributions/51612/attachments/38615/60758/HHatILC500.pdf"
}

@article{Bliewert:2024hed,
    author = "Bliewert, Bryan and Vernieri, Caterina and Ntounis, Dimitris and List, Jenny and Torndal, Julie Munch and Tian, Junping",
    title = "{Towards an update of the ILD ZHH analysis}",
    eprint = "2410.15323",
    archivePrefix = "arXiv",
    primaryClass = "hep-ex",
    reportNumber = "ILD-PHYS-PROC-2024-012, DESY-24-156",
    doi = "10.1051/epjconf/202431501010",
    journal = "EPJ Web Conf.",
    volume = "315",
    pages = "01010",
    year = "2024"
}

@article{Bechtle:2020uwn,
    author = "Bechtle, Philip and Heinemeyer, Sven and Klingl, Tobias and Stefaniak, Tim and Weiglein, Georg and Wittbrodt, Jonas",
    title = "{HiggsSignals-2: Probing new physics with precision Higgs measurements in the LHC 13 TeV era}",
    eprint = "2012.09197",
    archivePrefix = "arXiv",
    primaryClass = "hep-ph",
    reportNumber = "BONN-TH-2020-09, DESY-20-228, DESY 20-228, IFT-UAM/CSIC-20-081, LU TP 20-53",
    doi = "10.1140/epjc/s10052-021-08942-y",
    journal = "Eur. Phys. J. C",
    volume = "81",
    number = "2",
    pages = "145",
    year = "2021"
}

@article{Bechtle:2020pkv,
    author = "Bechtle, Philip and Dercks, Daniel and Heinemeyer, Sven and Klingl, Tobias and Stefaniak, Tim and Weiglein, Georg and Wittbrodt, Jonas",
    title = "{HiggsBounds-5: Testing Higgs Sectors in the LHC 13 TeV Era}",
    eprint = "2006.06007",
    archivePrefix = "arXiv",
    primaryClass = "hep-ph",
    reportNumber = "BONN-TH-2020-03, DESY 20-093, DESY-20-093, IFT-UAM/CSIC-20-072, LU 20-27",
    doi = "10.1140/epjc/s10052-020-08557-9",
    journal = "Eur. Phys. J. C",
    volume = "80",
    number = "12",
    pages = "1211",
    year = "2020"
}

@article{Bechtle:2015pma,
    author = "Bechtle, Philip and Heinemeyer, Sven and Stal, Oscar and Stefaniak, Tim and Weiglein, Georg",
    title = "{Applying Exclusion Likelihoods from LHC Searches to Extended Higgs Sectors}",
    eprint = "1507.06706",
    archivePrefix = "arXiv",
    primaryClass = "hep-ph",
    reportNumber = "BONN-TH-2015-08, DESY-15-093, SCIPP-15-05",
    doi = "10.1140/epjc/s10052-015-3650-z",
    journal = "Eur. Phys. J. C",
    volume = "75",
    number = "9",
    pages = "421",
    year = "2015"
}

@article{Bechtle:2014ewa,
    author = "Bechtle, Philip and Heinemeyer, Sven and St\r{a}l, Oscar and Stefaniak, Tim and Weiglein, Georg",
    title = "{Probing the Standard Model with Higgs signal rates from the Tevatron, the LHC and a future ILC}",
    eprint = "1403.1582",
    archivePrefix = "arXiv",
    primaryClass = "hep-ph",
    reportNumber = "DESY-14-026, BONN-TH-2014-05",
    doi = "10.1007/JHEP11(2014)039",
    journal = "JHEP",
    volume = "11",
    pages = "039",
    year = "2014"
}

@article{Bechtle:2013wla,
    author = "Bechtle, Philip and Brein, Oliver and Heinemeyer, Sven and St\r{a}l, Oscar and Stefaniak, Tim and Weiglein, Georg and Williams, Karina E.",
    title = "{$\mathsf{HiggsBounds}-4$: Improved Tests of Extended Higgs Sectors against Exclusion Bounds from LEP, the Tevatron and the LHC}",
    eprint = "1311.0055",
    archivePrefix = "arXiv",
    primaryClass = "hep-ph",
    reportNumber = "BONN-TH-2013-21, DESY-13-110",
    doi = "10.1140/epjc/s10052-013-2693-2",
    journal = "Eur. Phys. J. C",
    volume = "74",
    number = "3",
    pages = "2693",
    year = "2014"
}

@article{Bechtle:2013xfa,
    author = "Bechtle, Philip and Heinemeyer, Sven and St\r{a}l, Oscar and Stefaniak, Tim and Weiglein, Georg",
    title = "{$HiggsSignals$: Confronting arbitrary Higgs sectors with measurements at the Tevatron and the LHC}",
    eprint = "1305.1933",
    archivePrefix = "arXiv",
    primaryClass = "hep-ph",
    reportNumber = "BONN-TH-2013-07, DESY-13-078",
    doi = "10.1140/epjc/s10052-013-2711-4",
    journal = "Eur. Phys. J. C",
    volume = "74",
    number = "2",
    pages = "2711",
    year = "2014"
}

@article{Bechtle:2011sb,
    author = "Bechtle, Philip and Brein, Oliver and Heinemeyer, Sven and Weiglein, Georg and Williams, Karina E.",
    title = "{HiggsBounds 2.0.0: Confronting Neutral and Charged Higgs Sector Predictions with Exclusion Bounds from LEP and the Tevatron}",
    eprint = "1102.1898",
    archivePrefix = "arXiv",
    primaryClass = "hep-ph",
    reportNumber = "FR-PHENO-2011-002, BONN-TH-2011-02, DESY-11-016",
    doi = "10.1016/j.cpc.2011.07.015",
    journal = "Comput. Phys. Commun.",
    volume = "182",
    pages = "2605--2631",
    year = "2011"
}

@article{Bechtle:2008jh,
    author = "Bechtle, Philip and Brein, Oliver and Heinemeyer, Sven and Weiglein, Georg and Williams, Karina E.",
    title = "{HiggsBounds: Confronting Arbitrary Higgs Sectors with Exclusion Bounds from LEP and the Tevatron}",
    eprint = "0811.4169",
    archivePrefix = "arXiv",
    primaryClass = "hep-ph",
    reportNumber = "DCPT-08-172, IPPP-08-86, BONN-TH-2008-17",
    doi = "10.1016/j.cpc.2009.09.003",
    journal = "Comput. Phys. Commun.",
    volume = "181",
    pages = "138--167",
    year = "2010"
}

@article{CMS:2024awa,
    author = "Hayrapetyan, Aram and others",
    collaboration = "CMS",
    title = "{Constraints on the Higgs boson self-coupling from the combination of single and double Higgs boson production in proton-proton collisions at s=13TeV}",
    eprint = "2407.13554",
    archivePrefix = "arXiv",
    primaryClass = "hep-ex",
    reportNumber = "CMS-HIG-23-006, CERN-EP-2024-145",
    doi = "10.1016/j.physletb.2024.139210",
    journal = "Phys. Lett. B",
    volume = "861",
    pages = "139210",
    year = "2025"
}

@article{CEPCStudyGroup:2018ghi,
    author = "Dong, Mingyi and others",
    editor = "Guimar\~aes da Costa, Jo\~ao Barreiro and others",
    collaboration = "CEPC Study Group",
    title = "{CEPC Conceptual Design Report: Volume 2 - Physics \& Detector}",
    eprint = "1811.10545",
    archivePrefix = "arXiv",
    primaryClass = "hep-ex",
    reportNumber = "IHEP-CEPC-DR-2018-02, IHEP-EP-2018-01, IHEP-TH-2018-01",
    month = "11",
    year = "2018"
}

@article{CEPCStudyGroup:2023quu,
    author = "Abdallah, Waleed and others",
    collaboration = "CEPC Study Group",
    title = "{CEPC Technical Design Report: Accelerator}",
    eprint = "2312.14363",
    archivePrefix = "arXiv",
    primaryClass = "physics.acc-ph",
    reportNumber = "IHEP-CEPC-DR-2023-01, IHEP-AC-2023-01",
    doi = "10.1007/s41605-024-00463-y",
    journal = "Radiat. Detect. Technol. Methods",
    volume = "8",
    number = "1",
    pages = "1--1105",
    year = "2024"
}

@article{FCC:2018evy,
    author = "Abada, A. and others",
    collaboration = "FCC",
    title = "{FCC-ee: The Lepton Collider}: {Future Circular Collider Conceptual Design Report Volume 2}",
    reportNumber = "CERN-ACC-2018-0057",
    doi = "10.1140/epjst/e2019-900045-4",
    journal = "Eur. Phys. J. ST",
    volume = "228",
    number = "2",
    pages = "261--623",
    year = "2019"
}

@article{Ilyin:1995iy,
    author = "Ilyin, V. A. and Pukhov, A. E. and Kurihara, Y. and Shimizu, Y. and Kaneko, T.",
    title = "{Probing the H$^3$ vertex in e+ e-, gamma e and gamma gamma collisions for light and intermediate Higgs bosons}",
    eprint = "hep-ph/9506326",
    archivePrefix = "arXiv",
    reportNumber = "KEK-CP-030, KEK-PREPRINT-95-78, INP-MSU-95-16-380",
    doi = "10.1103/PhysRevD.54.6717",
    journal = "Phys. Rev. D",
    volume = "54",
    pages = "6717--6727",
    year = "1996"
}

@article{Moortgat-Pick:2015lbx,
    author = "Arbey, A. and others",
    editor = "Moortgat-Pick, G. and others",
    title = "{Physics at the e+ e- Linear Collider}",
    eprint = "1504.01726",
    archivePrefix = "arXiv",
    primaryClass = "hep-ph",
    reportNumber = "DESY-14-241, CERN-PH-TH-2015-042, FERMILAB-PUB-15-132-T",
    doi = "10.1140/epjc/s10052-015-3511-9",
    journal = "Eur. Phys. J. C",
    volume = "75",
    number = "8",
    pages = "371",
    year = "2015"
}

@article{Kanemura:2016lkz,
    author = "Kanemura, Shinya and Kikuchi, Mariko and Yagyu, Kei",
    title = "{One-loop corrections to the Higgs self-couplings in the singlet extension}",
    eprint = "1608.01582",
    archivePrefix = "arXiv",
    primaryClass = "hep-ph",
    reportNumber = "UT-HET-116",
    doi = "10.1016/j.nuclphysb.2017.02.004",
    journal = "Nucl. Phys. B",
    volume = "917",
    pages = "154--177",
    year = "2017"
}

@article{Arhrib:2015hoa,
    author = "Arhrib, Abdesslam and Benbrik, Rachid and El Falaki, Jaouad and Jueid, Adil",
    title = "{Radiative corrections to the Triple Higgs Coupling in the Inert Higgs Doublet Model}",
    eprint = "1507.03630",
    archivePrefix = "arXiv",
    primaryClass = "hep-ph",
    doi = "10.1007/JHEP12(2015)007",
    journal = "JHEP",
    volume = "12",
    pages = "007",
    year = "2015"
}

@article{Aoki:2012jj,
    author = "Aoki, Mayumi and Kanemura, Shinya and Kikuchi, Mariko and Yagyu, Kei",
    title = "{Radiative corrections to the Higgs boson couplings in the triplet model}",
    eprint = "1211.6029",
    archivePrefix = "arXiv",
    primaryClass = "hep-ph",
    doi = "10.1103/PhysRevD.87.015012",
    journal = "Phys. Rev. D",
    volume = "87",
    number = "1",
    pages = "015012",
    year = "2013"
}

@article{Chiang:2018xpl,
    author = "Chiang, Cheng-Wei and Kuo, An-Li and Yagyu, Kei",
    title = "{One-loop renormalized Higgs boson vertices in the Georgi-Machacek model}",
    eprint = "1804.02633",
    archivePrefix = "arXiv",
    primaryClass = "hep-ph",
    doi = "10.1103/PhysRevD.98.013008",
    journal = "Phys. Rev. D",
    volume = "98",
    number = "1",
    pages = "013008",
    year = "2018"
}

@article{Kon:2018vmv,
    author = "Kon, Tadashi and Nagura, Takuto and Ueda, Takahiro and Yagyu, Kei",
    title = "{Double Higgs boson production at $e^+e^-$ colliders in the two-Higgs-doublet model}",
    eprint = "1812.09843",
    archivePrefix = "arXiv",
    primaryClass = "hep-ph",
    doi = "10.1103/PhysRevD.99.095027",
    journal = "Phys. Rev. D",
    volume = "99",
    number = "9",
    pages = "095027",
    year = "2019"
}

@article{Sonmez:2018smv,
    author = "Sonmez, Nasuf",
    title = "{Measuring the triple Higgs self-couplings in two Higgs doublet model}",
    eprint = "1806.08963",
    archivePrefix = "arXiv",
    primaryClass = "hep-ph",
    reportNumber = "17-FEN-054",
    doi = "10.1007/JHEP10(2018)083",
    journal = "JHEP",
    volume = "10",
    pages = "083",
    year = "2018"
}

@article{Arhrib:2008jp,
    author = "Arhrib, Abdesslam and Benbrik, Rachid and Chiang, Cheng-Wei",
    title = "{Probing triple Higgs couplings of the Two Higgs Doublet Model at Linear Collider}",
    eprint = "0802.0319",
    archivePrefix = "arXiv",
    primaryClass = "hep-ph",
    doi = "10.1103/PhysRevD.77.115013",
    journal = "Phys. Rev. D",
    volume = "77",
    pages = "115013",
    year = "2008"
}

@article{Lopez-Val:2009xtx,
    author = "Lopez-Val, David and Sola, Joan",
    title = "{Neutral Higgs-pair production at Linear Colliders within the general 2HDM: Quantum effects and triple Higgs boson self-interactions}",
    eprint = "0908.2898",
    archivePrefix = "arXiv",
    primaryClass = "hep-ph",
    doi = "10.1103/PhysRevD.81.033003",
    journal = "Phys. Rev. D",
    volume = "81",
    pages = "033003",
    year = "2010"
}

@article{Asakawa:2010xj,
    author = "Asakawa, Eri and Harada, Daisuke and Kanemura, Shinya and Okada, Yasuhiro and Tsumura, Koji",
    title = "{Higgs boson pair production in new physics models at hadron, lepton, and photon colliders}",
    eprint = "1009.4670",
    archivePrefix = "arXiv",
    primaryClass = "hep-ph",
    reportNumber = "OCHA-PP-305, KEK-TH-1396, UT-HET-041, IC-2010-076",
    doi = "10.1103/PhysRevD.82.115002",
    journal = "Phys. Rev. D",
    volume = "82",
    pages = "115002",
    year = "2010"
}

@article{Ahmed:2021crg,
    author = "Ahmed, Ijaz and Nawaz, Ujala and Khurshid, Taimoor and Qazi, Shamona Fawad",
    title = "{Probing Triple Higgs Self-Coupling and Effect of Beam Polarization in Lepton Colliders}",
    eprint = "2110.03920",
    archivePrefix = "arXiv",
    primaryClass = "hep-ph",
    doi = "10.1155/2022/9735729",
    journal = "Adv. High Energy Phys.",
    volume = "2022",
    pages = "9735729",
    year = "2022"
}

@article{Lee:1973iz,
    author = "Lee, T. D.",
    editor = "Feinberg, G.",
    title = "{A Theory of Spontaneous T Violation}",
    doi = "10.1103/PhysRevD.8.1226",
    journal = "Phys. Rev. D",
    volume = "8",
    pages = "1226--1239",
    year = "1973"
}

@article{Osland:1998hv,
    author = "Osland, P. and Pandita, P. N.",
    title = "{Measuring the trilinear couplings of MSSM neutral Higgs bosons at high-energy e+ e- colliders}",
    eprint = "hep-ph/9806351",
    archivePrefix = "arXiv",
    reportNumber = "CERN-TH-98-189",
    doi = "10.1103/PhysRevD.59.055013",
    journal = "Phys. Rev. D",
    volume = "59",
    pages = "055013",
    year = "1999"
}

@article{Coimbra:2013qq,
    author = "Coimbra, Rita and Sampaio, Marco O. P. and Santos, Rui",
    title = "{ScannerS: Constraining the phase diagram of a complex scalar singlet at the LHC}",
    eprint = "1301.2599",
    archivePrefix = "arXiv",
    primaryClass = "hep-ph",
    doi = "10.1140/epjc/s10052-013-2428-4",
    journal = "Eur. Phys. J. C",
    volume = "73",
    pages = "2428",
    year = "2013"
}

@article{Bernon:2015qea,
    author = "Bernon, J\'er\'emy and Gunion, John F. and Haber, Howard E. and Jiang, Yun and Kraml, Sabine",
    title = "{Scrutinizing the alignment limit in two-Higgs-doublet models: m$_h$=125  GeV}",
    eprint = "1507.00933",
    archivePrefix = "arXiv",
    primaryClass = "hep-ph",
    doi = "10.1103/PhysRevD.92.075004",
    journal = "Phys. Rev. D",
    volume = "92",
    number = "7",
    pages = "075004",
    year = "2015"
}

@article{List:2024ukv,
    author = "List, Jenny and Bliewert, Bryan and Ntounis, Dimitris and Tian, Junping and Vernieri, Caterina and Torndal, Julie Munch",
    title = "{Higgs Self-coupling Strategy at Linear e+e- Colliders}",
    eprint = "2411.01507",
    archivePrefix = "arXiv",
    primaryClass = "hep-ex",
    reportNumber = "ILD-PHYS-PROC-2024-018, DESY-24-159",
    doi = "10.22323/1.476.0079",
    journal = "PoS",
    volume = "ICHEP2024",
    pages = "079",
    year = "2025"
}

@article{Cepeda:2019klc,
    author = "Cepeda, M. and others",
    editor = "Dainese, Andrea and Mangano, Michelangelo and Meyer, Andreas B. and Nisati, Aleandro and Salam, Gavin and Vesterinen, Mika Anton",
    title = "{Report from Working Group 2}: {Higgs Physics at the HL-LHC and HE-LHC}",
    eprint = "1902.00134",
    archivePrefix = "arXiv",
    primaryClass = "hep-ph",
    reportNumber = "CERN-LPCC-2018-04",
    doi = "10.23731/CYRM-2019-007.221",
    journal = "CERN Yellow Rep. Monogr.",
    volume = "7",
    pages = "221--584",
    year = "2019"
}

@article{CMS:2025hfp,
    collaboration = "ATLAS and CMS",
    author = "{ATLAS and CMS Collaborations}",
    title = "{Highlights of the HL-LHC physics projections by ATLAS and CMS}",
    eprint = "2504.00672",
    archivePrefix = "arXiv",
    primaryClass = "hep-ex",
    reportNumber = "ATL-PHYS-PUB-2025-018 CMS-HIG-25-002",
    month = "4",
    year = "2025"
}

@article{LinearCollider:2025lya,
    author = "Balazs, C. and others",
    collaboration = "Linear Collider",
    title = "{The Linear Collider Facility (LCF) at CERN}",
    eprint = "2503.24049",
    archivePrefix = "arXiv",
    primaryClass = "hep-ex",
    reportNumber = "FERMILAB-PUB-25-0239-CSAID",
    month = "3",
    year = "2025"
}

@article{LinearColliderVision:2025hlt,
    author = "Balazs, C. and others",
    collaboration = "Linear Collider Vision",
    title = "{A Linear Collider Vision for the Future of Particle Physics}",
    eprint = "2503.19983",
    archivePrefix = "arXiv",
    primaryClass = "hep-ex",
    reportNumber = "FERMILAB-PUB-25-0216-CSAID-TD",
    month = "3",
    year = "2025"
}

@article{Hermann:2012fc,
    author = "Hermann, Thomas and Misiak, Mikolaj and Steinhauser, Matthias",
    title = "{$\bar{B}\to X_s \gamma$ in the Two Higgs Doublet Model up to Next-to-Next-to-Leading Order in QCD}",
    eprint = "1208.2788",
    archivePrefix = "arXiv",
    primaryClass = "hep-ph",
    reportNumber = "SFB-CPP-12-60, TTP12-29, IFT-5-2012",
    doi = "10.1007/JHEP11(2012)036",
    journal = "JHEP",
    volume = "11",
    pages = "036",
    year = "2012"
}

@article{Misiak:2015xwa,
    author = "Misiak, M. and others",
    title = "{Updated NNLO QCD predictions for the weak radiative B-meson decays}",
    eprint = "1503.01789",
    archivePrefix = "arXiv",
    primaryClass = "hep-ph",
    reportNumber = "TTP15-007, SFB-CPP-14-121, SI-HEP-2015-08, QFET-2015-09, TTK-15-09, IFT-2-2015",
    doi = "10.1103/PhysRevLett.114.221801",
    journal = "Phys. Rev. Lett.",
    volume = "114",
    number = "22",
    pages = "221801",
    year = "2015"
}

@article{Misiak:2017bgg,
    author = "Misiak, Mikolaj and Steinhauser, Matthias",
    title = "{Weak radiative decays of the B meson and bounds on $M_{H^\pm }$ in the Two-Higgs-Doublet Model}",
    eprint = "1702.04571",
    archivePrefix = "arXiv",
    primaryClass = "hep-ph",
    reportNumber = "TTP17-004, IFT-1-2017",
    doi = "10.1140/epjc/s10052-017-4776-y",
    journal = "Eur. Phys. J. C",
    volume = "77",
    number = "3",
    pages = "201",
    year = "2017"
}

@article{Misiak:2020vlo,
    author = "Misiak, M. and Rehman, Abdur and Steinhauser, Matthias",
    title = "{Towards $ \overline{B}\to {X}_s\gamma $ at the NNLO in QCD without interpolation in m$_{c}$}",
    eprint = "2002.01548",
    archivePrefix = "arXiv",
    primaryClass = "hep-ph",
    reportNumber = "TTP20-001, P3H-20-005, IFT-01/2020",
    doi = "10.1007/JHEP06(2020)175",
    journal = "JHEP",
    volume = "06",
    pages = "175",
    year = "2020"
}

\end{document}